\documentclass[]{jfm}	
	\usepackage{graphicx}
	\usepackage{dcolumn}
	\usepackage{bm}
	\usepackage[utf8]{inputenc}
	\usepackage{subfig}
	\usepackage{graphicx}
	\usepackage{ragged2e}
	\usepackage{amsmath}
	\usepackage{mathtools}
	\usepackage{caption}
	\usepackage{color}
	\usepackage[most]{tcolorbox}
	\usepackage{xcolor}
	\usepackage{float}
	\usepackage{bigints}
	\usepackage[export]{adjustbox}
	\usepackage{tabularx}
	\usepackage{epstopdf, epsfig}
	\usepackage[toc,page]{appendix}
	\usepackage{nomencl}
	\usepackage{mathtools,amssymb,lipsum}
	\usepackage{newtxtext}
	\usepackage{newtxmath}
	\usepackage{natbib}
	\usepackage{hyperref}
	\usepackage[normalem]{ulem}
	\hypersetup{
	colorlinks = true,
	urlcolor   = blue,
	citecolor  = blue,
	linkcolor=blue,
	}
	\usepackage{multirow} 
	\usepackage{xcolor}
	
	\usepackage{makecell}

	\usepackage{siunitx}
	\usepackage{soul}

	\newcommand{\RomanNumeralCaps}[1]
	\linenumbers

	\newcommand{\mgt}[1]{\textcolor{black}{#1}}
	\newcommand{\mmgt}[1]{\textcolor{black}{#1}}
	
	\usepackage[normalem]{ulem}   
	\usepackage{xcolor}

	\title{Waves in a shear flow: transition between the KH, Holmboe and Miles instability}
	\author{Anil Kumar\aff{1}, S. Ravichandran\aff{2}\corresp{\email{sravichandran@iitb.ac.in}} \and Ratul Dasgupta\aff{1}\corresp{\email{dasgupta.ratul@gmail.com}}}
	
	\affiliation{\aff{1}Chemical Engineering, Indian Institute of Technology Bombay, India 400 076\aff{2}Centre for Climate Studies, Indian Institute of Technology Bombay, India 400 076}
	
\begin{document}
	\maketitle
	\newcommand{\mj}{{\mathrm{J}}}
	\begin{abstract}
	We study the generation of waves at the interface of two immiscible fluids, arising due to shear-driven, \mgt{inviscid} instabilities. The background profile is chosen to be the classic exponential velocity profile \citep{morland1993effect} with a sharp density interface (stable stratification). At low Froude \mgt{$(\sim O_m(1))$} and high Bond number \mgt{($\sim O_m(100)$)}, relevant to geophysical and astrophysical situations, we demonstrate a novel (\mgt{smooth}) transition for the fastest growing mode: from the Kelvin-Helmholtz (KH) instability at density ratio $\delta= 0.9$ \mgt{(ratio of upper to lower fluid densities)} to the Holmboe (H) instability as $\delta\rightarrow 0.5$ and onwards to the \cite{miles1957generation} critical layer instability as $\delta\rightarrow0.001$ (air-water). Notably, the Miles fastest growing mode with its (characteristic) sharp jump in (inviscid) Reynolds stress ($\tau$) at the critical location, persists up to $\delta =  0.01$ i.e. up to \textit{ten times} the air-water value. The vertical ($z$) variation of the Reynolds stress (fastest mode), displays a qualitative change with increasing $\delta$; transitioning from having a sharp jump at the critical location for $\delta << 1$ (\mgt{Miles}) to smooth variation through this location for $\delta \geq 0.5$ (\mgt{Holmboe}). A theoretical explanation for this is presented. In the \mgt{higher} density ratio regime ($0.5 \leq \delta \leq 0.9$), by comparing against stability results from the corresponding piece-wise linear (PL) velocity profile, we show the possibility of the H and the KH instability in the exponential model.
	
	Simulations, solving the incompressible Euler's equation with gravity and surface-tension are reported. In the Miles regime ($\delta=0.01$) excellent agreement with linear theory upto five wave periods is observed. \mgt{As these waves saturate, tiny surface ripples appear. Increasing $\delta$ to $0.1$, finite-amplitude waves resulting from the instability display, larger-amplitude ripples on their surface; these are reminiscent of Stokes waves with capillary effects.} With further increase ($\delta =0.5$), exponentially growing waves display a sheared cusp at the crest, in the nonlinear regime. They emit spume droplets and resemble the asymmetric Holmboe waves in the \mgt{simulations and} experiments of \cite{lawrence1998search}. At $\delta=0.9$, growing interfacial waves rapidly distort into classic KH spirals within five wave periods. Several comparisons reported between the PL and the exponential profile, clarify the role of background profile curvature and the critical location. To our knowledge, this is the first demonstration of three canonical instabilities, within these background states \mgt{and without usage of the Boussinesq approximation}.
	\end{abstract}
	
	
	\section{Introduction}\label{sec:intro}
	Shear generated surface or internal waves in density stratified flows, particularly at high Reynolds numbers, have occupied a central space in hydrodynamic stability research since the pioneers (\cite{helmholtz1868xliii}, \cite{rayleigh1895stability}, \cite{thomson1871xlvi}, \cite{taylor1931effect} and \cite{goldstein1931stability}). That important outstanding problems persist, attest to the complexity of the subject and highlight its relevance to wide ranging situations of geophysical, astrophysical or industrial interest. Adverse effects of climate change seemingly call for renewed attention into geophysical problems involving surface and internal waves; this is so because processes accompanying these often generate complicated, multiscale phenomena, both in the ocean and the atmosphere with long term climate consequences. Examples include wave breaking (surface or internal gravity waves in the ocean and atmosphere) and associated mixing (\cite{peltier2003mixing}, section $6.2.1$ in \cite{fritts2003gravity} etc.) or ocean spray generation \citep{deike2022mass}. Due to the multi-scale nature, parameterisation of some of these processes in large-scale numerical models of wave or climate is necessary \citep{cavaleri2007wave,janssen2008progress,perlin2013breaking,staquet2002internal} and in turn calls for accurate knowledge of the micro-physics. We refer the reader to figure $9$ in \cite{garrett1979internal} (figure $5$ in \cite{thorpe1975excitation}) for a caricature of some of these processes in the ocean.
	
\begin{figure}
	\centering
	\includegraphics[scale=0.4]{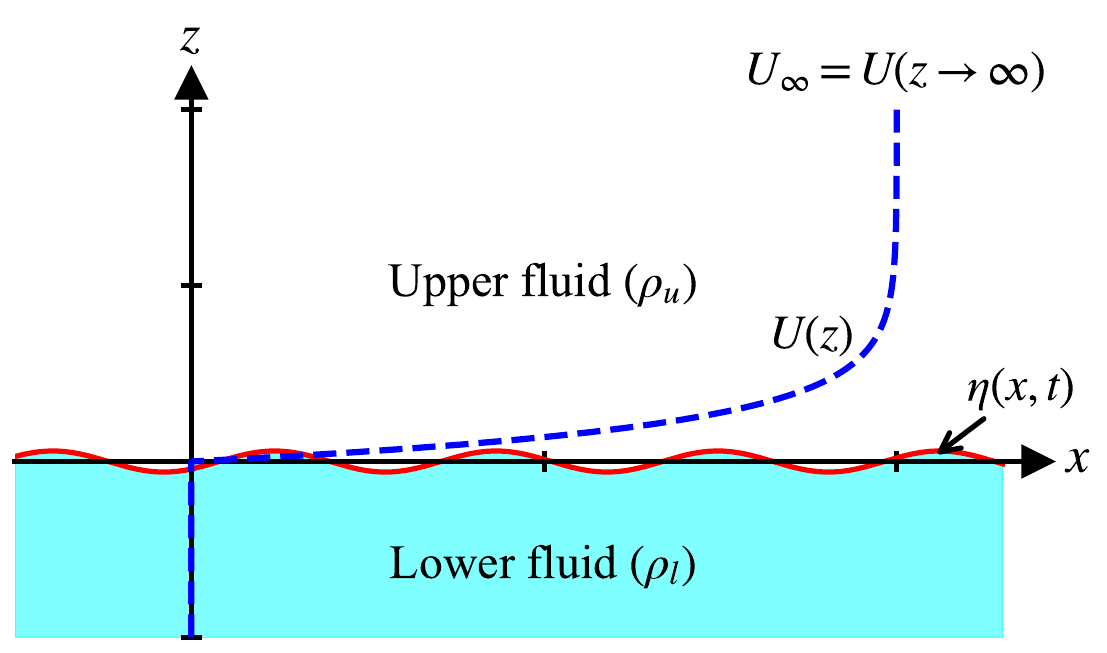}
	\caption{Exponential velocity profile with quiescent fluid below, in the base (background) state with upper fluid velocity varying as $U_u(z)=U_{\infty}\left[1-\exp(-z/\Delta)\right]$. A sharp interface separates two fluids of differing densities in a statically stable configuration: $\rho(z)=\rho_l,\; z< 0$ and $\rho(z)=\rho_u,\; z >0$ with $\rho_l > \rho_u$.}
	\label{fig1}
	\end{figure}

	The mechanism through which an overlying, moving gas or a liquid raises waves at the interface of an underlying, immiscible liquid layer (the common example being wind blowing over the deep ocean) has been of near constant interest since \cite{jeffreys1925formation}; see \citet{pizzo2021does} for a recent summary and \cite{mitsuyasu2002historical} for a historical overview of the wind-wave problem. The configuration also applies to industrial situations involving gas-liquid \citep{matas2011experimental} or liquid-liquid combinations involving thick or thin films \citep{cohen1965generation,boomkamp1996classification} subject to shear. Our interest here is restricted to the regime of geo/astro-physical interest involving two fluid layers (gas-liquid or liquid-liquid depending on the density ratio) with shear in the upper layer and a sharp density jump. Such a configuration is of oceanic interest at low density ratio (air-water) \citep{miles1957generation} and of astrophysical as well as geophysical interest at higher values \citep{alexakis2002shear,pouliquen1994propagating}. In the base-state (also referred to as the `background state') depicted in figure \ref{fig1}, a shear profile is imposed in the upper layer (the exponential velocity profile studied first in \cite{morland1993effect} with a sharp density interface at $z=0$, representative of a quasi-laminar wind profile), with quiescent lower fluid (stable density stratification). We study the linear stability of this configuration analytically and the time-evolution of the most unstable mode computationally, within the inviscid framework, accounting for gravity and surface tension. Our analysis is restricted to unstable waves sufficiently long compared to the shear layer thickness (i.e. the background vorticity layer) as well as relevant capillary and viscous length scales (we return to viscous effects at the end of the study), but sufficiently short such that the deep layer approximation remains applicable. Within these simplifying approximations, we study transition in the nature of the fastest growing mode in the non-dimensional space comprising Froude number ($Fr$) (ratio of inertia to gravity) and density ratio ($\delta$) (upper to lower fluid), at large Bond ($Bo>>1$) number (ratio of gravity to surface tension). The precise definition of these non-dimensional numbers are introduced in \S\ref{sec:lin_stab}.
	
	The key result of our study is the demonstration, at high Bond number and moderately low Froude number, of a novel transition (for the fastest growing mode) from the Kelvin-Helmholtz (KH hereafter) \citep{thomson1871xlvi,helmholtz1868xliii} to the Holmboe (H hereafter) instability \citep{holmboe1962behavior} and eventually to the Miles \citep{miles1957generation} critical layer instability, as the density ratio is reduced from $0.9$ through $0.5$ to the air-water value ($0.001$). While transition from the KH to the H instability \citep{lawrence1991stability,lawrence1998search,smyth1989transition,smyth2003turbulence} for other background profiles or from the Miles to the KH instability \citep{morland1993effect,alexakis2004nonlinear} for the current profile have been reported, we demonstrate here the occurrence of all three aforementioned, \textit{inviscid, shear instabilities} within the single base-state model of figure \ref{fig1}; to our knowledge this is the first such demonstration.
	
	Our analytical results receive support from comparison of the temporal evolution of the interface shape, obtained through solving the nonlinear, inviscid equations of motion. Simulations conducted using the open source code Basilisk \citep{popinet2025basilisk} show very good agreement with theoretical predictions in the linear regime. In the nonlinear regime, we note qualitative differences between the KH spirals (at $\delta=0.9$) and contrast these to sheared, cusp-like interface deformation emitting spume droplets for Holmboe-like ($\delta=0.5$) waves; these numerical observations are consistent with the numerical and experimental observations of \cite{lawrence1998search} distinguishing the KH from the H instability for asymmetric velocity profiles. At $\delta=0.1$, simulations display features in the nonlinear regime, reminiscent of ripples on finite-amplitude, capillary-gravity Stokes waves. Their appearance typically accompanies an abrupt reduction of the parent wave amplitude.  In the rest of this study, we refer to waves appearing for $\delta >> 0.1$, as \textit{internal waves} consistent with experimental literature \citep{mercier2011resurrecting}. The analysis of \cite{yih1960gravity} showed that waves on a surface of density  discontinuity are a limiting case of internal (gravity) waves appearing in continuously, stratified flows, thus supporting this nomenclature. In contrast, at very low density ratios ($<< 0.1$), we describe \textit{surface-waves}.
	\subsection{Literature review}
	In the low density ratio regime (air-water, $\delta=0.001$), linear stability of the base-state of figure \ref{fig1}, has been studied by several authors for a range of Froude and Bond numbers. For example, refer to \cite{morland1993effect,leblanc2007amplification,carpenter2017physical,bonfils2022asymptotic,kadam2023wind} for linear stability studies with distinct perspectives and to \cite{alexakis2004nonlinear,leblanc2007amplification} for weakly nonlinear investigations; these studies are relevant to \textit{wind-driven, water waves}  with $\delta=0.001$. Apparently, situations involving immiscible fluids subject to shear at density ratios significantly higher than the air-water value, also occur commonly. Fresh and salt-water, for example, have density ratio $>0.95$; a sharp interface is maintained on experimental time-scales between the two, see \cite{mercier2011resurrecting}. \cite{pouliquen1994propagating} studied the KH and H instability experimentally using fluids of high density ratio $\sim 0.92$ (water and equal mixture of silicone oil with 1-2-3-4-tetrahydronaphthalene, see their table $1$), comparing their experimental results against inviscid stability theory. That an inviscid theory suffices to describe their experimental results involving somewhat viscous fluids, is justified from the estimate of the viscous diffusion time scale on page $282$ \citep{pouliquen1994propagating} alongwith the Reynolds number (page $293$). Beyond geophysical scales, \cite{alexakis2002shear} report that density ratio of $0.1$ in shear flows occurs in astrophysical situations involving accretion from a normal star to a white dwarf star, see discussion in their pages $1$-$2$.
	
	To our knowledge, the only analytical study of the stability of the exponential profile of figure \ref{fig1} at larger density ratio (compared to air-water), has been carried out by \cite{alexakis2002shear,alexakis2004nonlinear}. These authors studied the moderately low Froude number and large wavenumber limit theoretically, obtaining an analytical expression for the growth-rate (in the linear regime) at arbitrary density ratio; no analysis of the nature of the instability or any transition therein with varying density ratio was reported, as is done here. \cite{alexakis2004weakly}  studied the weakly nonlinear regime for small Froude (the weak wind limit) and small density-ratio, obtaining amplitude equations which were then solved numerically. These solutions validated the prediction \citep{haberman1972critical} that for a growing mode \citep{miles1957generation}, the phase-change across the critical layer expected from linear theory, saturates to zero as the wave amplitude increased with time. The nonlinear simulations by \cite{alexakis2004nonlinear} were however in the moderate Froude number $Fr=2.3-9.5$ and low Mach number regime, at fixed density ratio of $0.1$. Qualitative agreement with linear theory was reported for the growth of gravitational potential energy of unstable Fourier mode(s), although significant scatter is apparent in their comparison, see their figure $4$. At these Froude numbers, the authors reported regimes separating the Miles from the KH instability, although their KH model is similar to the one in figure \ref{fig2b} with a jump at the interface, and not the more general continuous and piecewise linear (PL hereafter) model of figure \ref{fig2a}. We will demonstrate later on that at $\delta=0.1$ (as is the case for \cite{alexakis2004nonlinear}) and low Froude number, the instability in the exponential model of figure \ref{fig1}, begins to lose its signature at the critical location, indicative of a potential change in the instability mechanism.
	
	We note the important study by \cite{morland1993effect}, which explored transitions from the Miles instability to the discontinuous KH instability in the exponential model, at low (air-water) density ratio and for large $Fr$ and infinite Bond. Importantly, they proved the continuation of the Howard semi-circle theorem to shear flow with a sharp density interface; this had been done by \cite{yih1972surface} albeit for surface waves ($\delta=0$) as also by \cite{alexakis2002shear} for arbitrary $\delta$, subsequently. In \S\ref{sec:highFr}, we will see that even at $Fr \sim 3000 >> 1$, the growth rate of the fastest growing mode in the exponential model, agrees much better with that of the PL model (figure \ref{fig2a}) rather than the discontinuous KH model of figure \ref{fig2b} considered by \cite{morland1991waves} and  \cite{alexakis2002shear}. With this brief survey above, it becomes clear that the nature of the linear instability for the exponential model of figure \ref{fig1} as a function of density ratio and transitions if any, for the fastest growing mode are not known. This is demonstrated in the present study employing theory and numerical simulations.
	\subsection{Inviscid, shear instabilities and the exponential model}
	For the exponential profile in figure \ref{fig1}, the background shear layer thickness is approximately $\approx \Delta$. With a sharp density interface placed within this shear layer, three inviscid instabilities can potentially manifest, depending on the parameter regime. Two of these (the KH and Holmboe), are approximately understood using piecewise linear velocity profiles (their continuations to smooth profiles also being well-known) while for the third instability \citep{miles1957generation}, curvature of the background velocity profile is crucial. For completeness, we recap some salient features.
	
\begin{figure}
	\centering
	\subfloat[Piecewise linear profile]{\includegraphics[scale=0.37]{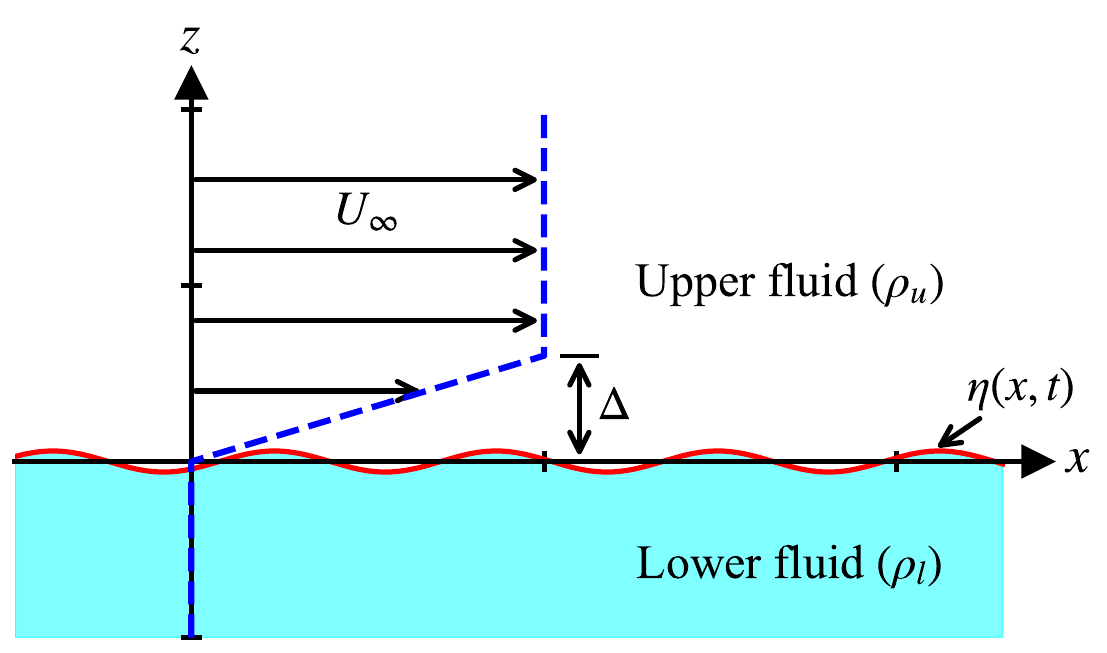}\label{fig2a}}
	\subfloat[Discontinuous KH profile]{\includegraphics[scale=0.37]{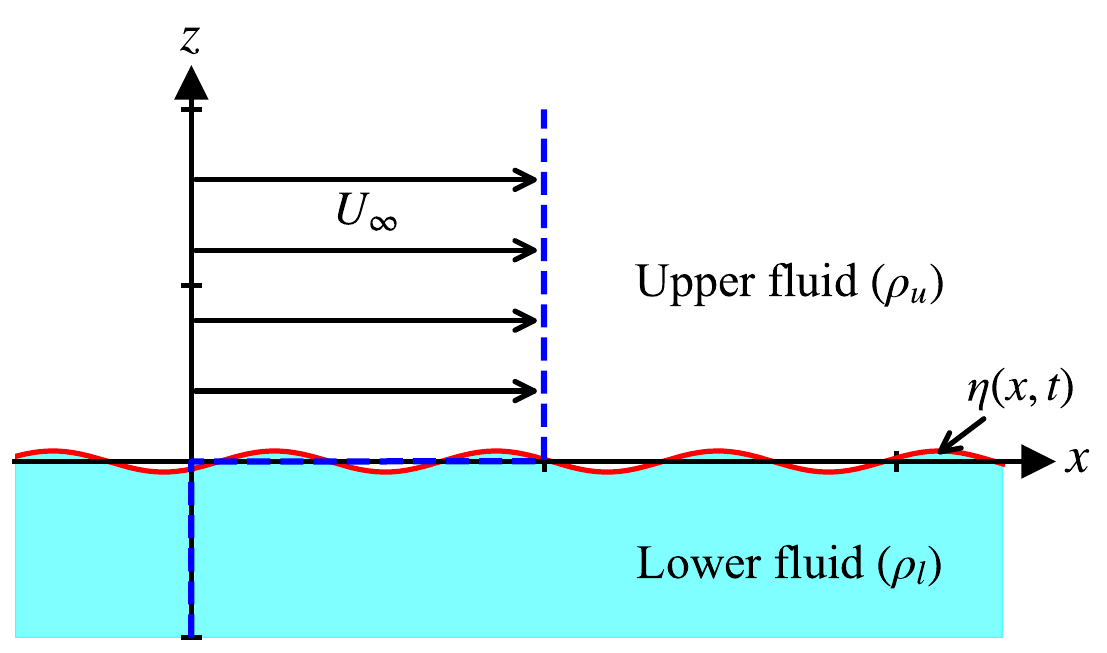}\label{fig2b}}
	\caption{Panel (a) The piecewise linear (PL) profile with a shear layer of thickness $\Delta$. The upper fluid background velocity is given by $U_u(z)=U_\infty\left( \dfrac{z}{\Delta}\right),\; 0\leq z \leq \Delta$ and $U_u(z)=U_{\infty},\; z \geq \Delta$. Panel (b) The classic KH discontinuous profile, obtained when $\Delta\rightarrow0$ in the PL profile on the left panel. In both panels, the lower fluid is quiescent in the base-state i.e. $U_l(z)=0,\; z<0$. The statically stable, base-state density profile is $\rho(z)=\rho_l,\; z< 0$ and $\rho(z)=\rho_u,\; z >0$ with $\rho_l > \rho_u$.}
	\label{fig2}
	\end{figure}

	Recall that the KH instability occurs through an inflexional instability mechanism (see discussion in page $277-278$ of \cite{pouliquen1994propagating}) applicable to a \mgt{homogeneous} fluid under shear (i.e. no stratification). The textbook KH profile is usually the base-state of figure \ref{fig2b} (a vortex sheet) representing the limiting case of an inflectional velocity profile \citep{michalke1964inviscid} such as a $\tanh$ profile; the vortex sheet of figure \ref{fig2b} is unstable to all wavenumbers when the density ratio ($\delta \equiv \rho_u/\rho_l$) is unity. Density stratification plays a \textit{stabilising} role for the sheet - only sufficiently large wavenumbers are unstable when $0 < \delta < 1$. The KH instability can also manifest in the PL model of figure \ref{fig2a} \mgt{for perturbations whose wavelength $\lambda$ is large compared with the shear-layer thickness ($\lambda \gg \Delta$)}. Replacing the smooth, background vorticity gradient ($d^2U_u/dz^2$) of figure \ref{fig1}, by the apparently simpler PL profile of figure \ref{fig2a} with a jump in background vorticity at $z=\Delta$, the KH instability can also be interpreted as resonance between waves on the two vorticity interfaces at $z=0$ and $z=\Delta$, see discussion around figure $1$ in \cite{baines1994mechanism} and around eqn. $33$ in \cite{carpenter2011instability} for smooth profiles.
	
	In contrast to the KH instability, the Holmboe instability \citep{holmboe1962behavior} occurs due to the \textit{destabilising influence} of stable stratification under shear, manifesting in situations where the thickness of the shear layer is much larger than that of the density layer. Employing the same PL profile of figure \ref{fig2a}, one can interpret the Holmboe (H) instability as resonance between a wave on the density and that on the vorticity interface(s), see figure $16d$ in \cite{carpenter2011instability}. The difference between the H and KH instability is seen most readily when the center of the shear layer coincides with the density interface (symmetric profiles) leading to stationary (KH) and travelling (H) modes. The profiles of figure \ref{fig1} and \ref{fig2a} however are asymmetric, the latter represents a limiting case of the profile in \cite{lawrence1991stability} and we will return to this asymmetry later in our study.
	
	The third instability admissible for the profile of figure \ref{fig1}, is the critical layer instability of \cite{miles1957generation}. \mmgt{Although the Rayleigh equation is strictly not singular for real values of the vertical coordinate $z$ and a growing mode with complex $c$}, for our present study we define the critical location as the (\mmgt{real}) location $z_c$ where $U_u(z=z_c)=c_r(k)$ ($c_r$ denotes the real part of the phase-speed $c$), this being done for unstable modes. Unlike the H \& KH instabilities, non-zero curvature of the background profile at the critical location is crucial to the instability (see below). This instability, more broadly, belongs to the `class B' family of instabilities of \cite{benjamin1960effects,benjamin1963threefold}. As explained in \cite{benjamin1960effects} (last paragraph, page $514$), the instability ``\textit{takes the form of waves progressing in the flow direction at very nearly the same speed as free surface waves in the flexible medium, the waves being amplified
	by the action of the flow which supplies sufficient energy to counterbalance internal dissipation}''. At low density ratio, \cite{miles1957generation} demonstrated that for this instability the growth rate of a Fourier mode is proportional to $(-d^2U_u/dz^2)_{z_c}$, $U_u(z)$ being the background shear flow in the upper fluid (subscript `u' for upper). In regimes where this critical location dominates the instability mechanism, we will verify that for the fastest growing Fourier mode, a near discontinuous jump in the (inviscid) Reynolds stress appears at the critical location and upto the interface (also see figure $2$ in \cite{boomkamp1996classification})\footnote{We recall that the critical layer plays a crucial role in the \textit{viscous} instability of non-inflectional, velocity profiles in unstratified, parallel shear flows e.g. the plane Poiseuille flow (figure $IX.\; 8.$ in ch. IX, \cite{lighthill1963laminar}) or TS instability of the Blasius boundary layer (ch. $2$, \cite{ruban2023fluid}).}. The associated work done by this shear stress, leads to extraction of energy from the background shear \textit{primarily} in the region between the critical location and the interface. This will be contrasted against situations where the energy extraction occurs in the entire shear layer with no distinct signature of the Reynolds stress at the critical location; this is despite the mathematical existence of a critical location for every growing Fourier mode, as ensured by the semi-circle theorem.
	
	While we have discussed the three instabilities above in an apparent mutually exclusive manner, critical locations appear within the inviscid framework (Taylor-Goldstein equation), for the KH as well as Holmboe instabilities with smooth velocity (and density) profiles. Refer to figure $5$ and figure $7$ in \cite{smyth1989transition}, where phase jumps and peaks are seen at the critical location for the eigenfunction and the Reynolds stress for unstable KH and H modes, respectively. We bear this aspect in mind when distinguishing between critical layer Miles instabilities at low density ratio and H and KH instability at higher values. In addition, comparison of instabilities obtained from curved background profiles (figure \ref{fig1}) with their straight line counterparts (figure \ref{fig2}), as is done throughout our study, despite being common, requires caution e.g. see \cite{morland1991waves} for the distinction in the transition mechanism between instability in the piecewise linear profile and that of the exponential profile, for a sheared lower fluid ($\delta=0$). \mmgt{For background profiles with a smooth second derivative, the logarithmic singularity in Tollmien's inviscid solution (applicable to a neutral mode, provided such a mode exists. Such a mode is singular at a (real) $z$ where the second derivative of the background profile is non-zero and the modal phase-speed matches the background speed) gets regularised and spread over a finite thickness for a growing mode. It is possible to describe this critical layer region via kinks present in piecewise linear profiles; a large number of kinks being necessary for an accurate representation, see \cite{carpenter2017physical}}. Closely related to our study, we also note the interesting study by \cite{eaves2019instability}, where several diagnostics, both analytical and numerical, in order to distinguish the Taylor-Caulfield instability from the H and the KH instability are discussed.
	
	This study is organised as follows: in \S\ref{sec:lin_stab}, we discuss the temporal, linear stability of the exponential profile in figure \ref{fig1} obtaining the dispersion relation. The same is also done for the PL profile of figure \ref{fig2a}. \S\ref{sec:explor_param} presents a detailed exploration of the parameter space for high and low Froude with increasing density ratio, for high Bond. \S\ref{sec:sim} presents numerical simulations validating linearised predictions but also presenting several novel results in the nonlinear regime. \mgt{In \S\ref{sec:EffectofDelta}, we revisit the validity of the background-state of figure \ref{fig1}, particularly for higher density ratios, close to unity}. We conclude with a summary in \S\ref{sec:concl}.
	
	\section{Temporal, linear stability analysis}\label{sec:lin_stab}
	The base-state (figure \ref{fig1}) comprises a stably stratified ($\rho_l > \rho_u$) configuration with a sharp density jump at $z=0$ and shear (only) in the upper fluid:
	\begin{subequations}\label{eq2.1}
	\begin{align}
	& U(z) = \left\{
	\begin{array}{ll}
	U_u(z)=U_{\infty}\left(1-\exp\left(-\dfrac{z}{\Delta}\right)\right), & \qquad z > 0  \vspace{0.2cm} \\
	U_l(z)=0,                                                            & \qquad  z < 0
	\end{array}\right. \label{eq2.1a} \\
	& \rho(z) = \left\{
	\begin{array}{ll}
	\rho_u, & \qquad z > 0  \vspace{0.2cm} \\
	\rho_l, & \qquad  z < 0
	\end{array}\right. \label{eq2.1b}
	\end{align}
	\end{subequations}
	In the base-state the interface is flat at $z=0$. The perturbed state is represented by the streamfunctions $\psi_u(x,z,t)$ and $\psi_l(x,z,t)$ (for upper and lower fluid respectively) while the deviation of the interface from its flat state is measured by $\eta(x,t)$ (see figure \ref{fig1}). The vertical variation of the perturbation streamfunction ($\Psi_{u/l}$) is governed by the Rayleigh equations below (subscripts $u$ or $l$ refer to upper and lower layers, respectively)
	\begin{subequations}\label{eq2.2}
	\begin{align}
	& \bigg\lbrace U_{u/l}(z)-c(k)\bigg\rbrace(D^2 - k^2)\Psi_{u/l}(z) - D^2U_{u/l}(z)\Psi_{u/l}(z) = 0 \tag{\theequation a,b}
	\end{align}
	\end{subequations}
	where $D\equiv \dfrac{d}{dz}$ and $\Psi_{u/l}(z)$ is related to the streamfunction $\psi_{u/l}(x,z,t)$ via a travelling wave, Fourier mode representation viz. (we omit the $u$ or $l$ subscripts below)
	\begin{eqnarray}
	&&\psi(x,z,t) = \dfrac{1}{2}\Psi(z)\exp\left[i k(x-ct)\right] + \text{c.c.},\quad u \equiv \dfrac{\partial\psi}{\partial z},\quad w\equiv -\dfrac{\partial\psi}{\partial x}. \label{eq2.3}
	\end{eqnarray}
	Here $i\equiv \sqrt{-1}$ and $u$ and $w$ are the horizontal and vertical, perturbed velocity components respectively while c.c. denotes the complex conjugate. The perturbed interface \mgt{(vertical displacement)} is similarly represented using complex notation as,
	\begin{eqnarray}
	\eta(x,t)=\dfrac{1}{2}\eta_0\exp\left[i k(x-ct)\right] + \text{c.c.}, \label{eq2.4}
	\end{eqnarray}
	where $\eta_0$ is a complex constant. Eqns. \ref{eq2.2} represent eigenvalue problems for determining complex functions $\Psi_{u}(z),\Psi_{l}(z)$ and $c(k)$ subject to boundary conditions (see below), for the chosen base-state profile in eqn. \ref{eq2.1} and for real $k$. The boundary conditions are the (linearised) kinematic boundary condition
	\begin{subequations}\label{eq2.5}
	\begin{align}
	c\eta_0 = \Psi_u(0)= \Psi_l(0),
	\tag{\theequation a,b}
	\end{align}
	\end{subequations}
	and the jump in normal stress (pressure in the inviscid framework here) due to surface-tension ($T$)
	\begin{eqnarray}
	P_u(0) - P_l(0) = \left(\rho_u-\rho_l\right)g\eta_0- Tk^2\eta_0,  \label{eq2.6}
	\end{eqnarray}
	where the perturbation pressure (with appropriate subscripts) is $p(x,z,t)=\dfrac{1}{2}P(z)\exp\left[i (kx-ct)\right]+\text{c.c.}$. The expressions relating the perturbation streamfunction $\Psi(z)$ to the vertically varying part of the perturbation pressure ($P(z)$) are:
	\begin{subequations}\label{eq2.7}
	\begin{align}
	P_u(0)=-\rho_u\left[(U_u(z)-c)D\Psi_{u} - \left(DU_u\right)\Psi_{u}\right]_{z=0}
	,\quad
	P_l(0)=c\rho_l\left(D\Psi_{l}\right)_{z=0}. \tag{\theequation a,b}
	\end{align}
	\end{subequations}
	In addition to the boundary conditions above, it is also necessary to ensure that the solution to eqns. \ref{eq2.2} do not diverge as $z\rightarrow\pm\infty$ and this determines two constants of integration in expressions for $\Psi_{l}(z)$ and $\Psi_{u}(z)$. As shown by J. Miles in \cite{morland1993effect}, (see \S$3.3$ in \cite{russell1994survey} for details), the Rayleigh equation for the upper fluid may be converted into the Gauss hypergeometric equation whose analytical properties are well-known. The Rayleigh equation for the lower fluid has constant coefficients and is thus straightforward to solve. We thus obtain expressions for the perturbation streamfunction in both fluids as,
	\begin{subequations}\label{eq2.8}
	\begin{align}
	\Psi_u(z) = M\,\, \exp\left[-k\left(z-z_c\right)\right]\, _2F_1\left(\alpha,\beta,\gamma;\dfrac{\exp\left(-\dfrac{z}{\Delta}\right)}{1 - \dfrac{c}{U_{\infty}}}\right), \quad \quad
	\Psi_{l}(z) = N\exp\left(kz\right), \tag{\theequation a,b}
	\end{align}
	\end{subequations}
	with $\alpha \equiv k\Delta + \sqrt{1 + k^2\Delta^2}, \beta \equiv k\Delta - \sqrt{1 + k^2\Delta^2}$ and $\gamma \equiv 1 + 2k\Delta$ while $M$ and $N$ are (complex) constants of integration. In eqn. \ref{eq2.8}a, $\;_2F_1(\cdot,\cdot,\cdot;\cdot)$ is the Gauss hypergeometric function defined by the series:
	\begin{eqnarray}
	_2F_1(\alpha,\beta,\gamma;t) = 1 + \dfrac{\alpha\beta}{\gamma}\dfrac{t}{1!} + \dfrac{\alpha\left(\alpha+1\right)\beta\left(\beta+1\right)}{\gamma\left(\gamma+1\right)}\dfrac{t^2}{2!} + \ldots,\quad\quad |t| < 1 \label{eq2.9}
	\end{eqnarray}
	Note that with $t \equiv \dfrac{\exp(-z/\Delta)}{1-(c/U_{\infty})}$, as $z$ varies in the range $\infty < z \leq z_c$, we have the variation $0< t \leq 1$ assuming $z_c$ is real (for a neutral mode) and determined using $U_u(z_c)=c(k)$. For the range $0 \leq z \leq z_c\Rightarrow\left(\dfrac{U_{\infty}}{U_{\infty}-c_r}\right) \geq t \geq 1$ assuming that $c < U_{\infty}$. This necessitates the continuation of $\;_2F_1(\cdot,\cdot,\cdot;t)$ for $|t|>1$ for which, for example, the Euler's integral representation of $_2F_1$ may be used (see further discussion in page $65$ in \cite{andrews1999special}):
	\begin{align}
	& _2F_1(\alpha,\beta,\gamma;t) \equiv  \dfrac{\Gamma(\gamma)}{\Gamma(\beta)
	\Gamma (\gamma-\beta)}\int_{0}^{1}
	\dfrac{r^{\beta-1}(1-r)^{\gamma-\beta-1}}{\left( 1-
	rt\right)^{\alpha}} dr.\label{eq2.10}
	\end{align}
	Using the kinematic boundary condition eqn. \ref{eq2.5}a,b in eqn. \ref{eq2.8} we obtain
	\begin{subequations}\label{eq2.11}
	\begin{align}
	\Psi_{u}(z) = c\eta_0 \exp\left(-kz\right)\left[\dfrac{_2F_1\left(\alpha,\beta,\gamma;\dfrac{\exp\left(-\dfrac{z}{\Delta}\right)}{1 - \dfrac{c}{U_{\infty}}}\right)}{_2F_1\left(\alpha,\beta,\gamma;\dfrac{1}{1-\dfrac{c}{U_{\infty}}}\right)}\right], \quad \Psi_l(z) = c\eta_0\exp\left(kz\right) \tag{\theequation a,b}
	\end{align}
	\end{subequations}
	We also note that $t=1$ is a branch-point of $_2F_1(\alpha,\beta,\gamma;t)$; this generates a (near) jump in phase of the stream-function $\Psi_u(z)$ in eqn. \ref{eq2.11}a for a Miles growing mode (i.e. a $k$ for which $Im(c)= c_i>0$), as the critical location $z_c$ is crossed ($t=1^{+}$). This in turn also generates the phase shift in perturbation pressure at $z=0^{+}$ compared to the interface $\eta$. This phase shift produces the \cite{jeffreys1925formation} mechanism of generating wind waves, extensively referred to in \cite{miles1957generation}.
	
	Using expressions for $\Psi_{u}(z)$ and $\Psi_{l}(z)$ from eqns. \ref{eq2.11} in eqns. \ref{eq2.7} and the normal stress condition in eqn. \ref{eq2.6}, we obtain the following dispersion relation. In non-dimensional form, this may be expressed as:
	\begin{align}
	\begin{split}
	& \left[\delta\left\lbrace \kappa - \left(\dfrac{1}{1+2\kappa}\right)\left(\dfrac{Fr}{Fr-\tilde{c}}\right)\dfrac{_2F_1\left(\alpha+1,\beta+1,\gamma+1;\dfrac{Fr}{Fr-\tilde{c}}\right)}{_2F_1\left(\alpha,\beta,\gamma;\dfrac{Fr}{Fr-\tilde{c}}\right)}\right\rbrace + \kappa\right]\tilde{c}^2
	\\
	& \qquad \qquad -\delta Fr\tilde{c} - \left(1-\delta\right) - \dfrac{1}{Bo}\kappa^2 = 0, \label{eq2.12}
	\end{split}
	\end{align}
	where
	\begin{eqnarray}
	\tilde{c} \equiv \dfrac{c}{\sqrt{g\Delta}},\; \kappa \equiv k\Delta,\; Fr \equiv \dfrac{U_{\infty}}{\sqrt{g\Delta}},\; \delta \equiv \dfrac{\rho_u}{\rho_l},\;Bo \equiv \dfrac{\rho_lg\Delta^2}{T}. \nonumber
	\end{eqnarray}
	Here $Fr$, $Bo$ and $\delta $ represent Froude number, Bond number and the density ratio respectively. For a given $\kappa$ (wavenumber), the complex roots ($\tilde{c}$) to eqn. \ref{eq2.12} are obtained numerically using the \textit{findroot} function from the \textit{mpmath} library \citep{mpmath} in Python, which provides arbitrary-precision root finding based on Muller's method \citep{muller1956}. In Appendix A, we provide benchmarking with the results of \cite{young2014generation} and \cite{morland1993effect} for validation of the numerical procedure of solving eqn. \ref{eq2.12}. 
	\mgt{In the following sections, we extensively use the symbols $c_g,c_{gc}$ and $c_{KH}$ whose formulae are provided below in the deep limit:
	\begin{subequations}\label{eq2.13}
		\begin{align}
		&\text{Unsheared surface-gravity wave speed:} && c_g^2 \equiv {\left( \dfrac{1-\delta }{1+\delta} \right) \dfrac{g}{k}} \label{eq:cg} \\
		&\text{Unsheared capillary-gravity wave speed :} && c_{gc}^{2} \equiv 	\left(\dfrac{1-\delta}{1+\delta}\right)\dfrac{g}{k} + \dfrac{Tk}{\rho_l\left(1+\delta\right)} \label{eq:cgc}\\
		& \text{KH wave speed:} && c_{_{KH}} \equiv \left(\dfrac{\delta}{1+\delta}\right)U_\infty \label{eq:ckh}   
	\end{align}
	\end{subequations}}	
	\section{Exploring the parameter space}\label{sec:explor_param}
	In the following analysis, we explore the nature of the instability in the non-dimensional space of Froude number ($Fr$) and density ratio ($\delta$), for given value(s) of Bond number ($Bo$). Wherever relevant, we compare the results of the piecewise linear (PL) profile of figure \ref{fig2a} with that of the exponential profile of figure \ref{fig1}, both profiles chosen to have the same vertical scale $\Delta$. \mgt{We investigate four representative cases by varying  $\delta$ and $Fr$ at fixed $Bo$:
		(A) high $Fr = 3194.38$, low $\delta = 0.001$;
		(B) high $Fr = 3194.38$, high $\delta = 0.5$;
		(C) moderately low $Fr = 4.79$, low $\delta = 0.001$; and
		(D) moderately low $Fr = 4.79$, high $\delta = 0.5$. For the benefit of the reader, table \ref{tab1} summarises key observations, as will be inferred from following analysis. We commence our parametric exploration, with the high Froude number case with zero surface tension i.e. the infinite Bond case.}
	
	\begin{table}
	\centering
	\caption{\mgt{Summary of observations based on fastest growing modes}}
	\resizebox{\textwidth}{!}{
	\mgt{
	\begin{tabular}{ccp{8cm}}
		\hline
		\textbf{Section} & \textbf{Profile \&  parameter range} & \textbf{Key result(s)} \\
		\hline
		\multirow{1}{*}{\makecell{\ref{sec:highFr}: High Froude                                                                                         \\
		$\begin{pmatrix} Fr=3194.38, \\ Bo \rightarrow \infty \end{pmatrix}$}} & \makecell{\\\\\ref{sec:3.1.1}: PL and Exponential model\\ Low $\delta$ \\ $(\delta = 0.001<<1)$} &  Growth rate from exponential model matches that of the PL model, but not the discontinuous KH model. Energy extraction of KH mode across entire shear layer.\\\\ 
		\multirow{2}{*}{\makecell{\ref{sec:3.2}: Moderately low Froude \\
		$\begin{pmatrix} Fr=4.79, \\ Bo \rightarrow \infty \end{pmatrix}$}} & \makecell{\\ \ref{sec:3.2.1}: PL model \\ $(0.001 \le \delta \le 1)$} & Transition from H ($\delta=0.001$) to KH ($\delta=0.9$) instability with increasing $\delta$ - wave resonances as metric. \\ 
	    & \makecell{\\\\\ref{sec:3.2.2}: Exponential \\ model\\ $(0.001 \le \delta \le 1)$} &  Low $\delta$ $(= 0.001)$: Miles' instability with phase speed $c_{g}$. Sharp jump in Reynolds stress and peak in perturbation kinetic energy at critical location. Energy extraction between critical location and density interface. \\ 
	    & \makecell{ \\ \;}  & Increasing $(\delta \gtrsim 0.1):$ Reynolds stress smoothens and kinetic energy peak at critical location becomes subdominant. Energy extraction across entire shear layer. Good match with PL model at $\delta=0.5$ indicative of Miles to H instability transition.\\
		\multirow{1}{*}{\makecell{\ref{sec:3.3}: Moderately low Froude\\
		$\begin{pmatrix} Fr=4.79, \\ Bo = O_m(88) \end{pmatrix}$}}& \makecell{ \\ \\Exponential \\ model \\ $(0.001 \le \delta \le 1)$} & Finite but large $Bo$ instabilities: qualitatively similar to those with $Bo\rightarrow\infty$\\[0.7cm] \\
		\multirow{1}{*}{\makecell{\ref{sec:sim}: Nonlinear simulations\\
		$\begin{pmatrix} Fr=4.79, \\ Bo = O_m(88) \end{pmatrix}$}}& \makecell{ \\ Exponential \\ model } & Finite-amplitude waves exhibit surface ripples $(0.01 \leq \delta \leq 0.1)$. Finite amplitude H waves with cusp-like crests emitting droplets ($\delta =0.5$). Typical KH spirals ($\delta=0.9$). \\\\[0.7cm]
	\end{tabular}
	}}
     \label{tab1}
	\end{table}

	\subsection{Infinite Bond ($Bo\rightarrow\infty$) and high Froude ($Fr >> 1$)}\label{sec:highFr}
	\subsubsection{Low density ratio ($\delta << 1$):}\label{sec:3.1.1}
	To our knowledge the first study which investigated transition in the nature of the instability for the exponential profile of eqn. \ref{eq2.1} with a sharp density jump at $z=0$ was by \cite{morland1993effect}. These authors solved the Rayleigh equation numerically for the exponential profile (quiescent water in base-state) for air-water ($\delta=0.001$) taking into account gravity but setting surface-tension to zero ($Bo\rightarrow\infty$). Among other conclusions, \cite{morland1993effect} noted a change in the nature of the instability at $Fr >>1$ (for constant $\delta$), transitioning from `Miles unstable' (critical layer induced instability) gravity waves with approximate phase-speed $c_g^2$ (in deep-water, \mgt{eqn. \ref{eq:cg}}) to Kelvin-Helmholtz (KH) unstable waves (i.e. waves appearing on the base state of figure \ref{fig2b}) with phase speed given by $c_{_{KH}}$ \mgt{(eqn.~\ref{eq:ckh})}. Such KH like behaviour is anticipated for unstable modes of wavelength $\lambda$ satisfying $\Delta << \lambda < \dfrac{2\pi U_{\infty}^2\delta}{g\left(1-\delta^2\right)}$ (see discussion below eqn. $3.2$ in \cite{morland1993effect}).
	\begin{figure}
	\centering
	\subfloat[Growth rate]{\includegraphics[scale=0.4]{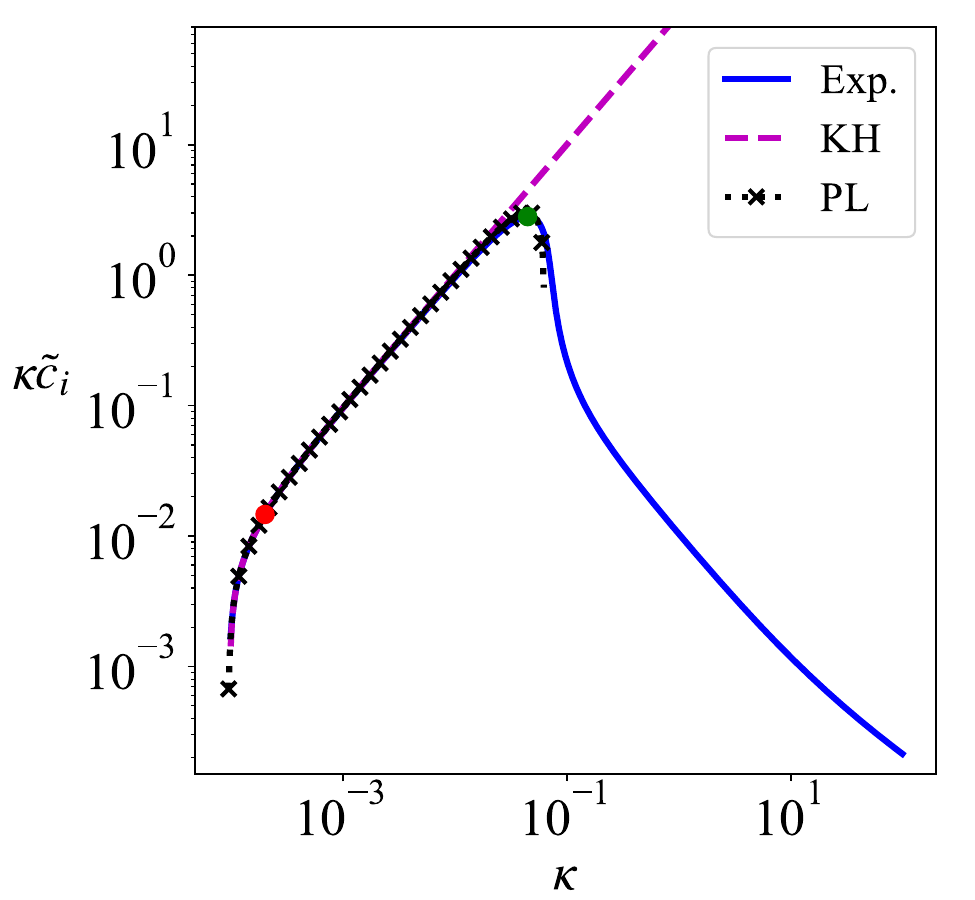}	\label{fig3a}}
	\subfloat[Phase speed]{\includegraphics[scale=0.4]{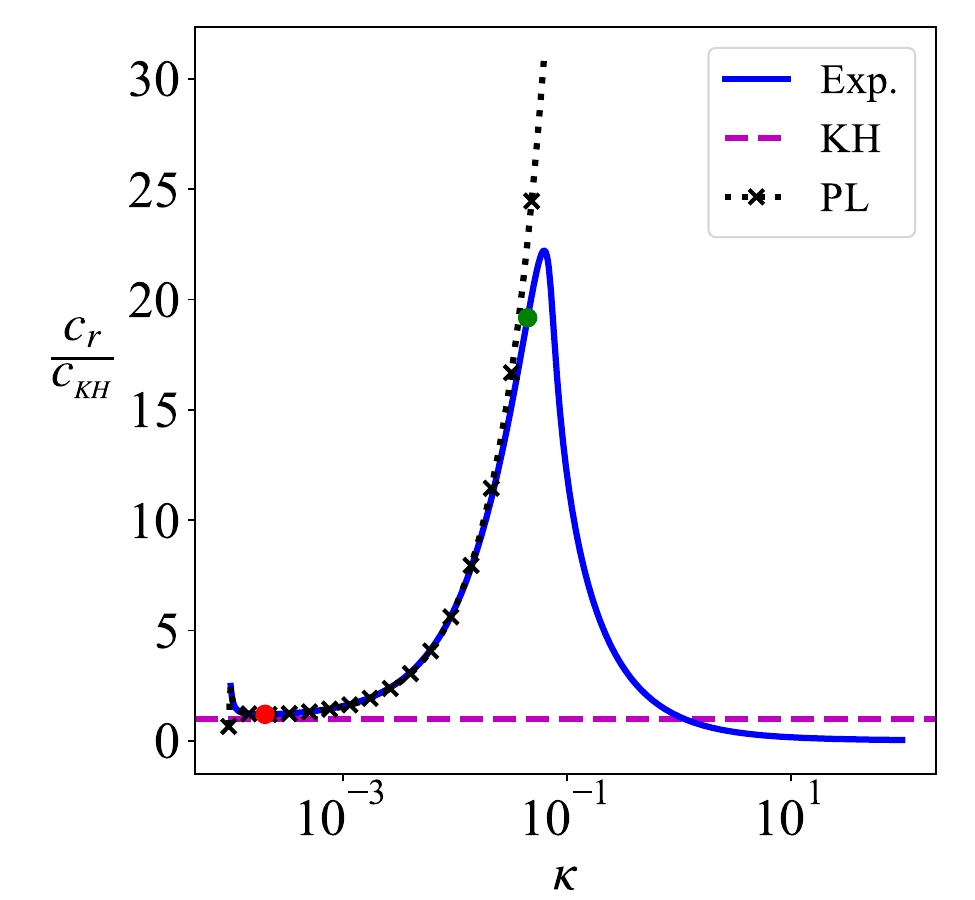}	\label{fig3b}}
	\caption{$Bo\rightarrow\infty$, low (air-water) density ratio ($\delta=0.001$) and high Froude ($Fr=3194.38$). (Panel (a)) Growth rate ($\kappa \tilde{c}_i$) and (panel (b)) phase speed ($c_r/c_{_{KH}}$) as a function of the non-dimensional wavenumber $\kappa$ for the Kelvin-Helmholtz (KH) dispersion relation from eqn. \ref{eq3.1} (pink dashed curve), for the exponential profile (solid blue curve) from eqn. \ref{eq2.12} and for the piecewise-linear (PL) profile from the dispersion relation in eqn. \ref{eq3.2} (`$\cdot\times\cdot$' symbols) for the base-states of figures \ref{fig1} and \ref{fig2}. We see that the growth rate and phase speed are very similar for $\kappa\rightarrow0$. The KH dispersion relation of eqn. \ref{eq3.1} does not contain the Froude number as a parameter and represents the $Fr\rightarrow\infty$ limit. The red and green dots in both panels depict a small wavenumber and the fastest growing wavenumber respectively.}
	\label{fig3}
	\end{figure}
	It is also intuitively clear from examining the exponential profile in eqn. \ref{eq2.1} that taking the limit $\Delta\rightarrow 0$ ($Fr\rightarrow\infty$)\footnote{for the exponential profile, the $Fr\rightarrow\infty$ limit for fixed $\delta << 1$ as discussed here, is qualitatively different from $\delta\rightarrow 0,Fr\rightarrow\infty$, the limits taken in that order. See discussion below eqn. $7.17$ in \cite{young2014generation}.} for fixed $U_{\infty},\;g$ and density ratio $\delta$, deforms the exponential profile in figure \ref{fig1} towards a uniform velocity profile ($U_{\infty}$) in the upper fluid; in other words the classic KH base-state profile of figure \ref{fig2b}. In dimensional variables, the dispersion relation for this KH base-state is (eqn. $4.20$ in \cite{drazin2004hydrodynamic}):
	\begin{eqnarray}
	\dfrac{c(k)}{U_{\infty}} = \dfrac{\delta}{1+\delta} \pm \left[\left(\dfrac{1-\delta}{1+
	\delta}\right)\dfrac{g}{kU_{\infty}^2} - \dfrac{\delta}{(1+\delta)^2}\right]^{1/2}. \label{eq3.1}
	\end{eqnarray}
	The relation \ref{eq3.1} predicts that unstable KH modes travel with the speed $c_{_{KH}}$, independent of wavenumber. At any $U_{\infty}$, sufficiently long waves are stable (due to gravity), while shorter waves are destabilised by the shear and grow exponentially in time.
	
	We also compare predictions from the dispersion relation, eqn. \ref{eq2.12} for the exponential profile in the high Froude number limit, with that of the PL velocity profile indicated in figure \ref{fig2a}. In terms of the non-dimensional variables and parameters introduced earlier, the dispersion relation \mgt{(cubic polynomial in $\tilde{c}$)} for the PL profile of figure \ref{fig2a} can be shown to be:
	\begin{align}
	\begin{split}
		& \Bigl( Fr(1 - 2 \kappa) + 2\tilde{c}\kappa \Bigr) \cdot
		\left( \dfrac{1 - \delta}{1 + \delta}(-2 + \tilde{c}Fr) - \tilde{c} Fr + 2\tilde{c}^2 \kappa \right) \\
		& \qquad + e^{-2\kappa} \cdot Fr \cdot \left\{ \tilde{c} Fr \left(1 - \dfrac{1 - \delta}{1 + \delta}\right) + 2\left(\dfrac{1 - \delta}{1 + \delta}\right)(1 - \tilde{c}^2 \kappa) \right\}=0, \quad \tilde{c}\neq 0. \label{eq3.2}
	\end{split}
	\end{align}
	
	Plotted in figure \ref{fig3a} and \ref{fig3b} are the growth-rate and the phase-speed of the exponential profile (labelled as `Exp') as obtained from eqn. \ref{eq2.12}, for the KH profile from \ref{eq3.1} and for the PL profile from eqn. \ref{eq3.2}, at high Froude number viz. $Fr=3194.39$. A few qualitative features are discernible readily. On account of having an additional length-scale $\Delta$ compared to the classic KH model in figure \ref{fig2b}, the stability characteristics of the PL profile (`$\cdot\times\cdot$' symbols) differ qualitatively at large wavenumber from the KH model (dashed pink line) as seen in figure \ref{fig3}. Wavenumbers with $\kappa \geq O_m(0.1)$\footnote{$\mathcal{O}_m$ denotes order-of-magnitude} are stable in the PL model - this is the reason for the abrupt termination of the `$\cdot\times\cdot$' symbols (PL model) around $\kappa\sim 10^{-1}$ in figure \ref{fig3a}, above this wavenumber the growth-rate goes to zero rapidly. For the exponential profile, also recall that a lower cutoff in wavenumber exists at infinite depth (\cite{young2014generation,kadam2023wind}); wavenumbers less than this cutoff are stable \citep{young2014generation} as seen from the abrupt decay in growth-rate, in figure \ref{fig3a} at low $\kappa$.
	
	Figures \ref{fig3a} and \ref{fig3b} also show that as $\kappa\rightarrow 0$ (long wave indicated by the red dot), the growth rate and the phase-speed of the exponential as well as the PL modes, asymptote to the KH values provided by expression \ref{eq3.1}. The fastest growing mode for the exponential profile, is indicated by the green dot in both figures. The red dot corresponds to a Fourier mode of wavelength $\approx 230$ times the wavelength of the green dot. It is clear that both are long wavelength modes compared to the shear layer thickness ($\lambda/\Delta\approx 3.1\times 10^4$ for the red dot and $\approx 1.4\times 10^2$ for the green dot). A qualitative difference is apparent between these two, long wavelength modes: the fastest growing mode (green dot) of the exponential profile grows somewhat slower than the KH growth estimate of eqn. \ref{eq3.1}, travelling at a speed which is nearly twenty times the KH prediction, cf. figure \ref{fig3b}. Interestingly, its growth rate and phase speed both agree quite well with that of the PL model (`$\cdot\times\cdot$' symbol).
	
	Having established that the fastest growing mode of the exponential profile in the high Froude number (and infinite Bond and low density ratio) limit has growth rate and phase speed agreeing with the corresponding mode of the PL profile, it becomes necessary to inquire about the nature of this mode in the PL profile. For this, we recall a few essential stability properties of the PL profile. This base-state of figure \ref{fig2a}, by virtue of having a density interface (zero thickness) much smaller in thickness than the shear layer, admits the \textit{Holmboe instability} (interaction between a vorticity and a density interface) as well as the \textit{KH instability} (interaction between two vorticity interfaces, see \cite{carpenter2011instability} and references therein). Under the Boussinesq approximation, depending on the choice of parameters on the Richardson number $J\equiv \left(\dfrac{1-\delta}{1+\delta}\right)\dfrac{g\Delta}{U_{\infty}^2}$ versus wavenumber ($\kappa$) space, the instability can be dominantly one of the two types or be even of mixed nature. As noted by \cite{lawrence1991stability}, the PL profile that we employ is an \textit{asymmetric} profile, implying that the center of its shear layer ($z = \Delta/2 > 0$, see figure \ref{fig2a}) does not coincide with the location of the density interface ($z=0$). Indeed, the velocity profile of figure \ref{fig2a} may be recovered from that of \cite{lawrence1991stability}, by setting $u_2=0$ and their asymmetry parameter $\epsilon$ to unity (refer to figure $1$ in \cite{lawrence1991stability}). Unlike symmetric profiles, for asymmetric profiles the transition between the KH and the Holmboe instability is not readily apparent on the $J-\kappa$ plane, as both modes have non-zero phase speeds, see page $9$, section $B$ in \cite{carpenter2010identifying}; \mgt{also see the related study by \cite{barros2011holmboe} for non-Boussinesq effects}. Using wave resonance ideas reviewed in \cite{carpenter2011instability}, one can use the phase speed versus wavenumber plane to conclude if the dominant contribution comes from the KH (vorticity-vorticity resonance) or the Holmboe instability (vorticity-density resonance) for the fastest growing mode of figure \ref{fig3a}. We defer this exercise to \S\ref{subsec:wave_interac} (the low Froude number limit), where we show that at low Froude number and low as well as moderately high density ratio, growing modes in the PL model arise due to resonance between a surface-gravity mode and a vorticity wave (i.e. the Holmboe instability dominates over KH).
	
	\begin{figure}
	\centering
	\subfloat[Reynolds stress ]{\includegraphics[scale=0.4]{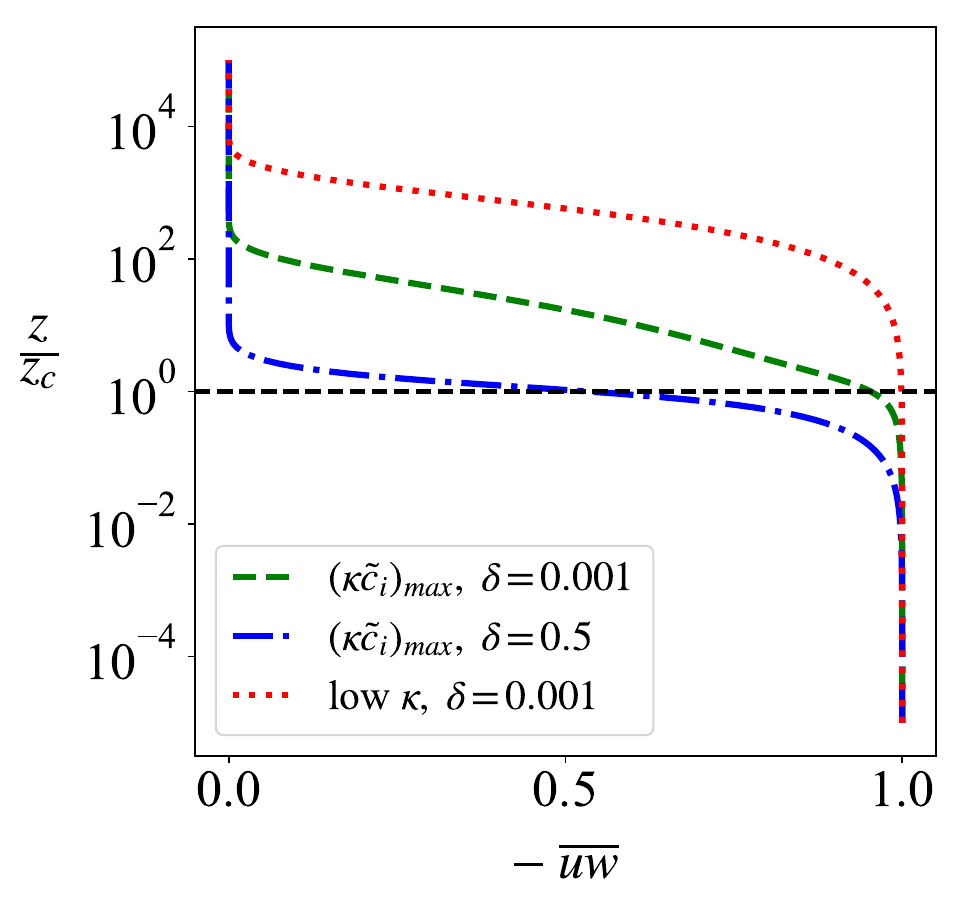}	\label{fig4a}}
	\subfloat[Perturbation kinetic energy]{\includegraphics[scale=0.4]{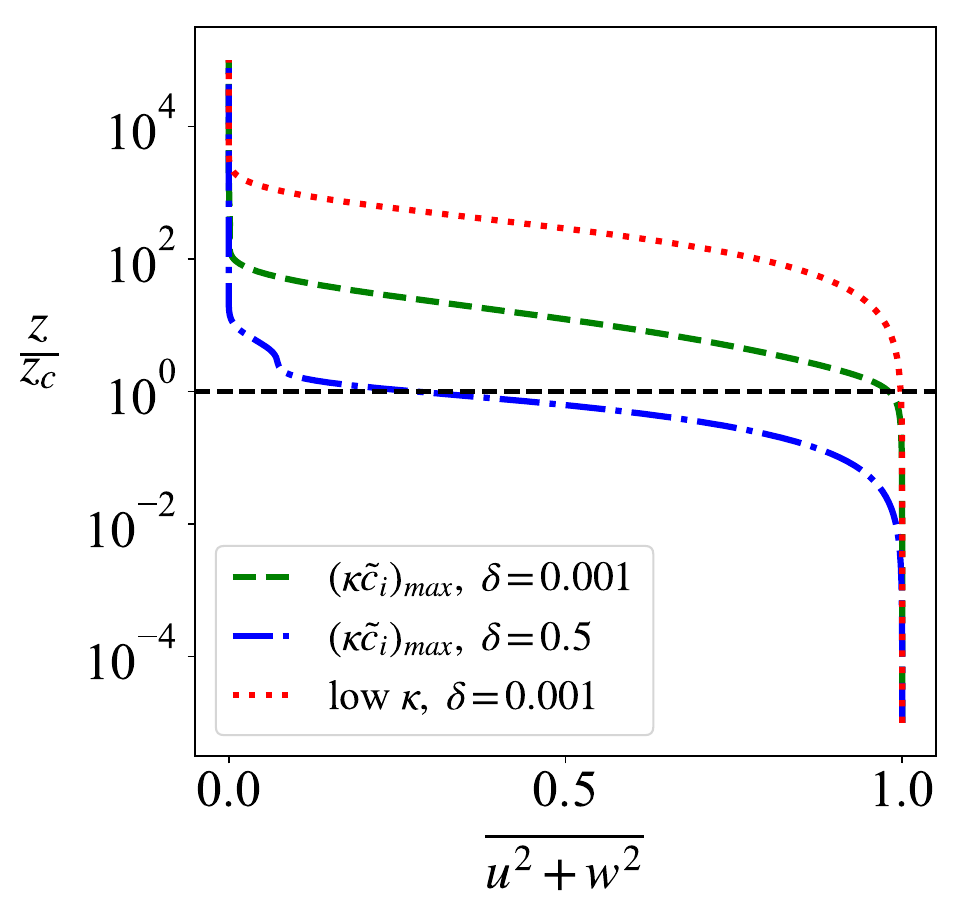}	\label{fig4b}}
	\caption{$(Bo\rightarrow\infty,Fr=3194.8)$ Vertical ($z$) profiles of the (Panel (a)) Reynolds stress and (Panel (b)) perturbation kinetic energy in the upper fluid, normalised by their values at the location of the undisturbed interface at $z=0$, for the exponential profile. In panels (a) and (b), the red and green curves correspond to the low wavenumber and fastest growing modes from figure \ref{fig3}, respectively. The vertical coordinate $z$ is scaled here by the coordinate $z_c$ of the critical layer where
	the base velocity $U_{u}\left(z_{c}\right)$ matches the real part of the phase speed of an unstable mode i.e. $c_{r}\left(k\right)$. The blue curve is for the fastest growing mode with $\delta=0.5$.}
	\label{fig4}
	\end{figure}
	Figures \ref{fig4a} and \ref{fig4b} present the vertical variation of the (inviscid) Reynolds stress $-\overline{\rho uw}(z)$ and (twice) the perturbation kinetic energy $\overline{u^2+w^2}(z)$ for both wavenumbers (green and red dot) of figure \ref{fig3}. In these (and subsequent figures), the Reynolds stress and perturbation kinetic energy have been normalised by their respective values at $z=0$. The over bar indicates averaging along the horizontal coordinate over one wavelength i.e.
	\begin{eqnarray}
	\overline{uw}(z) \equiv \dfrac{1}{\lambda}\int_{x}^{x+\lambda}u(x,z,t)\;w(x,z,t) dx \label{eq3.3}
	\end{eqnarray}
	Recall that the base-state vorticity ($DU_u(z)$ from eqn. \ref{eq2.1}) is non-zero and positive in the approximate range $0 < z < \Delta$ in the upper fluid. This corresponds to $0< z/z_c < \Delta/z_c = 51.4,(\text{wavelength}\; \lambda/z_c\approx 7.2\times10^3)$ for the green curves and $0<z/z_c < \Delta/z_c =832.2, (\lambda/z_c\approx 2\times10^7)$ for the red curves, in both panels of figure \ref{fig4}. It is clear that the perturbation kinetic energy $\left(\overline{u^2+w^2}\right)(z)$ as well as Reynolds stress ($\left(-\overline{\rho uw}\right)(z)$) becomes non-zero close to the onset of the shear layer for both modes. While the fastest growing mode (green curves) reaches its plateau very close to its maximum value at the critical layer ($z/z_c = 1$) in both panels, the KH mode (red) plateaus significantly \textit{before} the critical layer is reached i.e. $z/z_c \sim 100$.
	One may derive an equation for the rate of growth of perturbation kinetic energy viz.  $E_k(z) \equiv \dfrac{1}{2}\rho\left(\overline{u^2+w^2}\right)(z)$ which is (see eqn. $4$ in \cite{smyth1989transition}):
	\begin{eqnarray}
	\dfrac{\partial E_k(z)}{\partial t} = -\bigg(DU(z)\bigg)\overline{\rho uw}(z) - D\overline{wp}(z). \label{eq3.4}
	\end{eqnarray}
	
	In eqn. \ref{eq3.4} the first term on the right-hand side is the rate of work done by the Reynolds stress and has the physical interpretation of a \textit{source} term involving energy (per unit volume) extraction from the base-state, when it is positive. The second term however, represents the ($z$) divergence of the flux of perturbation kinetic energy/volume and acts to \textit{re-distribute} the available perturbation kinetic energy (see discussion in page $3705$ of \cite{smyth1989transition}). The vertical variation of Reynolds stress in figure \ref{fig4a} shows that for the fastest growing mode (green curve), the dominant energy extraction from the base-state occurs in the region between the critical layer and the interface i.e. $0<z/z_c<1$. In contrast, for the low wavenumber KH mode (red curve), a significant proportion of the energy extracted from the base-state shear \textit{also} occurs in the region where the shear commences $z\approx \Delta$ and up to the critical location i.e. $z/z_c = 1$.
	
	\subsubsection{Higher density ratio ($\delta < 1$)}\label{sec:3.1.2}
	
	\begin{figure}
	\centering
	\subfloat[]{\includegraphics[scale=0.4]{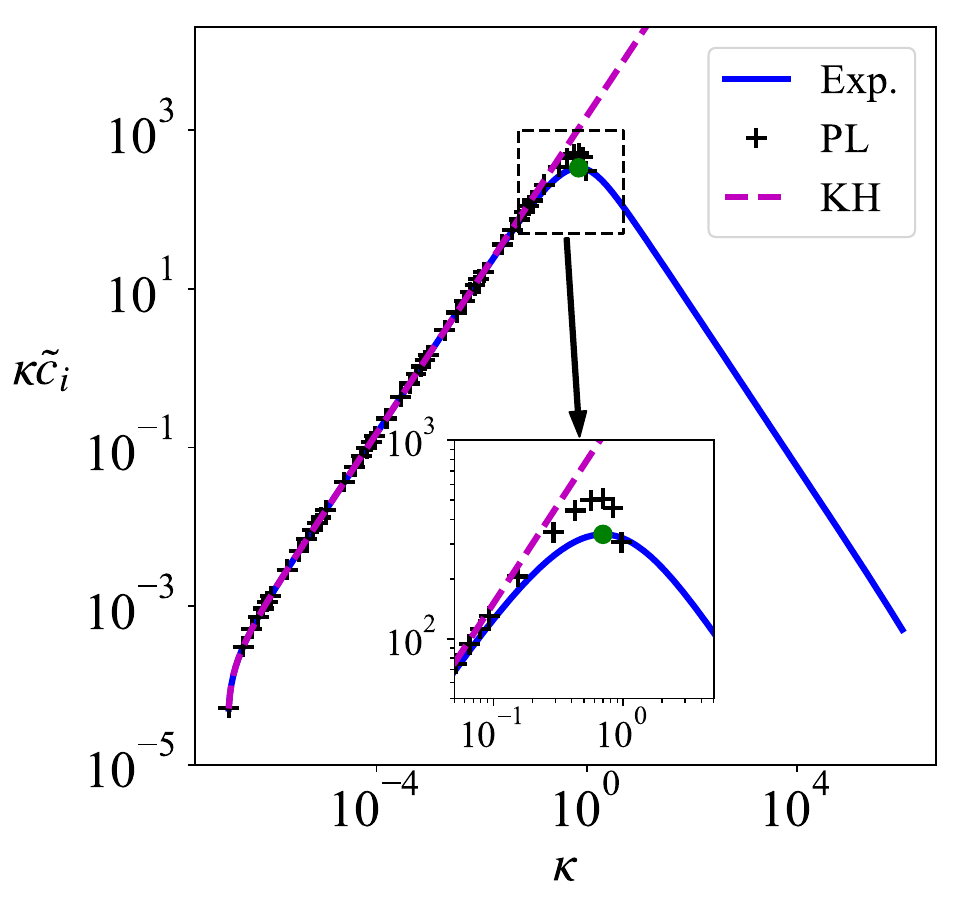}\label{fig5a}}
	\subfloat[]{\includegraphics[scale=0.4]{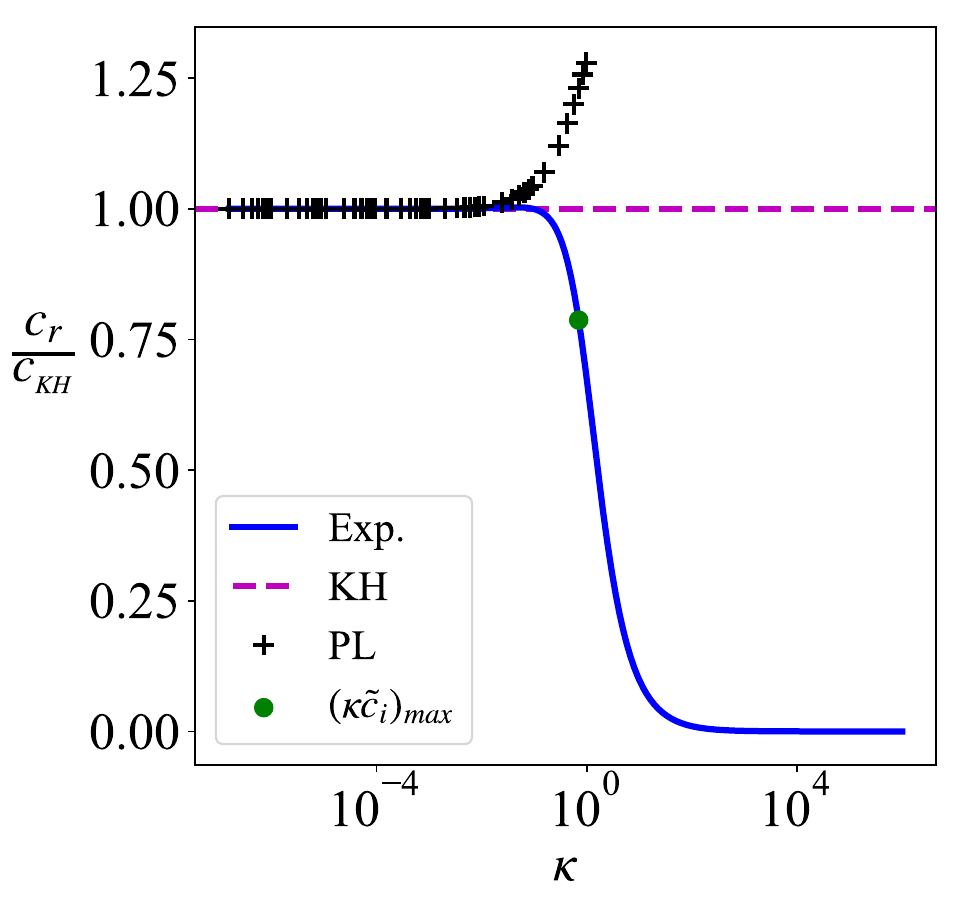}\label{fig5b}}
	\caption{$Bo\rightarrow\infty$, high density ratio ($\delta=0.5$) and high Froude ($Fr=3194.38$). (Panel (a)) Growth rate ($\kappa \tilde{c}_i$) and (Panel (b)) phase speed ($c_r/c_{_{KH}}$) as a function of the non-dimensional wavenumber $\kappa$ for the Kelvin-Helmholtz (KH) dispersion relation from eqn. \ref{eq3.1} (pink dashed curve), for the exponential profile (solid blue curve) from eqn. \ref{eq2.12} and for the piecewise-linear (PL) profile from the dispersion relation in eqn. \ref{eq3.2} (`+' symbols) for the base-states of figures \ref{fig1} and \ref{fig2}. Green dot indicates the fastest growing mode in both panels. The phase speeds of the PL and the exponential profile are markedly different.}
	\label{fig5}
	\end{figure}
\begin{figure}
	\centering
	\includegraphics[scale=0.48]{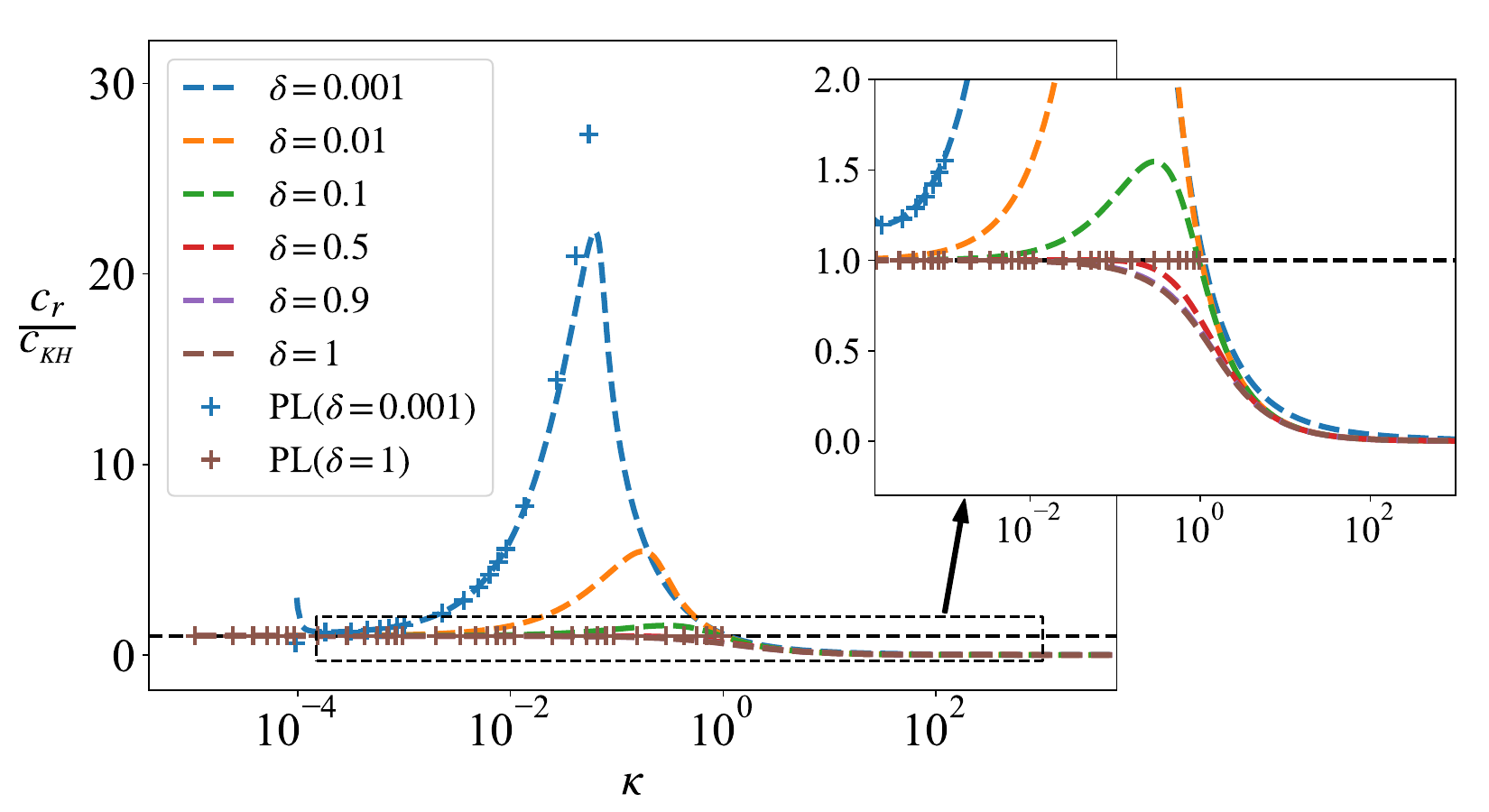}
	\caption{$(Bo\rightarrow\infty,\;Fr=3194.38$ for all cases). Variation of phase-speed with wavenumber, for varying density ratio. All dashed curves are for the exponential model of eqn. \ref{eq2.1}a,b.}
	\label{fig6}
	\end{figure}

	We now move towards higher density ratio and the high Froude number regime. As discussed, the PL profile admits instabilities even for the unstratified case $(\delta=1)$ due to resonance between the vorticity interfaces at $z=0$ and $z=\Delta$ (figure $3d$ of \cite{lawrence1991stability} in the Boussinesq approximation). While the growth rates of the fastest growing mode in the exponential model agree reasonably well with the corresponding PL mode (figure \ref{fig5a}), the phase-speed differs noticeably, see panel (b) of same figure. The corresponding Reynolds stress and perturbation kinetic energy for the fastest growing mode are plotted in blue in both panels of figure \ref{fig4}. Note the appearance of a peak in the perturbation kinetic energy at the location where the shear layer commences i.e. $\Delta/z_c\approx 3.29$, we have checked that no such peak is detected for the fastest growing mode at air-water density ratio (green curves corresponding to $\delta=0.001$ in the same figure). This change in the nature of variation of $c_r/c_{_{KH}}$ with wavenumber $\kappa$ for unstable wavenumbers as the density ratio is varied, is seen both in the PL model and in the exponential model in figure \ref{fig6}, where a systematic increase of density ratio up to unity is reported.
	
	\subsection{Infinite Bond ($Bo\rightarrow\infty$) and low Froude $\left(Fr \sim 1\right)$}\label{sec:3.2}
	We next turn to the case of infinite Bond, and moderately low Froude, examining the effect of variation of density ratio on the instability mechanism in the exponential profile. Before doing this, it is useful to digress and highlight the instability mechanisms in the PL profile of figure \ref{fig2a}, within this parameter regime of interest as a function of density ratio; these results will serve as reference for the exponential profile emphasizing the role of background curvature. In this context, we note the study by \cite{morland1991waves} who compared free-surface waves subject to shear (only in lower layer with $\delta=0$) generated by an exponentially decaying velocity profile (infinite depth) against those from a piecewise linear profile. The instability for their piecewise linear profile \citep{morland1991waves} was shown to be a Holmboe-like instability, arising due to resonance (referred to as `collision' in \cite{morland1991waves}) between a retrograde capillary-gravity mode and a prograde vorticity mode. Here `prograde' and `retrograde' modes are defined (following \cite{carpenter2011instability}) as those which travel leftward (negative velocity) or rightward (positive velocity) respectively, in a frame of reference which moves at the local background speed at the vorticity or density interface; see explanation around eqns. $25$ in \cite{carpenter2011instability}. We will see in the next subsection that an analogous Holmboe instability occurs in our PL profile of figure \ref{fig2a}, at low density ratio and low Froude.
	
	\subsubsection{The instabilities in the PL profile at low $Fr$: density ratio ($\delta$) variation}\label{subsec:wave_interac}\label{sec:3.2.1}
\begin{figure}
	\centering
	\subfloat[$\delta=0.001$ - low density ratio]{\includegraphics[scale=0.4]{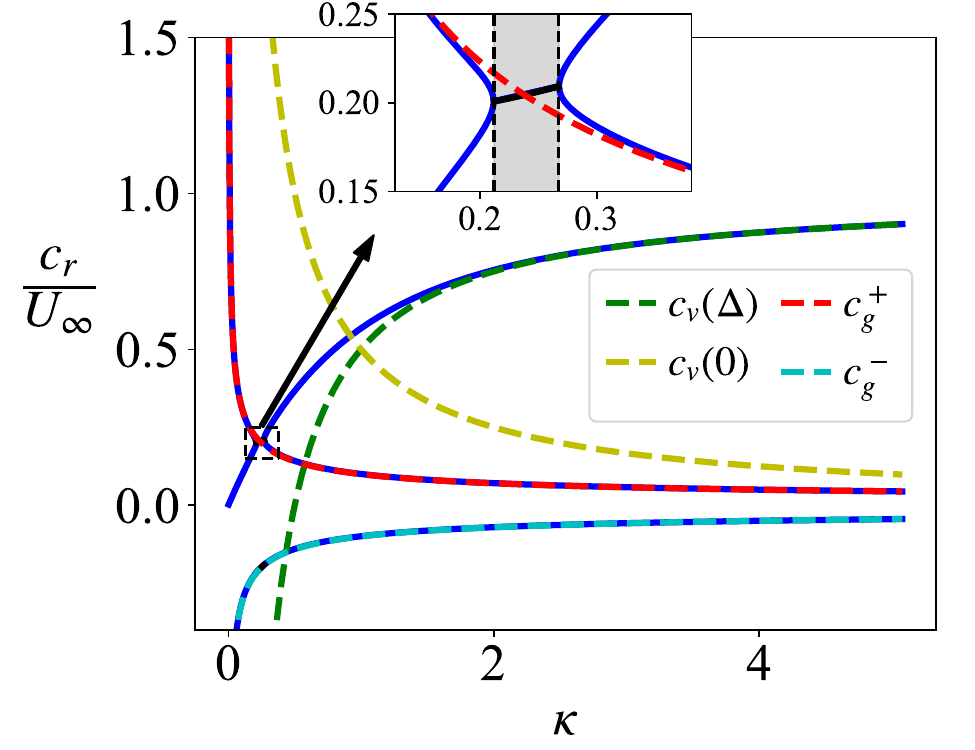}\label{fig7a}}
	\subfloat[$\delta=0.9$ - high density ratio]{\includegraphics[scale=0.4]{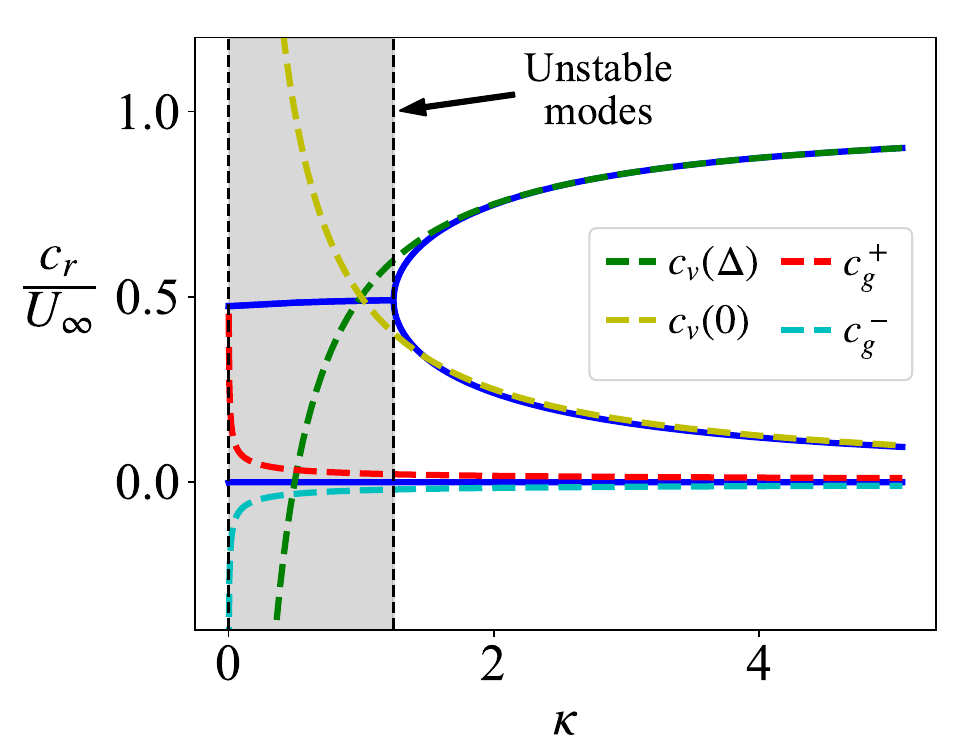}\label{fig7b}}
	\caption{($Bo\rightarrow\infty,\;Fr\approx 10$) The normalised phase-speed ($c_r/U_{\infty}$) variation with wavenumber $\kappa$ for the isolated prograde ($c_v(0)$) and retrograde ($c_v(\Delta)$) vorticity modes as estimated using eqn. $25$ in \cite{carpenter2011instability} and isolated surface-gravity modes with $c_{g}^{+/-}$ \mgt{(eqn. \ref{eq:cg})}. The three solid blue curves are the roots of the dispersion relation given by eqn. \ref{eq3.2}. Grey shaded region represent unstable wavenumbers. In panel (a) inset, unstable wavenumbers are indicated as black solid curve. Panel (a) At low $\delta=0.001$, the instability	is the result of resonance between the prograde gravity-capillary wave and the retrograde vorticity wave. Panel (b) At higher $\delta=0.9$, the instability is the result of resonance between the prograde and the retrograde vorticity waves. Consistent with the Howard semi-circle theorem \citep{morland1993effect}, unstable wavenumbers have their phase-speeds satisfying $0< c_r/U_{\infty} < 1$ in both panels.}
	\label{fig7}
	\end{figure}

	Here, we interpret the instabilities admitted by the PL profile of figure \ref{fig2a}, at low Froude (infinite Bond) and for varying density ratio, employing wave resonance ideas reviewed in \cite{carpenter2011instability}. Figure \ref{fig7a} shows the phase-speed versus wavenumber (refer to the inset for the narrow range of wavenumbers, which are unstable and shaded) for the PL profile at $\delta=0.001$ and $Fr\approx 10$. The three solid blue lines in the figure are obtained from the numerical solution to the dispersion relation given by eqn. \ref{eq3.2}. Recall that the density interface at $z=0$ (refer figure \ref{fig2a}) supports stable prograde and retrograde surface-gravity waves in the absence of shear (labelled as $c_{g}^{+}$ and $c_{g}^{-}$ in dashed red and dashed cyan respectively in figure \ref{fig7a}). The vorticity interface when present alone at $z=\Delta$ and in the absence of stratification, supports a stable retrograde vorticity wave ($c_v(\Delta)$, green dashed line in figure \ref{fig7a}). Similarly, the vorticity interface at $z=0$ when present alone, supports a stable prograde vorticity wave ($c_v(0)$, yellow dashed line); note that one of the blue curves computed from the numerical solution to eqn. \ref{eq3.2}, asymptotes to the green dashed line in figure \ref{fig7a}. However, the other two blue curves from the numerical solution to eqn. \ref{eq3.2} are indistinguishable from the two $c_{g}$ modes, indicating the negligible effect of shear on the phase-speed of these.
	
	The inset in figure \ref{fig7a} makes it clear that the instability in the PL model (in this regime), arises due to resonance between the prograde gravity wave ($c_{g}^{+}$) at $z=0$ and the slightly modified version of the retrograde vorticity wave at $z=\Delta$ (blue curve asymptotic to the green dashed line). Refer to the inset where it's seen that the coalescence of these aforementioned curves produces the band of unstable wavenumbers indicated as a shaded region; we conclude that these wavenumbers in figure \ref{fig7a} are \textit{Holmboe unstable} following the definition of the Holmboe instability \citep{carpenter2011instability}. In figure \ref{fig7b}, a qualitatively different behaviour is seen at a much higher density ratio of $\delta=0.9$ ($Fr\approx 10$). Now the band of unstable wavenumbers (hashed region) arise from resonance between the two vorticity waves at $z=0$ and $z=\Delta$, as might be intuitively expected from a consideration of the PL configuration, excluding stratification i.e. at $\delta=1$. These wavenumbers in figure \ref{fig7b} are \textit{KH unstable}, propagating nearly at the mean background speed $U_{\infty}/2$, consistent with eqn. \ref{eq3.1}. It is thus established that low Froude, the PL model of figure \ref{fig2a} admits the Holmboe instability at low density ratio, transitioning to the KH instability at higher density ratio closer to unity. \textit{How does inclusion of background profile curvature affect this Holmboe instability?} We answer this next.
	
	\subsubsection{Low density ratio ($\delta << 1$) - \cite{miles1957generation} instability in the exponential profile}\label{sec:3.2.2}
	Having discussed the two instabilities in the PL profile, we now turn to studying the effect of background curvature at the low Froude and low density ratio regime where the PL profile was studied in the previous subsection. Because of the presence of a critical location where the background profile curvature is non-zero, instead of the Holmboe instability we now expect to recover the classic \cite{miles1957generation}, wind-wave instability, dominated by critical location dynamics. In this context, it is relevant to mention the recent study by \cite{carpenter2017physical}, which extends the wave interaction mechanism discussed in \S\ref{sec:intro}, in the context of the KH and H instability, to the Miles instability as well. Interestingly, they show that the Miles critical layer instability (at air-water density ratio) can be interpreted as resonance between a vortex sheet at the \text{critical layer} and a surface wave at the density interface.
	
\begin{figure}
	\centering
	\subfloat[Dispersion relation]{\includegraphics[scale=0.4]{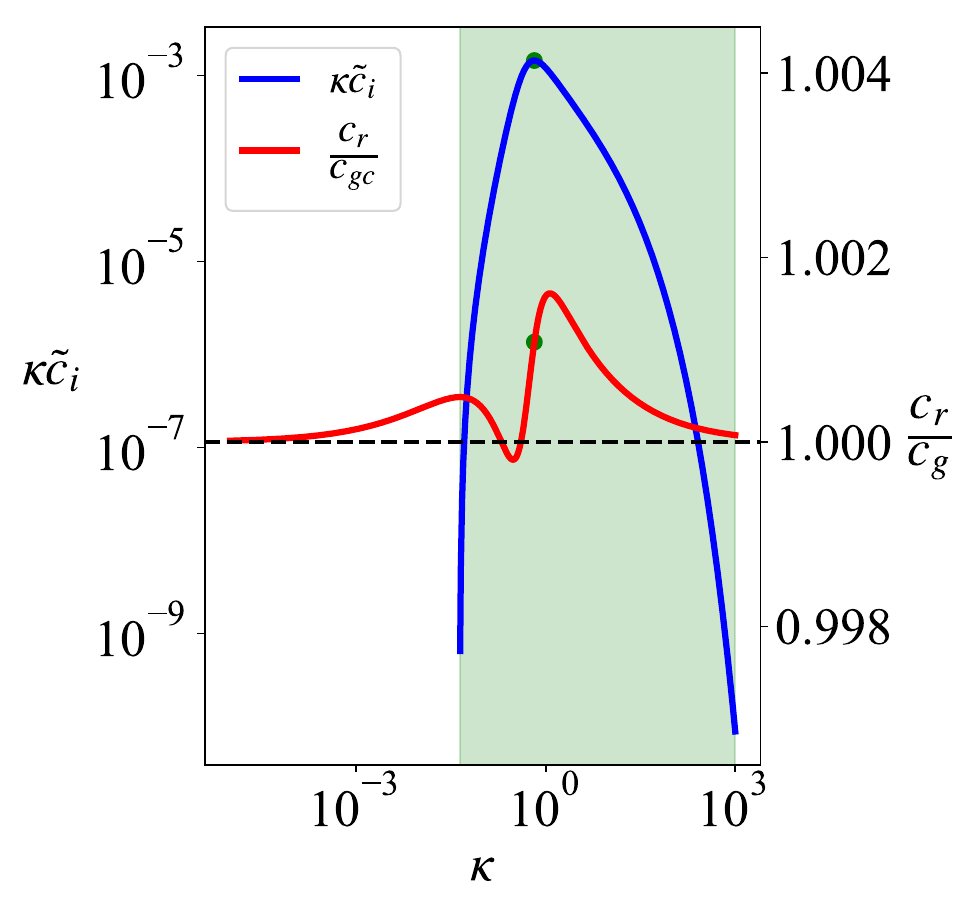}\label{fig8a}}
	\subfloat[Reynolds stress and kinetic energy]{\includegraphics[scale=0.4]{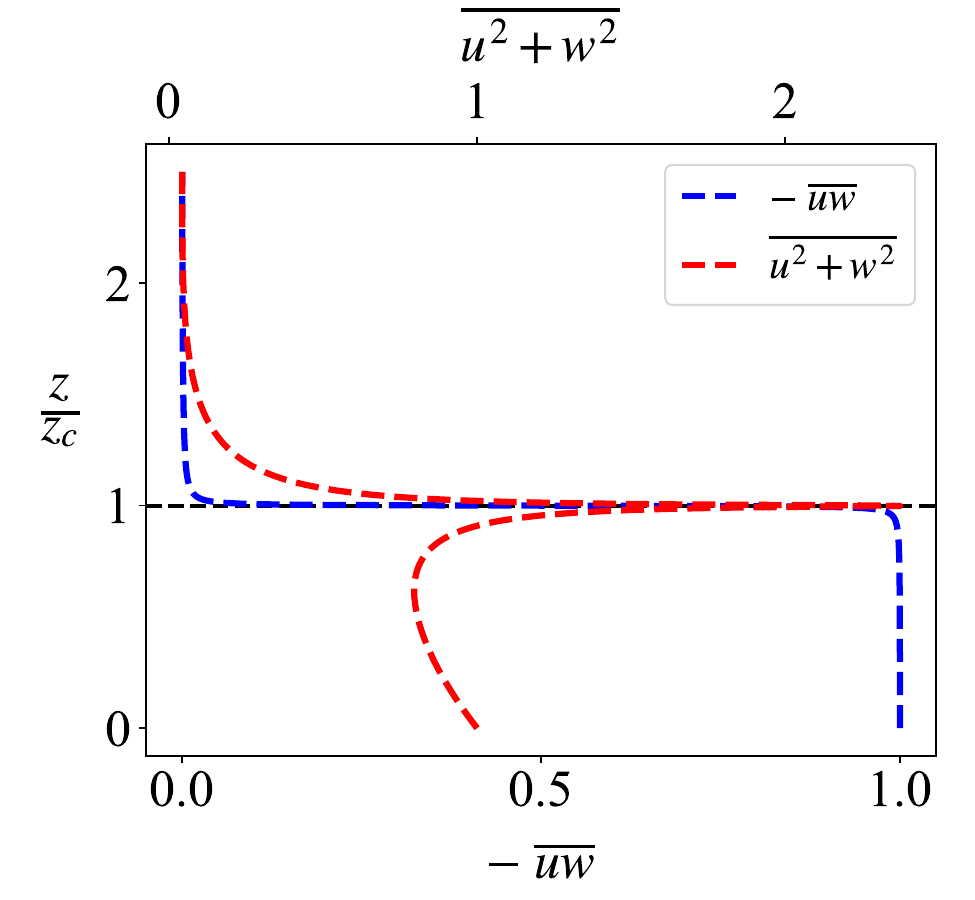}\label{fig8b}}
	\caption{$Bo\rightarrow\infty$,  low (air-water) density ratio ($\delta=0.001$), low Froude ($Fr = 4.79$). Panel (a) Growth rate ($\kappa \tilde{c}_i$) and phase speed ($c_r/c_{\text{g}}$) as a function of the wavenumber $\kappa$. The shaded region is that beyond which, it is numerically challenging to detect growth, due to very small growth rate ($< 10^{-12}$). Panel (b) Vertical variation of Reynolds stress and perturbation kinetic energy for the fastest growing mode represented by dots in panel (a) with $\lambda/z_c\approx32$.}
	\label{fig8}
	\end{figure}

	We note that our choice of low $Fr \approx 4.8$ in this sub-section, is in the same ballpark range as \cite{young2014generation}, see their figure $11$ where their smallest Froude $\sim 1.4$ corresponds to the air-water oceanic situation. In this parameter regime, the fastest growing mode has features of a classic `Miles mode' of instability \citep{miles1957generation} dominated by the critical layer \citep{miles1957generation,young2014generation,kadam2023wind}; figure \ref{fig8} presents data to support this. Unlike the high Froude cases studied previously, the fastest growing mode in this moderate Froude case in figure \ref{fig8a}, travels nearly at the surface-gravity wave speed viz. $c_{g}$. Note that $c_{g}/c_{_{KH}}=\left(\dfrac{1-\delta^2}{\delta^2}\right)^{1/2}\kappa^{-1/2}Fr^{-1} >> 1$ for air-water density ratio, as for the fast growing mode $\kappa = O_m(1)$ with $Fr = O_m(1)$, see figure \ref{fig8a}. That at sufficiently low $\delta$, unstable, prograde modes travel nearly at the corresponding surface-gravity wave speed in still air/water, is originally due to \cite{miles1957generation}. This has been derived formally by several subsequent studies following Miles procedure of expanding the eigenfunction and eigenvalue in the Rayleigh equation, employing density ratio as a small perturbation parameter e.g. see below eqn. $3.10$ in \cite{alexakis2002shear} or figure $3$ in \cite{young2014generation} where capillarity has also been taken into account. In what follows, we illustrate several attributes of the fastest growing Miles mode and use these as reference to contrast the change observed, as the density ratio is increased significantly beyond that of air-water.
	
	Figure \ref{fig8b}, presents the vertical variation of perturbation kinetic energy and the Reynolds stress for the fastest growing `Miles mode' (dots in figure \ref{fig8a}). Note the classic near discontinuous behaviour of the Reynolds stress at the critical location $z/z_c=1$. This is consistent with \cite{lin1954some} (their eqn. $3.10$) who showed, following  \cite{tollmien1935allgemeines}, that the Reynolds stress $\tau(z) \equiv \overline{-\rho u w}$ across the critical location $z_c$ (at which $D^2U_u < 0$ and $DU_u\neq 0$), jumps by the amount
	\begin{eqnarray}
	\tau(z=z_c^{+}) - \tau(z=z_c^{-}) = \dfrac{\pi}{k}\left(\dfrac{D^2U_u}{|DU_u|}\right)_{z=z_c}\rho\overline{w_c^2} \label{eq3.5}
	\end{eqnarray}
	for a \textit{singular neutral mode} i.e. a neutral solution to the Rayleigh equation for which there is a critical layer with $D^2U_u(z_c), DU_u(z_c)\neq 0$ (refer to Tollmien's inviscid solution in eqns. $22.25$ and $22.26$ in \cite{drazin2004hydrodynamic} or around eqn. $2.7$ in \cite{stewartson1981marginally}). In eqn. \ref{eq3.5}, $w_c$ is the vertical perturbation velocity at the critical location $z_c$. Although, figure \ref{fig8b} has been plotted for the fastest growing Miles mode (and not the neutral mode to which eqn. \ref{eq3.5} applies), even for the fastest growing Miles mode $c_i/c_r \approx 0.001 << 1$, leading to the near discontinuous behaviour seen in the Reynolds stress (see further discussion on this around eqn. \ref{eq3.6}). In figure \ref{fig8b}, the shear layer is between $0 < z/z_c < \Delta/z_c\approx 3.4$ and it is clear that the energy extraction (work done by the Reynolds stress term in eqn. \ref{eq3.4}) from the base-state occurs only within the thin region between the critical layer and the interface. We also contrast figure \ref{fig8b} (where $Fr \sim O(1)$) with its counterparts presented earlier in figure \ref{fig4} $(Fr>>1)$. Unlike figure \ref{fig8b}, sizable contributions to both perturbation kinetic energy and Reynolds stress in figure \ref{fig4} were noted in the region between the start of the shear layer and the critical location ($1 < z/z_c < \Delta/z_c$).
	
\begin{figure}
	\centering
	\includegraphics[scale=0.4]{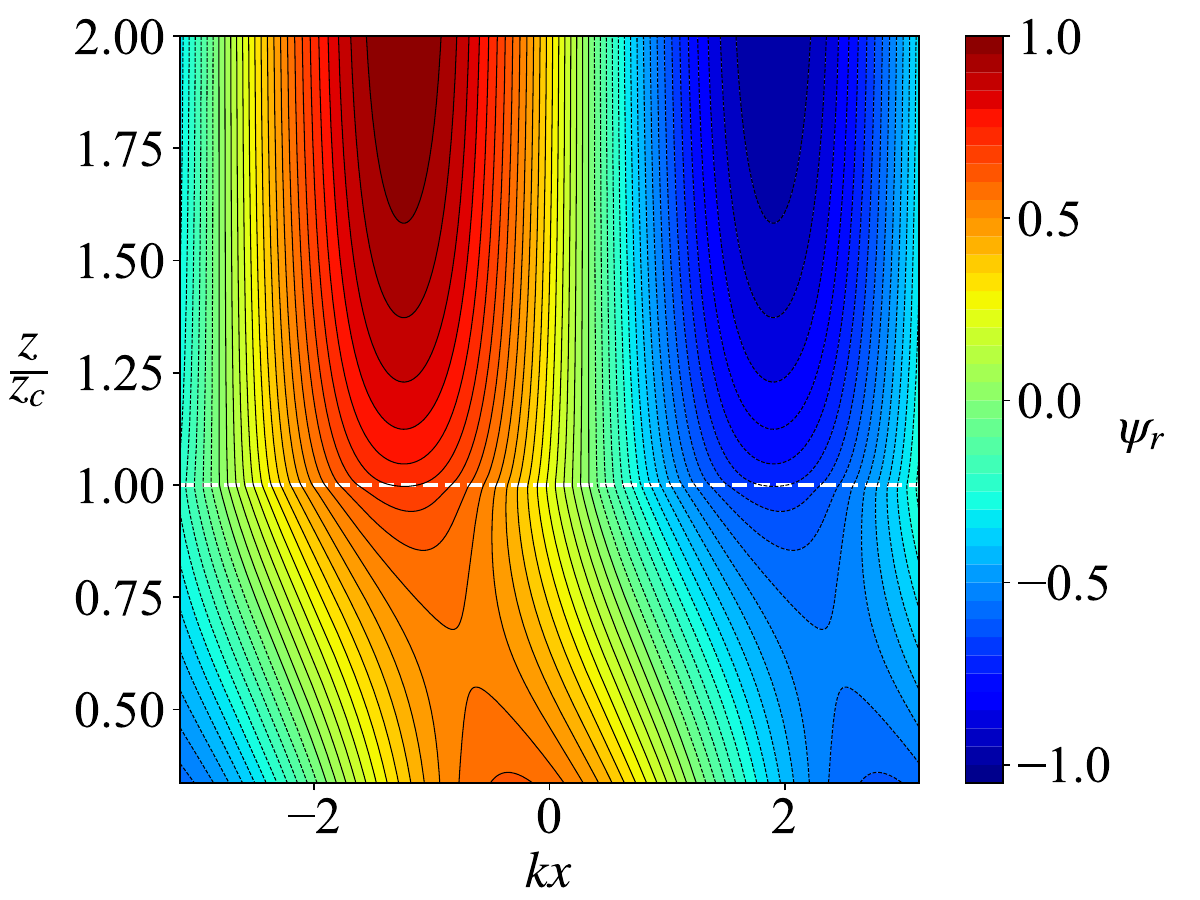}
	\caption{Real part of the perturbation streamfunction $\psi_u(x,z,t=0)$ (eqn. \ref{eq2.3}) for the fastest growing mode showing the tilt in the isocontours indicating a phase-shift. The non-dimensional parameters are same as figure \ref{fig8}. The dotted white line is the critical location.}
	\label{fig9}
	\end{figure}

	Figure \ref{fig9} shows the (real part) of the perturbation stream function for the fastest growing mode in figure \ref{fig8a}. A clear bend in the streamfunction contours is visible at $z=z_c$ (figure $13$a in \cite{young2014generation} notes the same behaviour) indicating the phase-change in the perturbation streamfunction, characteristic of the critical layer. This in turn generates the phase change in perturbation pressure across $z/z_c=1$, causing the perturbation pressure (see eqn. \ref{eq2.7} relating the streamfunction $\Psi(z)$ to perturbation pressure) to have a component in phase with $\dfrac{\partial\eta}{\partial x}$. As lucidly explained in \cite{miles1959generation3}, at air-water density ratio the pressure component in-phase with $\dfrac{\partial\eta}{\partial x}$ is more efficient than the one in phase with $-\eta$ in generating instability; this leads to a much lower threshold of instability for the \textit{critical wind speed} when gravity and capillarity are both accounted for in the exponential model as compared to the corresponding discontinuous KH model (figure $1$ in \cite{kadam2023wind}).
	
\begin{figure}
	\centering
	\subfloat[Perturbation pressure at $z=0^{+}$]{\includegraphics[scale=0.4]{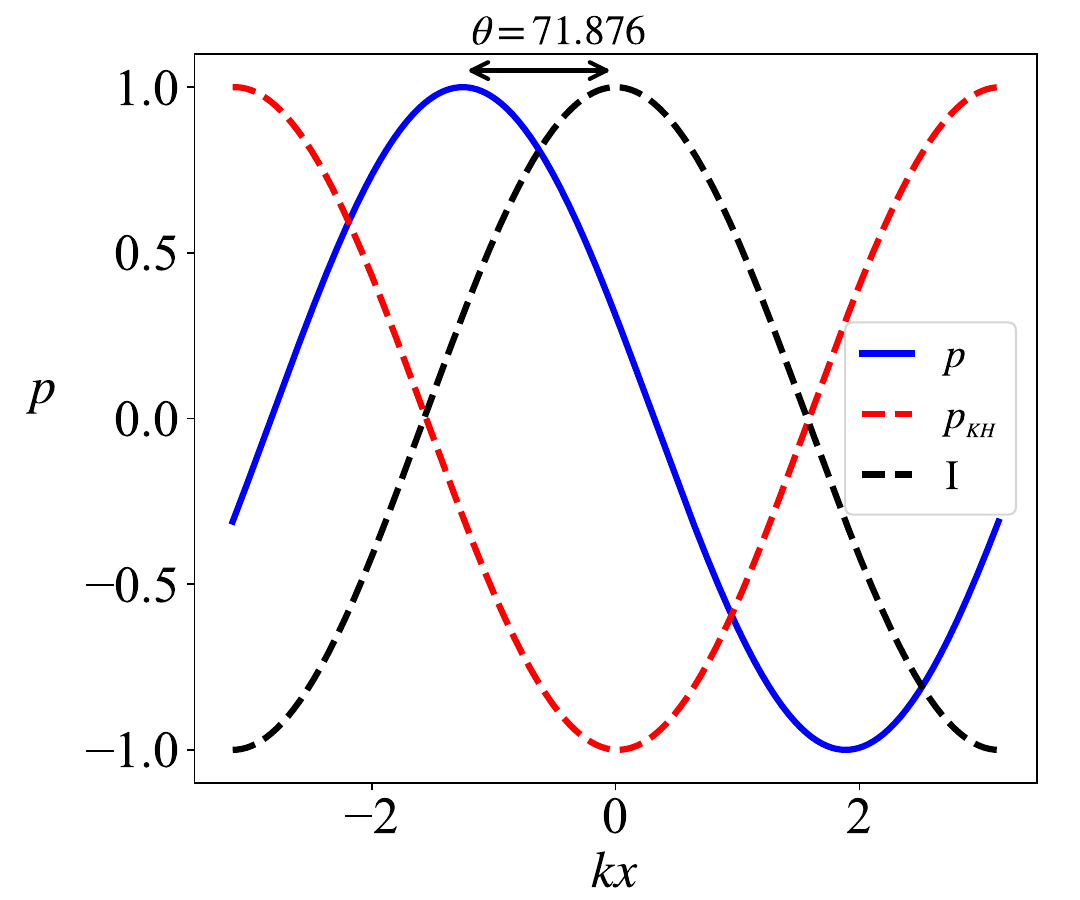}\label{fig10a}}
	\subfloat[Kinetic energy flux in eqn. \ref{eq3.4}]{\includegraphics[scale=0.4]{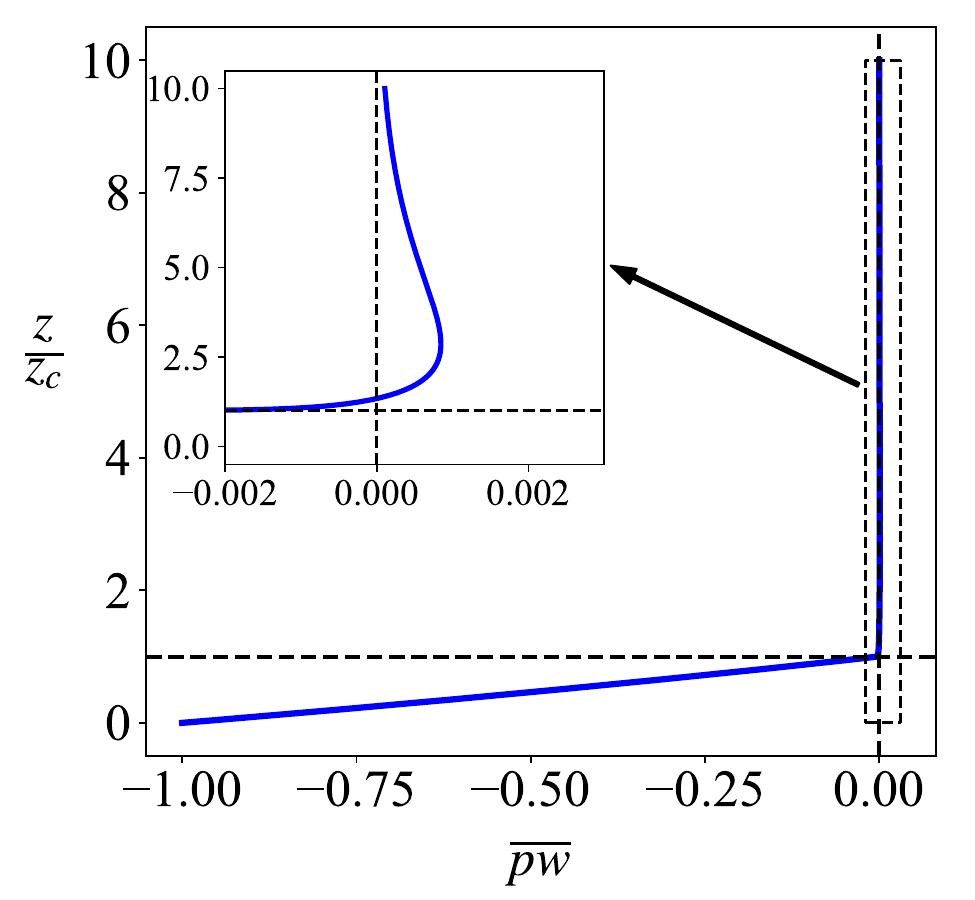}\label{fig10b}}
	\caption{$Fr = 4.79$, $\delta=0.001$, $Bo\rightarrow\infty$. Panel (a) Variation of perturbation pressure for fastest growing mode at $z=0^{+}$ with horizontal coordinate $x$, in figure \ref{fig8a}. The dotted black line represents the real part of the perturbed interface given by $\eta$ in eqn. \ref{eq2.4}. The blue curve corresponds to perturbation pressure for the fastest growing Miles mode in figure \ref{fig8a}. Red dashed line represents the perturbation pressure $p(z=0^{+})$ obtained from the KH model of figure \ref{fig2b} for $\delta=0.001$ and (in CGS units) $U_{\infty}=1000$. Panel (b) The vertical variation of the second term on the right hand side of equation \ref{eq3.4}.}
	\label{fig10}
	\end{figure}

	Figure \ref{fig10a} notes the variation of the perturbation pressure $p_u(x,z=0^{+},t=0)$. If the perturbation pressure just above the interface (air side) were to be entirely in phase with $\dfrac{\partial\eta}{\partial x}$, the phase difference with $\eta$ (black dashed curve in figure \ref{fig10a}) would be $\pi/2$ as would be the case for the Jeffrey model discussed earlier (see discussion around their eqn. $1.1$ in \cite{bonfils2022asymptotic}). For the fastest growing Miles mode in figure \ref{fig10a}, the phase angle ($\approx 72^{\circ}$) is much closer to $\pi/2$ compared to the angle $\pi$ of the KH model (red dashed curve) - we note in passing that the angle $\pi$ for the KH model changes with increasing $\delta$. Figure \ref{fig10b} notes the vertical variation of the kinetic energy flux (second term in eqn. \ref{eq3.4}) for the fastest growing Miles mode, note the sign change in the flux around $z_c$ indicating that a thin layer centered around the critical layer, redistributes bulk of the generated perturbation kinetic energy towards the interface \citep{smyth1989transition} and a much smaller fraction away from it. In the budget for wave energy integrated vertically over the entire domain, the integration of this flux-divergence term in eqn. \ref{eq3.4} equals the potential energy of the interface (purely gravitational in the current case) (eq. $3.2$ in \cite{young2014generation}).
	
	We next turn to examining the effect of increase of density ratio ($\delta$) on the fastest growing mode in the low Froude regime of current interest. In order to rationalise the behaviour of the dispersion relation of eqn. \ref{eq2.12}, it is useful to compare the numerical solution to this against the approximation obtained in the low density ratio limit by expanding $\tilde{c}(\kappa;\delta)$ as $\tilde{c}=\tilde{c}_0(\kappa) + \delta \tilde{c}_1(\kappa) + \mathcal{O}(\delta^2)$ for fixed Froude and Bond number.
	\begin{figure}
	\centering
	\subfloat[Growth rate]{\includegraphics[scale=0.4]{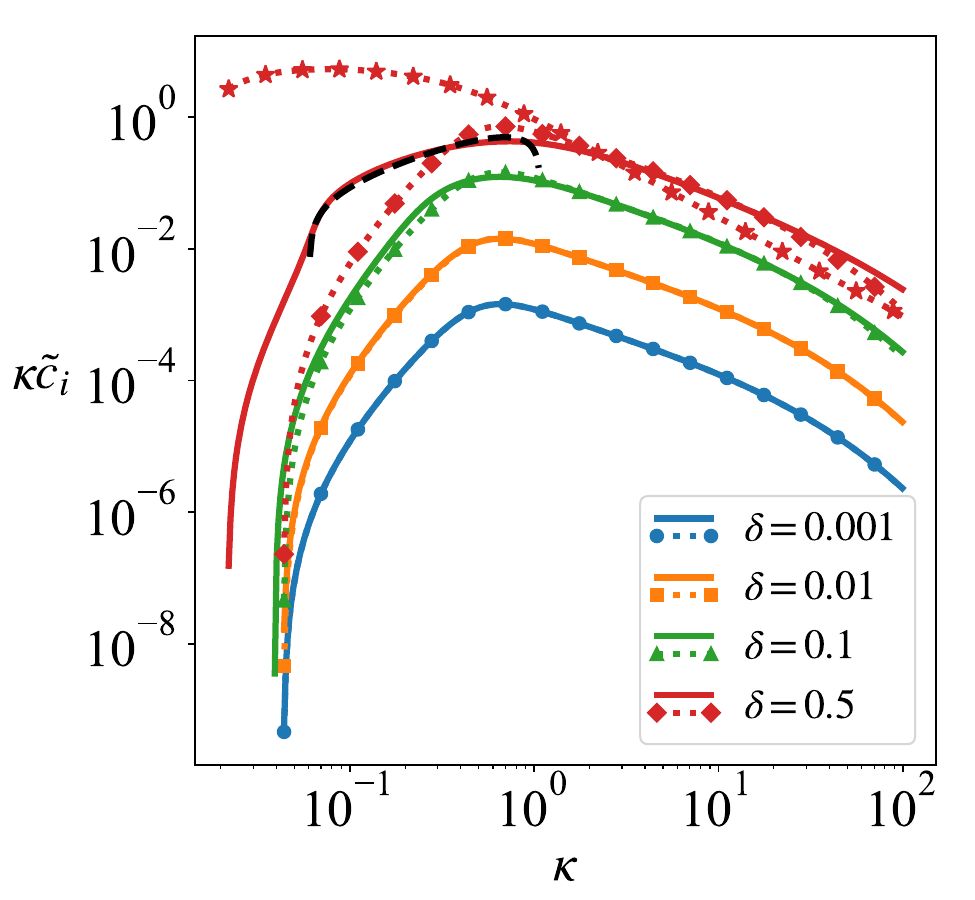}\label{fig11a}}
	\subfloat[Phase-speed]{\includegraphics[scale=0.4]{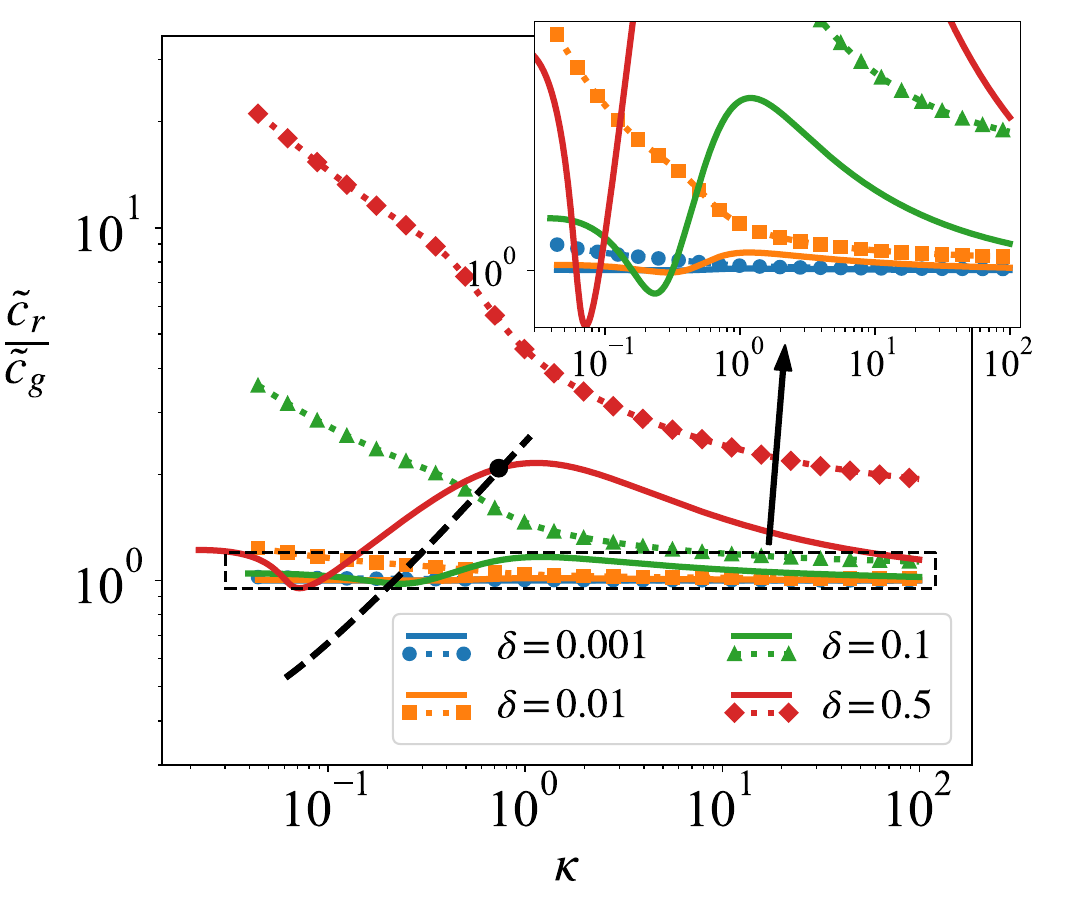}\label{fig11b}}
	\caption{Panel (a) Growth rate and panel (b) phase speed variation with $\kappa$ for the exponential profile. Non-dimensional parameters: $Fr = 4.79$, $Bo\rightarrow\infty$. In both panels, solid lines are from the full dispersion relation in \ref{eq2.12} while big \& small dots are from the approximation \ref{eq3.6}. The dashed black line in both panels is from the PL model in \ref{fig2a} for the same non-dimensional parameters, as above. The curve with stars \& dots, in the left panel, represents a comparison with formulae A31 in Appendix A of \cite{alexakis2004weakly}, valid for small Froude, large wavenumbers and assuming $c_i << c_r$. The black dot in panel (b) is for the fastest growing mode for the exponential model (solid red curve).}
	\label{fig11}
	\end{figure}
	\begin{figure}
	\centering
	\subfloat[Reynolds stress]{\includegraphics[scale=0.4]{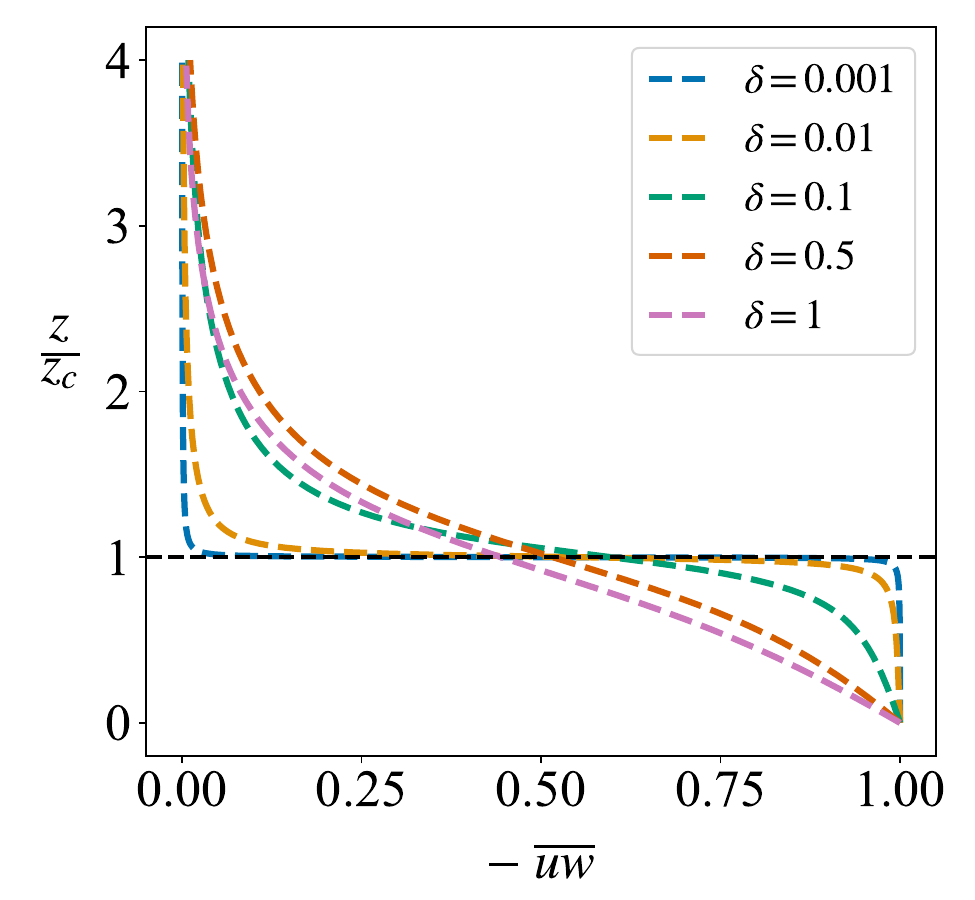}\label{fig12a}}
	\subfloat[Kinetic energy]{\includegraphics[scale=0.4]{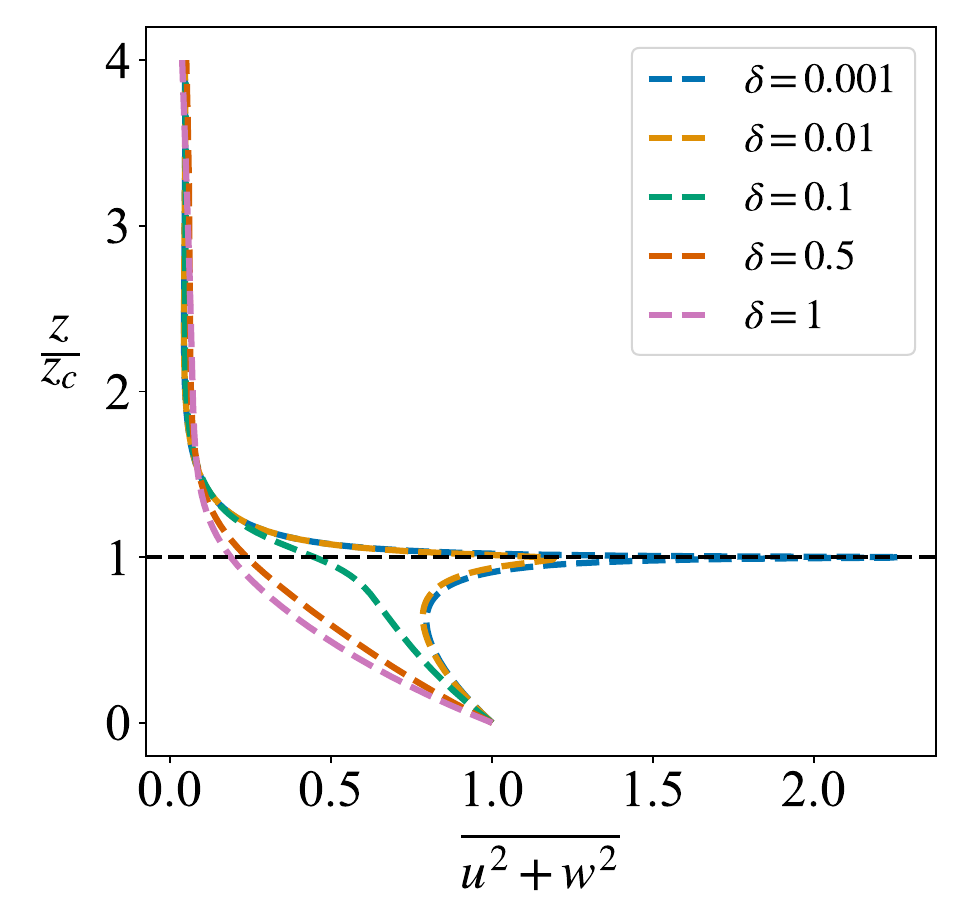}\label{fig12b}}
	\caption{Vertical variation of panel (a) Reynolds stress and panel (b) perturbation kinetic energy for the fastest growing mode for every $\delta$. The non-dimensional parameters: $Fr = 4.79$, $Bo\rightarrow\infty$. In panel (b) note the transition from having a dominant peak at the critical layer to one only at the interface.}
	\label{fig12}
	\end{figure}
	For the zero surface-tension limit relevant here, we use the expression upto $\mathcal{O}(\delta)$ obtained by \cite{carpenter2017physical} (see eqn. $15$ in their study). This, expressed in our notation is
	\begin{align}
	\tilde{c} \approx \tilde{c}_0 + \delta \tilde{c}_1 = \tilde{c}_0 + \dfrac{\delta}{2\kappa}\left[Fr + \tilde{c}_0 \bigg(\Xi\left(\tilde{c}_0, \kappa;Fr\right) - \kappa\bigg)\right], \label{eq3.6}
	\end{align}
	with ($\tilde{c}_0 \neq Fr$)
	\begin{align}
	\tilde{c}_0 \equiv \kappa^{-1/2},\; \Xi(\tilde{c}_0, \kappa;Fr) \equiv -\kappa + \left(\dfrac{1}{1+2\kappa}\right)\left(\dfrac{Fr}{Fr-\tilde{c}_0}\right)\dfrac{_2F_1\left(\alpha+1,\beta+1,\gamma+1;\dfrac{Fr}{Fr-\tilde{c}_0}\right)}{_2F_1\left(\alpha,\beta,\gamma;\dfrac{Fr}{Fr-\tilde{c}_0}\right)}, \nonumber
	\end{align}
	the definitions of $\alpha,\beta$ and $\gamma$ being available below eqn. \ref{eq2.8}. \mgt{We remind the reader that the perturbative expansion \ref{eq3.6} assumes that $\delta\rightarrow0,\;Fr=\mathcal{O}(1)$, thus excluding the high Froude ($Fr\rightarrow \infty$) limit, studied earlier.}
	
	Figures \ref{fig11a} and \ref{fig11b} compare the numerical solutions to the complete dispersion relation from eqn. \ref{eq2.12} (solid lines) with the approximation in eqn. \ref{eq3.6} (big and small dots). One notes the increase in growth rate beyond the linear dependence (for $\delta=0.5$) on $\delta$ predicted by eqn. \ref{eq3.6}. Note from figure \ref{fig11b} that as $\delta$ is increased, the correction to the phase-speed (Doppler shifting) compared to $\tilde{c}_{g}$ (in still air-water \mgt{$\tilde{c}_0 \approx \tilde{c}_g$}) for the fastest growing mode, becomes significant. We assert that this increase in $\tilde{c}_r$ in figure \ref{fig11b} indicates a \textit{change in the nature of the instability for the fastest growing mode}, as $\delta\rightarrow0.5$. To support this argument, we compare the growth rate ($\kappa\tilde{c}_i$) at $\delta=0.5$ for the exponential model (solid red curve in figure \ref{fig11a}) with the PL model (dashed black line in the same figure), noting the good agreement between the two around a region of wavenumbers close to the fastest growing mode; note also the match in the phase-speed in the right panel in figure \ref{fig11b}, although there is evident mismatch for wavenumbers centered around the fastest growing mode (dashed black curve and solid red curve). We have noted in the last section that at low Froude and low density ratio, the instability in the PL model is the Holmboe instability. It is further shown in Appendix B, that the Holmboe instability persists until $\delta \approx 0.5$ at low $Fr$ for the PL model (although KH contributions are also apparent, see figure \ref{figAppB-2}). Given the good agreement seen in the growth rates between the PL and the exponential model in figure \ref{fig11a} at $\delta=0.5$, we thus intuitively anticipate that the exponential model also admits the Holmboe instability, around $\delta=0.5$.
	
	Our aforemention assertion of a change from the Miles instability with increasing $\delta$, is further supported by the vertical variation in Reynolds stress (figure \ref{fig12a}) and perturbation kinetic energy (\ref{fig12b}) plotted for the fastest growing mode at every $\delta$. Note in figure \ref{fig12a} the near discontinuous jump for $\delta=0.001$ at the critical layer noted earlier, giving way at $\delta=0.5$, to a smooth transition through $z=z_c$ bereft of any distinguishing signature at this location; notice also that around $\delta=0.5$, significant contributions appear to the Reynolds stress from the range $z/z_c>1$ unlike the Miles case discussed earlier in figure \ref{fig8b}. \textit{Why does the vertical variation of Reynolds stress in figure \ref{fig12a}, for the fastest growing mode, show a qualitative change, as $\delta$ is increased?} We will address this question in \S\ref{sec:rey_stress}; a noteworthy observation from this section is as $\delta$ increases from $0.001$ to $0.5$ in \ref{fig12a}, the ratio $c_i/c_r$ (\mgt{for the fastest mode}) increases from $\approx0.001$ to $\approx 0.42$.
	\subsection{Effect of capillarity}\label{sec:3.3}
	We discuss the effect of capillarity in this section, restricting ourselves to the low Froude number regime where interesting transitions were recorded. As remarked earlier, the fastest growing mode studied so far are long waves (compared to the shear layer thickness as well as the air-water capillary length) and thus have subdominant contribution from surface-tension. For consistency, we only present high Bond number results here ($Bo=\mathcal{O}_m(88)$) relevant to geophysical and astrophysical situations. A discussion of low Bond number instabilities is eschewed because these can involve shorter waves where the viscous contribution to Reynolds stresses as well as the Tollmien-Schlichting instability can become important \citep{miles1962generation}. In addition, for shorter wavelengths one also expects the inviscid rippling instability \citep{stern1973capillary,morland1991waves, shrira1993surface,young2014generation,kadam2023wind} when shear is also included in the water layer as discussed earlier for \cite{morland1991waves}; the growth rates for this can be significantly more than that of the longer wavelength modes, especially at low density ratio. An important difference here, compared to the infinite Bond number case is the existence of a critical Froude number $Fr_c>0$ for instability. This is easily rationalised at $\delta<<1$. Recall that prograde, capillary-gravity waves propagating at the interface of still water/air (which are destabilised by shear in the upper fluid), have a non-zero, minimum phase speed (unlike the case of surface-gravity modes), thereby leading to a critical Froude number based on this minimum speed.
	\begin{figure}
	\centering
	\subfloat[Growth rate for varying $Fr$]{\includegraphics[scale=0.4]{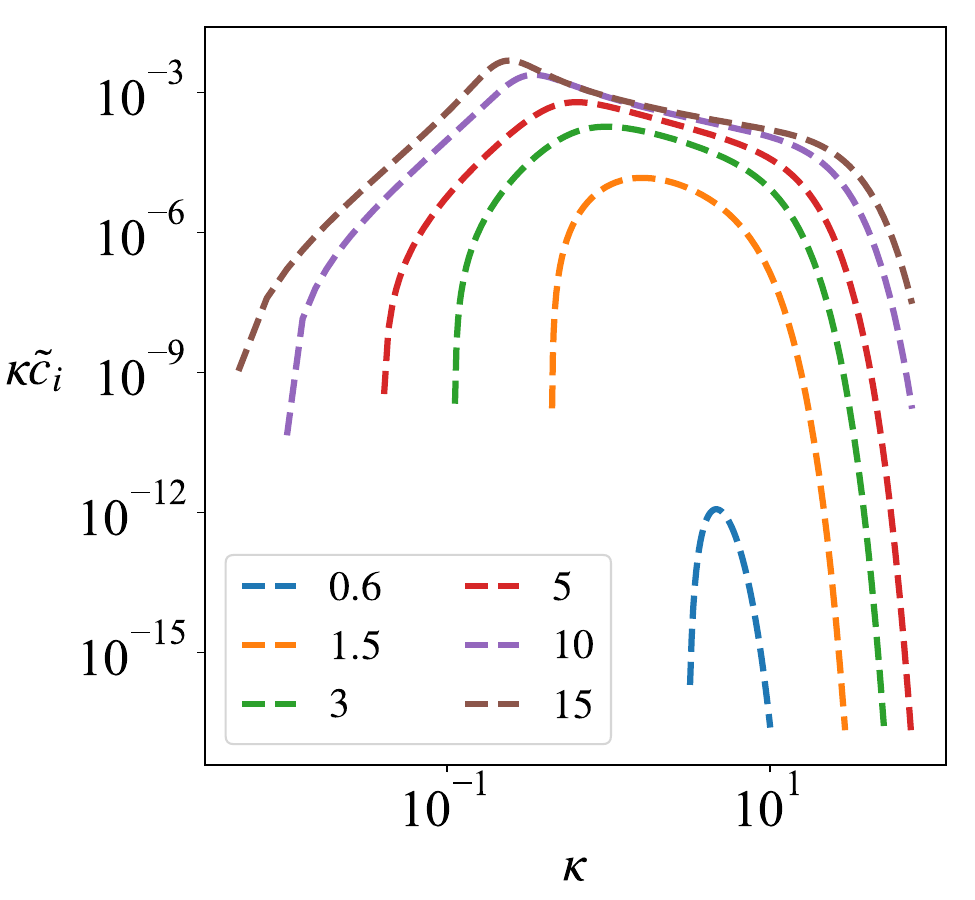}\label{fig13a}}
	\subfloat[Growth rate, phase speed for $Fr =  4$]{\includegraphics[scale=0.4]{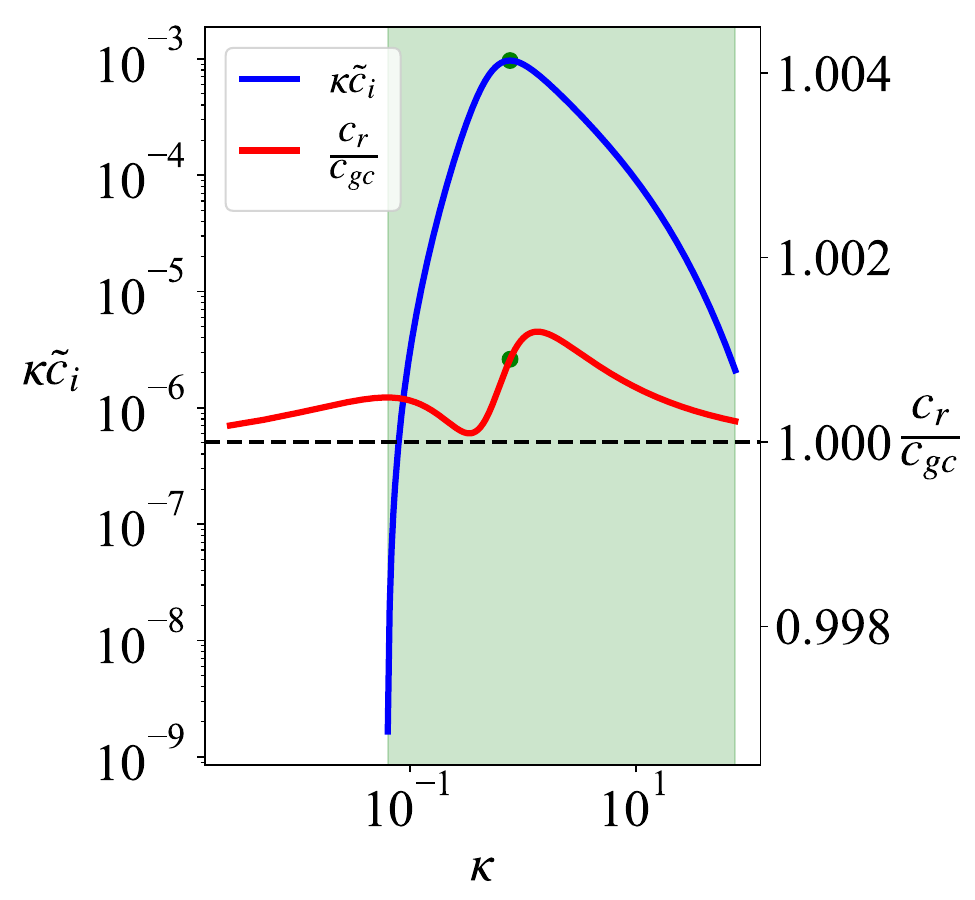}\label{fig13b}}
	\caption{($\delta=0.001,Bo = 87.813$ for both panels), Panel (a) Growth-rate versus wavenumber for varying $Fr>Fr_c\approx 0.5$. The shaded region, indicates the range of wavenumbers, beyond which the growth rate is numerically very small. (Panel (b)) Growth rate and phase speed for  $Fr = 4$. The Froude number ($Fr = 4$) here, has approximately the same value as figure \ref{fig8a}, which however is for infinite Bond compared to $Bo=87.813$ here.}
	\label{fig13}
	\end{figure}
	
\begin{figure}
	\centering
	\subfloat[Reynolds stress]{\includegraphics[scale=0.4]{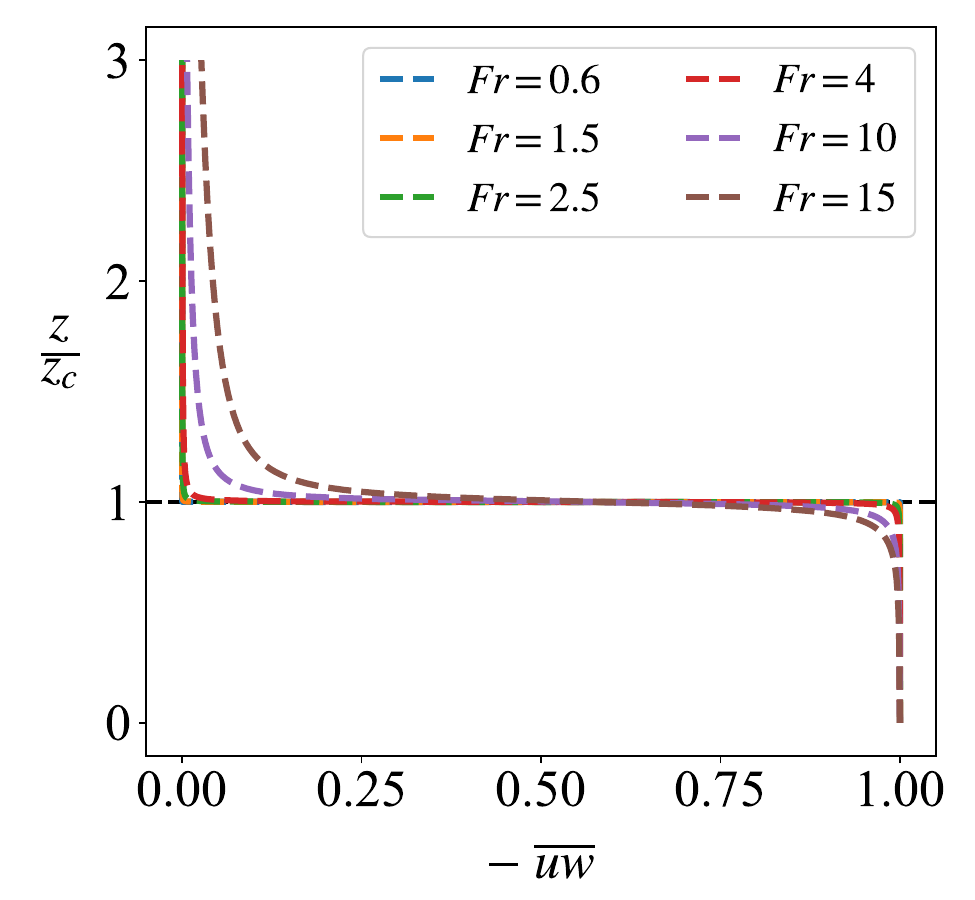}\label{fig14a}}
	\subfloat[Kinetic energy]{\includegraphics[scale=0.4]{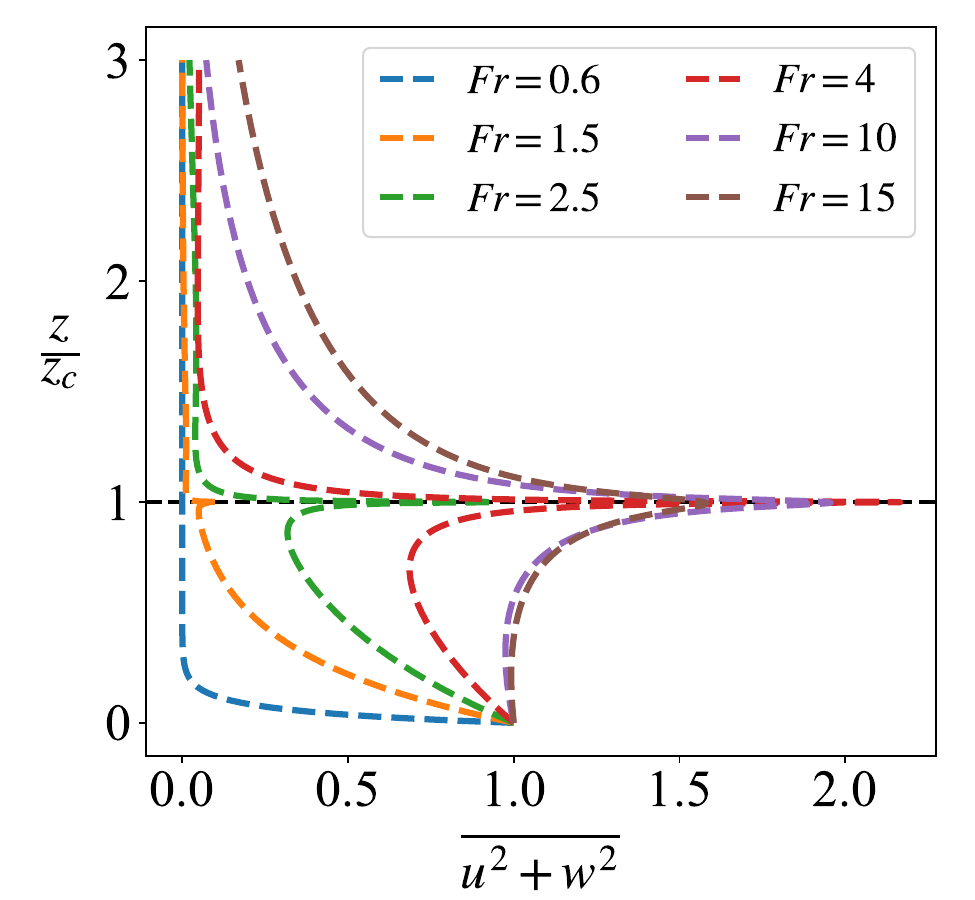}\label{fig14b}}
	\caption{($\delta=0.001,Bo = 87.81$) Panel (a) Reynolds stress and panel (b) kinetic energy variation, for fastest growing mode. In panel (b), note the non-monotonic behaviour of the peak at the critical layer with increasing $Fr$.}
	\label{fig14}
	\end{figure}

\begin{figure}
	\centering
	\subfloat[Reynolds stress]{\includegraphics[scale=0.4]{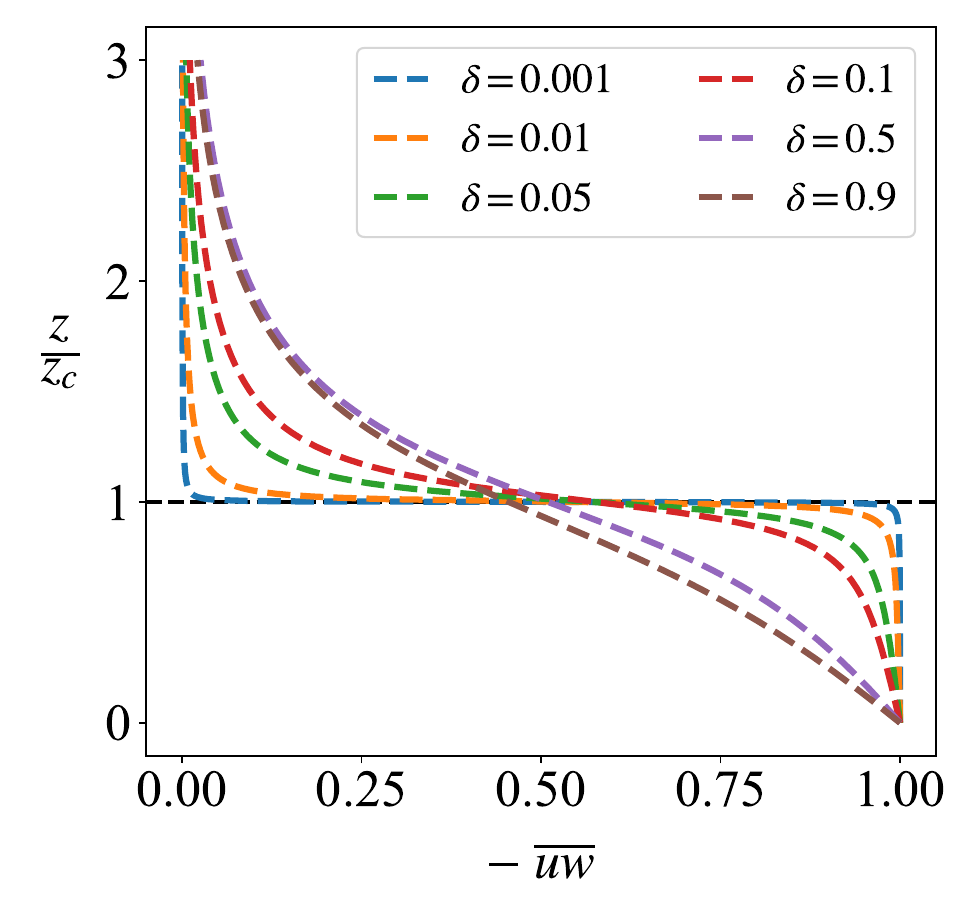}\label{fig15a}}
	\subfloat[Kinetic energy]{\includegraphics[scale=0.4]{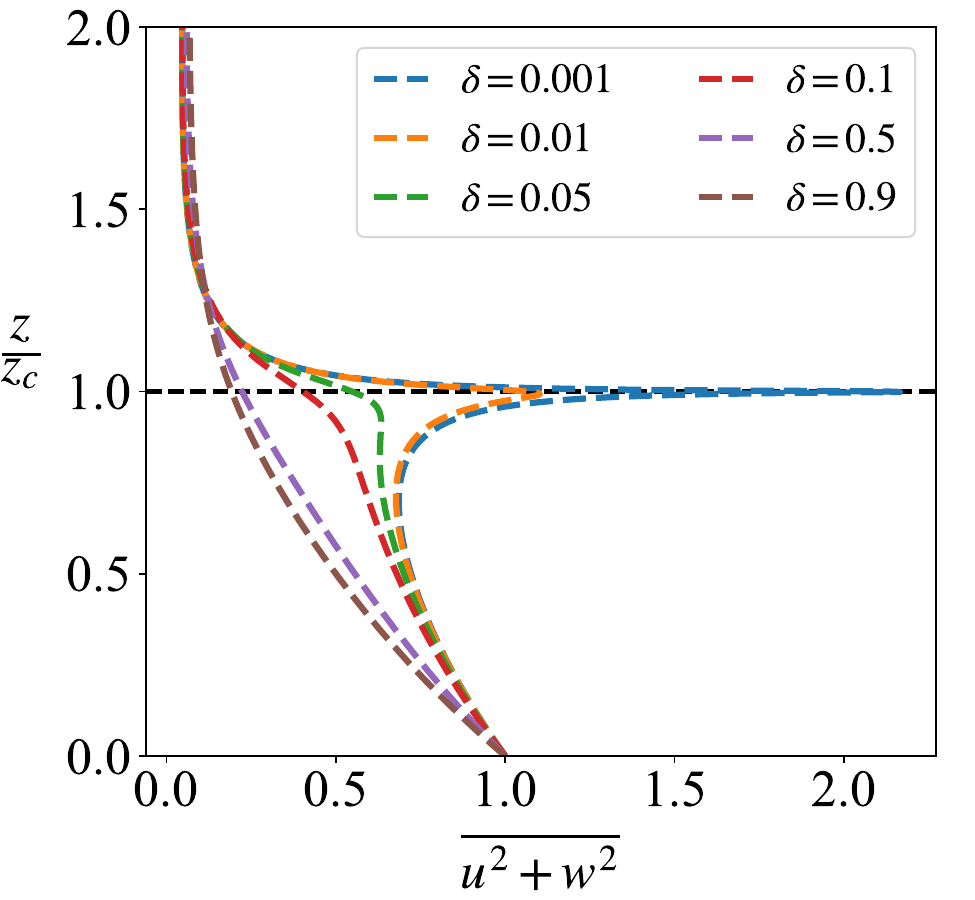}\label{fig15b}}
	\caption{($Fr=4,\;Bo= 87.81$) Panel (a) Vertical variation of Reynolds stress  and panel (b) kinetic energy for fastest growing mode. In panel (b), note the transition from having a dominant peak at the critical layer to one at the interface, with increasing $\delta$.}
	\label{fig15}
	\end{figure}

	Figure \ref{fig13a} presents the growth-rate curves for varying Froude numbers at low density ratio ($\delta=0.001$). We note that $Fr_c\approx 0.5$ and the case with $Fr=0.6$ in the figure, is very close to the instability onset. Figure \ref{fig13b} for $Fr=4$, shows that the fastest growing mode in this case travels almost at $c_{gc}$. Examination of the Reynolds stress and perturbation kinetic energy for $Fr=4$ in figures \ref{fig14a} and \ref{fig14b} establish that these correspond to a classic Miles mode with distinct (near) discontinuities and peaks, respectively at $z_c$. Note the interesting behaviour of the fastest growing mode for $Fr=0.6$ in these figures. There is a sharp near discontinuity at $z_c$ (cf. figure \ref{fig14a}) for $\tau\equiv -\overline{\rho u w}$ but no corresponding signature at $z_c$ in figure \ref{fig14b} for the kinetic energy. To conclude this section, figures \ref{fig15a} and \ref{fig15b} note the $z$ variation of the Reynolds stress and perturbation kinetic energy (for the fastest growing mode in each case) with increasing density ratio; similar to figures \ref{fig12} a clear change in the nature of variation of both metrics at $z/z_c=1$ is observed.

	\subsection{Why does the Reynolds stress become continuous at $z=z_c$ with increasing $\delta$?}\label{sec:rey_stress}
	
	We return to the question: why does the Reynolds stress for the fastest growing mode, show qualitative change with increasing $\delta$? This behaviour is observed both for figure \ref{fig12a} (infinite Bond) and in figure \ref{fig15a} ($\text{Bond}= 87.81$). It turns out that this can be understood from Lin's relation for the vertical variation of Reynolds stress $\tau(z)$ for an unstable mode with phase speed $c = c_r + ic_i$. Using background velocity as the independent variable (in place of $z$), this is:
	\begin{eqnarray}
	\dfrac{d\tau}{dU_u} = F(U_u)\left[\dfrac{c_i}{\bigg(U_u(z)-c_r\bigg)^2+c_i^2}\right], \label{eq3.7}
	\end{eqnarray}
	where $F(U_u(z))\equiv \dfrac{1}{k}\left(\dfrac{dZ}{dU_u}\right)\overline{\rho_uw^2(z)},\; Z(z)\equiv DU_u$. For our exponential profile \ref{eq2.1}, $DU_u$ can be rewritten in terms of $U_u$ as
	\begin{eqnarray}
	Z \equiv DU_u= \dfrac{U_{\infty}}{\Delta}\left[1-\dfrac{U_u}{U_{\infty}}\right] \label{eq3.8}
	\end{eqnarray}
	Using eq. \ref{eq3.8} in eqn. \ref{eq3.7} simplifies to the following non-dimensional form
	\begin{eqnarray}
	-\left(\dfrac{1}{\rho_u\overline{w^2(z)}}\right)\left(\dfrac{d\tau}{d\tilde{U}_u}\right) = \dfrac{1}{\kappa}\left[\dfrac{\tilde{c}_i}{\bigg(\tilde{U}_u(z)-\tilde{c}_r\bigg)^2+\tilde{c}_i^2}\right], \label{eq3.9}
	\end{eqnarray}
	where $\tilde{U}_u \equiv U_u/\sqrt{g\Delta}$.
	Importantly, even for the fastest growing Miles mode in figure \ref{fig8}, $c_i/c_r << 1$. Using now the observation that $\displaystyle\lim_{y\rightarrow0}\dfrac{1}{\pi}\left[\dfrac{y}{(x-x_0)^2+y^2}\right]\rightarrow \delta(x-x_0)$ (\cite{lin1954some}) \mgt{($\delta (x)$ being the \emph{Dirac delta function})}, we see that the quantity inside the square bracket in expression \ref{eq3.9}, for the fastest growing Miles mode has a \textit{near} delta function behaviour and this generates the jump in the Reynolds stress in figure \ref{fig8b}.
	
	\begin{figure}\label{fig16}
	\centering
	\subfloat[]{\includegraphics[scale=0.45]{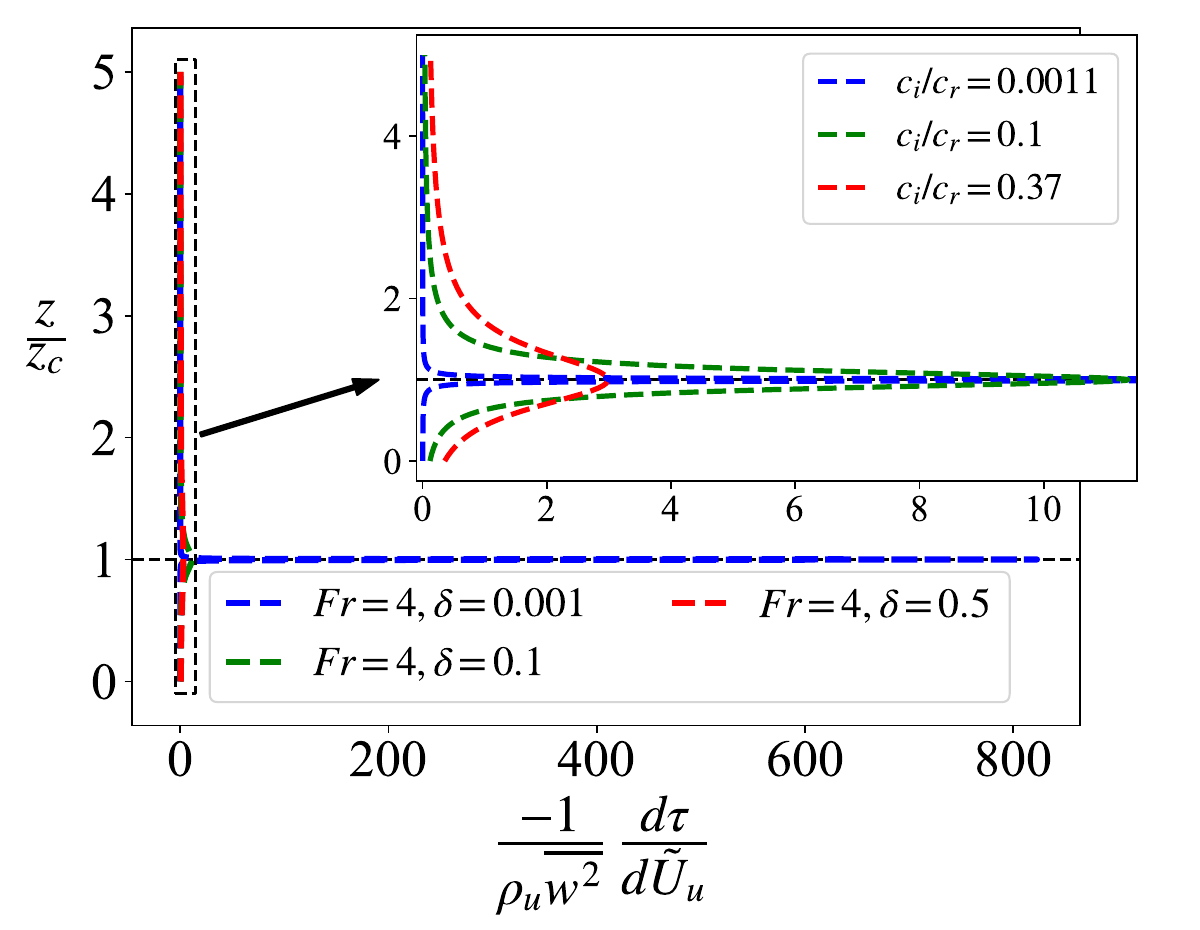}\label{fig16a}}\\
	\subfloat[]{\includegraphics[scale=0.4]{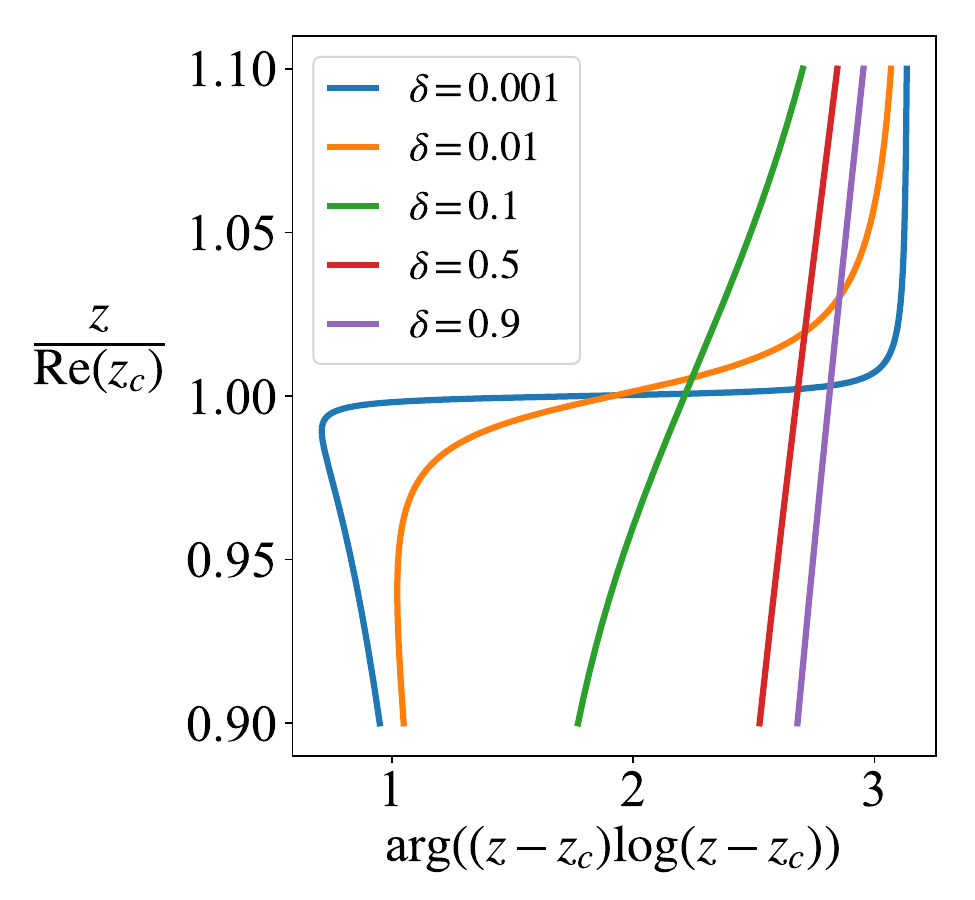}\label{fig16b}}
	\caption{(Panel (a)) Vertical variation of the right hand side of eqn. \ref{eq3.9}, for the exponential profile for different density ratio. In each case, the fastest growing mode has been chosen. Panel (b) Change in the phase of $\left(z-z_c\right)\log\left(z-z_c\right)$. This term appears in Tollmien's singular inviscid solution, see expressions $22.25$ and $22.26$ in \cite{drazin2004hydrodynamic}. For panel (b) $Fr=4,Bo= 87.81$.}
	\end{figure}
	
	To visualize this, in figure \ref{fig16a} we plot the right hand side of eqn. \ref{eq3.9} as function of the vertical coordinate ($z$), for the fastest growing mode in each case. As expected, the low Froude ($Fr=4$) and low density ratio ($\delta=0.001$) has the signature of a Dirac delta function (blue dashed curve) leading to the jump behaviour of $\tau$ in figure \ref{fig15a}. For the dashed green and red curves in figure \ref{fig16a}, notice the widening as the ratio $c_i/c_r$ increases significantly. For $\delta=0.5$ (red curve with $c_i/c_r=0.37$), a large fraction of the area under the curve comes from $z/z_c>1$ with a smaller contribution from $z/z_c < 1$. This generates the observed \textit{smooth} behaviour of $\tau$ through $z_c$ in figure \ref{fig15a}. Figure \ref{fig16b} justifies the transition seen with increasing density ratio from a slightly different viewpoint. Recall that Tollmien's inviscid solutions have a logarithmic singularity when the curvature at the critical location is finite (with $\left(DU_u\right)_c\neq 0$). Figure \ref{fig16b} computes the phase-shift in such a term in a small region around $z/\text{Re}{\left(z_c\right)}=1$ ($z_c$ is a complex number with $Im(z_c)>0$, and is computed for the fastest growing mode using $U_u(z_c)=c(k)$) with $c(k)$ being complex for the fastest growing mode. It is seen that a qualitative change occurs with increasing $\delta$ viz. the abrupt jump in phase seen at $z/\text{Re}\left(z_c\right)$ for the argument of the logarithmic term gives way to a smooth variation in figure \ref{fig16b}. These aforementioned arguments pinpoint the source of the differences seen in the vertical variation of $\tau(z)$ between the low and the high density ratio cases.
	\subsection{Distinguishing between the H and the KH instability for asymmetric profiles}\label{sec:3.5}
	Given the good comparison between the growth rates at $\delta=0.5$ between the PL profile and the exponential profile in figure \ref{fig11a}, we hypothesize that at low Froude and high Bond number, the nature of the fastest growing mode in the exponential profile undergoes a transition - it changes from being dominated by the critical location for $\delta << 1$, to the Holmboe instability for $\delta \approx 0.5$ and eventually the KH instability at even larger $\delta$. We emphasize that the KH to H transition seen clearly in the PL model occurs \textit{smoothly}, in particular at $\delta=0.5$ there seem to be contributions from both (see Appendix B). For the asymmetric, exponential background profile of figure \ref{fig1}, the vertical variation of the perturbation kinetic energy and the Reynolds stress do not show any qualitative change as $\delta$ is changed from $0.5$ to $0.9$, this may be ascertained from figures \ref{fig15a} and \ref{fig15b}. This lack of distinction however is not indicative; even for symmetric curved background profiles, where the KH and the H instability can be distinguished clearly based on phase-speed, the corresponding Reynolds stress and the perturbation kinetic energy do not show qualitative differences in their vertical distribution. \cite{smyth1989transition} comment on this in page $9$, first paragraph (italics and word inside bracket added by us) ``\textit{the two (Holmboe) modes illustrated in figures $\ldots$ are qualitatively very similar to Kelvin-Helmholtz modes$\ldots$}'', emphasizing the challenge of distinction between the two based on these metrics. One may, for example, use the diagnostic proposed in \cite{carpenter2010identifying} to partition the growth rate from the exponential profile, into a Holmboe and a KH contribution - see their figure $9$. We do not do this here, instead taking up numerical simulations of the incompressible, Euler's equations with surface tension and gravity; in the nonlinear regime captured through these simulations, clear differences will be seen to emerge with increasing density ratio. \mgt{These simulations will be seen to clearly distinguish between the H and the KH instability}, as described next.
	
	\section{Numerical simulations - Linear and nonlinear regime}\label{sec:sim}
	\begin{figure}
		\centering
		\subfloat[\mgt{Simulation domain - $\big(u(x,z,t=0),w(x,z,t=0)\big) = \text{Background state} + \text{small amplitude perturbation velocity field}$. }]{\includegraphics[scale=0.3]{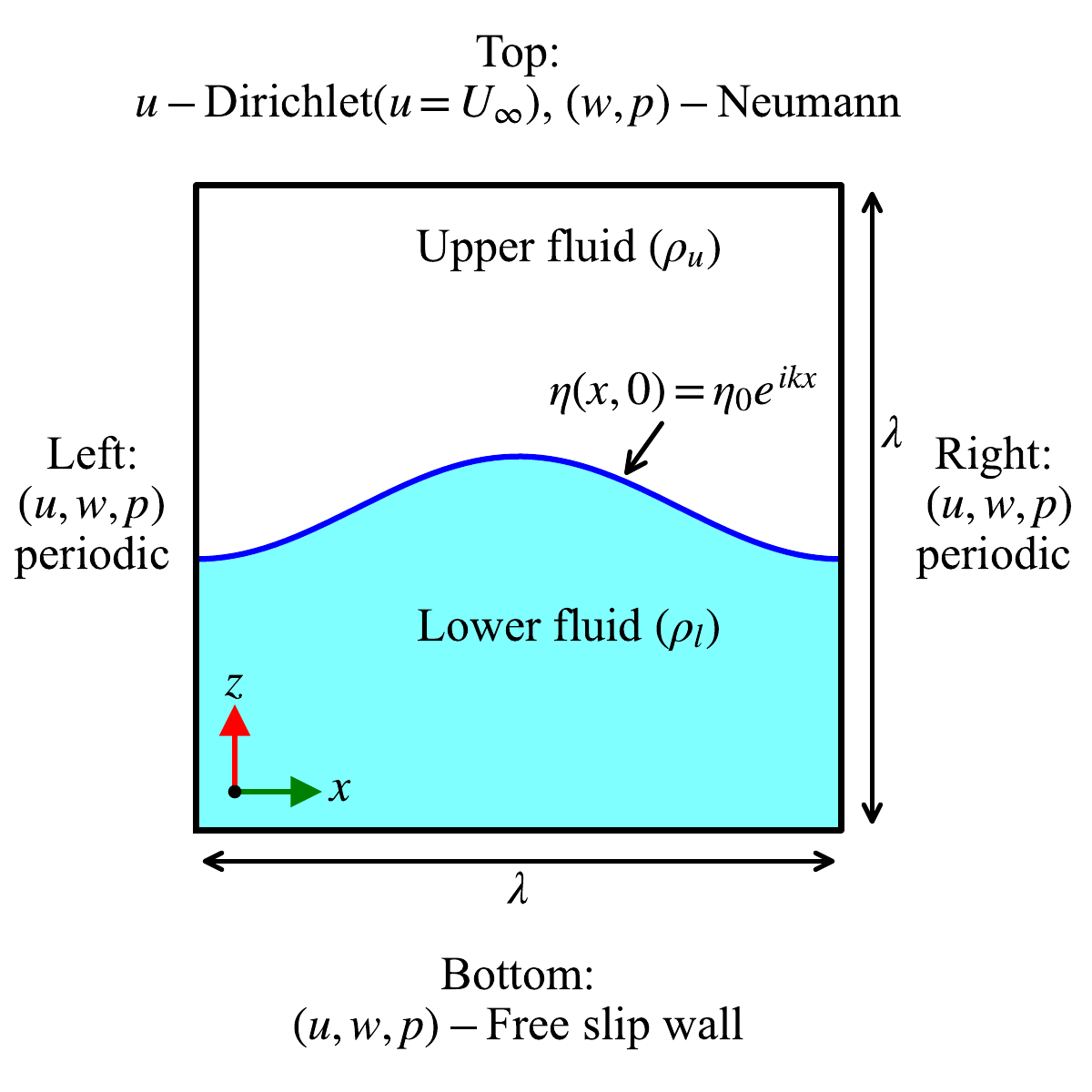}\label{fig17a}}\hfil
		\subfloat[\mgt{Background state and perturbation velocity fields}]{	\includegraphics[scale=0.3]{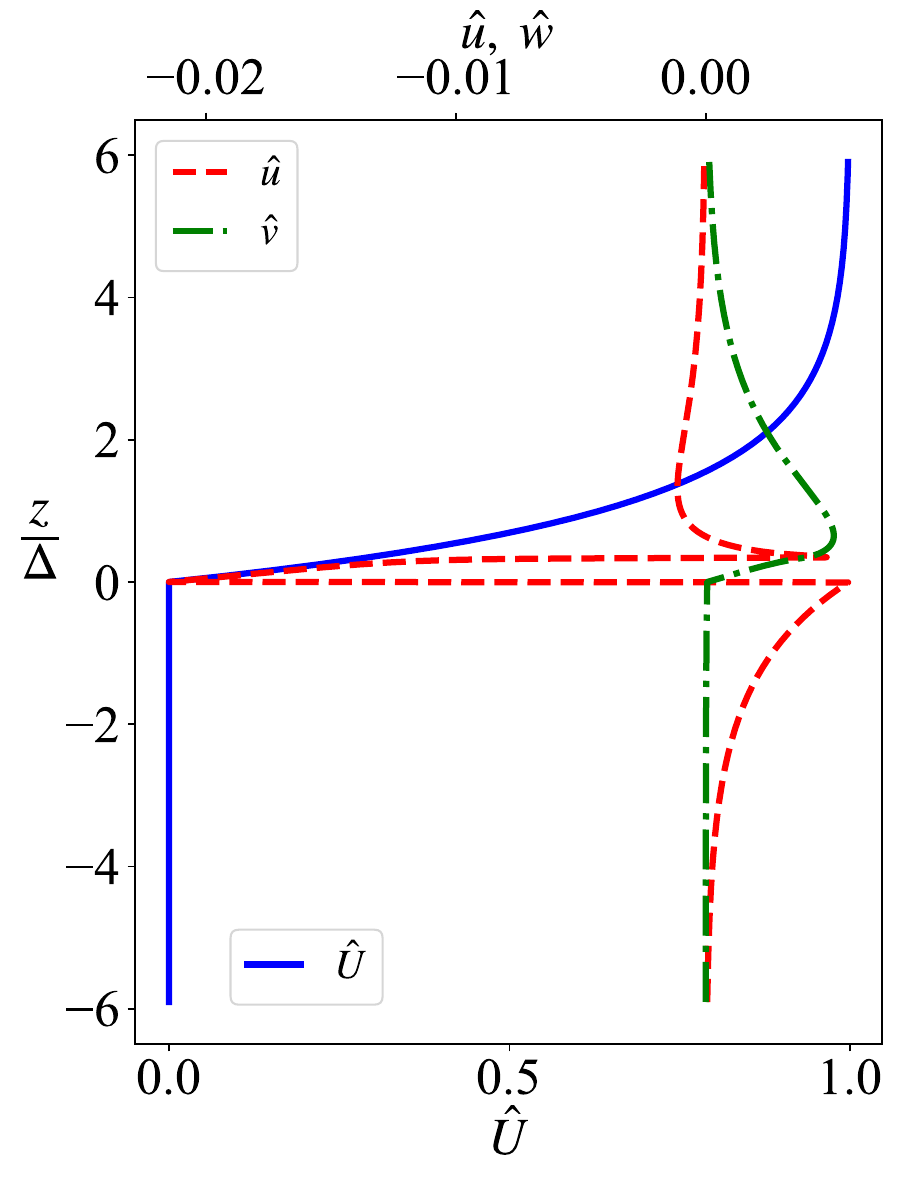}\label{fig17b}}\\
		\subfloat[Adaptive grids used in simulations]{\includegraphics[scale=0.2]{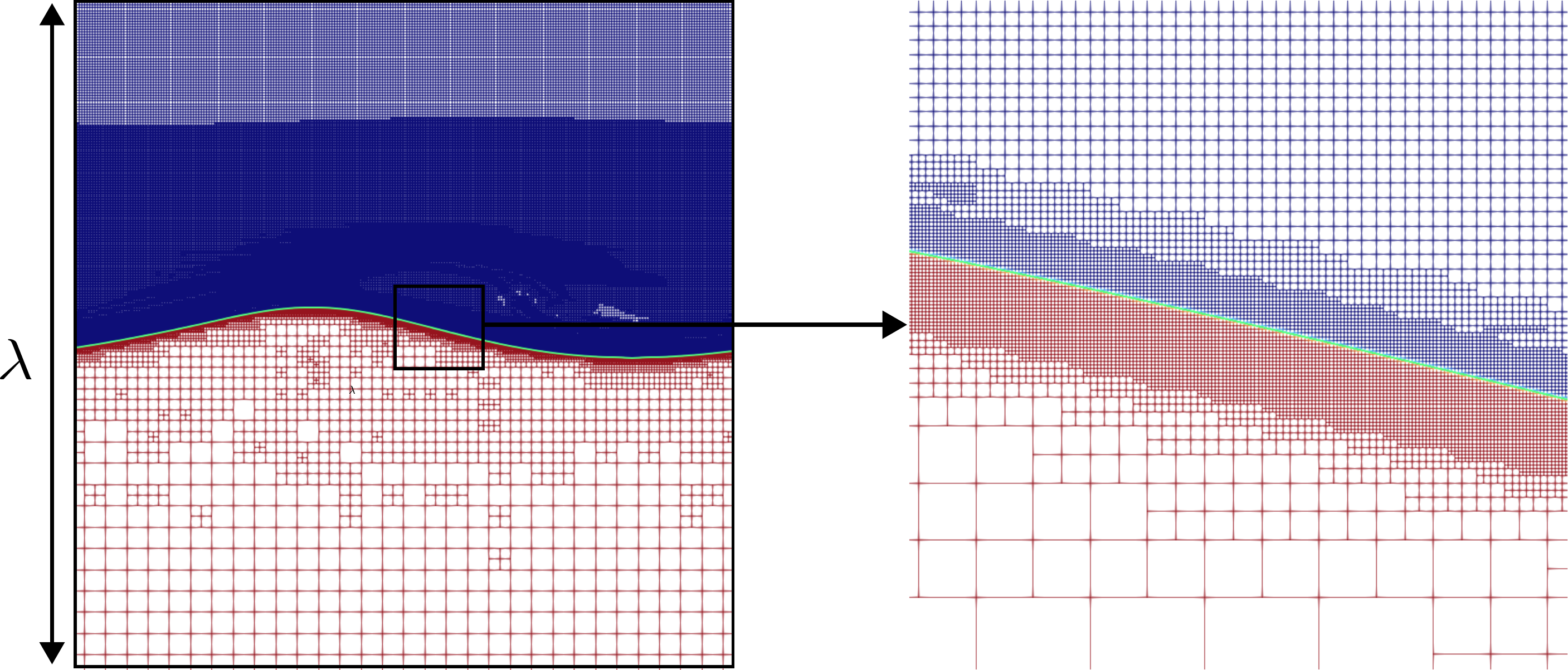}\label{fig17c}} \\
		\caption{Panel (a) A schematic of the computational domain alongwith boundary conditions implemented. (b) \mgt{Vertical variation of fastest growing eigenmodes and background shear. A superposition of background shear with the fastest eigenmode with small amplitude, is used as the initial velocity field in all simulations. $\left(\hat{U},\hat{u},\hat{w}\right) \equiv \left(U/U_{\infty}, u/U_{\infty},w/U_{\infty}\right)$} (c) Sample adaptive grid.}
		\label{fig17}
	\end{figure}

	In this section, we present results from numerical simulations obtained by solving the incompressible, Euler's equations using the open-source code Basilisk \citep{popinet2025basilisk}. We restrict ourselves to the low Froude and high Bond number regime discussed earlier. To be consistent with the inviscid stability model discussed so far, we solve the incompressible Euler's equations with gravity and surface-tension, in Basilisk, with an interface \mgt{separating two fluids}. The equations which are solved read as:
	\begin{subequations}\label{eq4.1}
	\begin{align}
		& \bm{\nabla.u}=0,\label{eq4.1a}\\
		& \dfrac{\partial\bm{u}}{\partial t}+\bm{\nabla}.(\mathbf{u\otimes u}) = - \dfrac{\bm{\nabla}p}{\rho} + g + \dfrac{T}{\rho}\kappa \delta_s \bm{n},\label{eq4.1b} \\
		& \dfrac{\partial f}{\partial t}+\bm{\nabla.}(f\bm{u})=0, \label{eq4.1c}
	\end{align}
	\end{subequations}
	where $\mathbf{u},\,p,\,\kappa$, $T$ and $f$ are the velocity field, pressure field, interface curvature, surface-tension and the volume-fraction field, respectively. \mgt{The velocity field is $\mathbf{u} = (u,w)$, where $u$ and $w$ are horizontal and vertical components, respectively. Basilisk implements the volume-of-fluid (VoF) algorithm in a one fluid formulation, for reconstruction and advection of the interface separating the two fluids. This is done alongwith time marching of the accompanying velocity and pressure fields. A volume fraction field $f(x,z,t)$ is defined which takes values \mgt{$\in [0,1]$}. Specifically, in all computational cells containing only the upper or lower fluid, $f$ takes the value $0$ or $1$. By definition thus, the interface is present in those cells where $0 < f < 1$, the value representing the fraction of the cell that is occupied by one of the two fluids. Given an $f$ field at any time instant, a reconstruction algorithm computes the interface using a piece-wise linear approximation for each computational cell where $0 < f < 1$. The $f$ field is advanced in time employing a discrete version of eqn. \ref{eq4.1c}; volume fluxes through cell walls being computed geometrically ensuring high accuracy of volume conservation, see \cite{tryggvason2011direct} and references therein. The density and viscosity fields are computed as weighted average of their respective values in the two phases, employing the local value of $f$ as the weight.} 
	
	\mgt{It is important to note that the VoF algorithm constructs a sharp representation of the interface. This representation is not just for visualization, but is also explicitly employed to estimate the volume fluxes, for each computational cell, while time-stepping eqn. \ref{eq4.1c}. Surface tension is incorporated as a body force into the momentum equations and curvature is computed using a generalised, height-function algorithm (sec. $5$ in \cite{popinet2025basilisk}). The code has been benchmarked extensively, particularly at high Reynolds numbers including inviscid test cases \citep{kayal2025focussing,popinet2003gerris}. We have, in particular, extensively employed the adaptive grid feature of this code in our current simulations. The criteria for adaptivity is based on changes to the volume fraction field $f$, curvature $\kappa$ and the velocity field, when subject to local grid refinement. The computational domain length has been chosen to match the wavelength $\lambda$ of the fastest growing mode, obtained from theory. The smallest local grid size is $\text{(Domain size)}/2^{13}=\lambda/2^{13}$. This level of adaptivity is referred to as $2^{13}$, for example, in figure \ref{fig19a} and similarly in other figures. In some of the simulation results which follow, qualitative features are observed to remain unchanged under grid refinement, although quantitative changes are apparent in certain cases. A sample simulation file for the simulation reported in figure \ref{fig20}, is also provided \citep{basilisk_shear_flow}.}
	
	Figure \ref{fig17a} depicts the simulation domain alongwith boundary conditions while \ref{fig17b} shows a sample adaptive grid employed. The initial conditions in our simulations are obtained from linear theory presented in \S\ref{sec:lin_stab}. In both upper and lower fluids, we initialize the velocity field using a velocity-field represented as base-state (eqn. \ref{eq2.1}) plus perturbation \mgt{(eqn. \ref{eq2.3})}. The perturbation field is obtained from the analytically computed eigenfunctions in eqn. \ref{eq2.11}. Note that there is only one free-parameter in our linear theory, viz. the value of $\eta_0$. The value of this is chosen such that the steepness $k\eta_0=0.02$ for all simulations, consistent with the linear theory assumed (\mgt{small steepness}). The interface is initialised using eqn. \ref{eq2.4} employing the fastest growing wavenumber in each case, as computed from the dispersion relation in eqn. \ref{eq2.12}, \mgt{for a Fourier mode} travelling rightwards.

	\begin{figure}
		\centering
		\subfloat[Evolution of interface shape]{\includegraphics[scale=0.4]{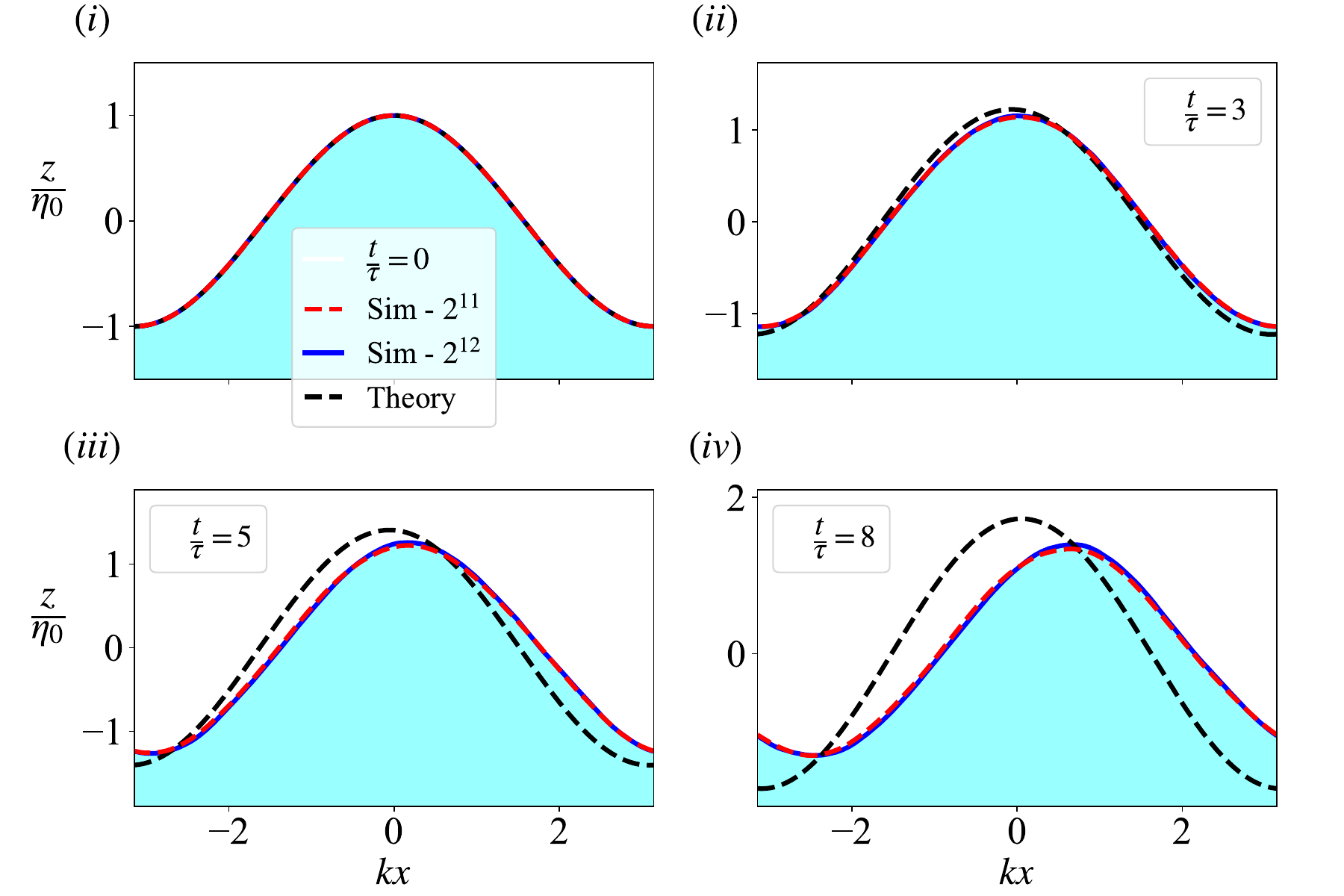}\label{fig18a}}\\
		\subfloat[Growth with time]{\includegraphics[scale=0.4]{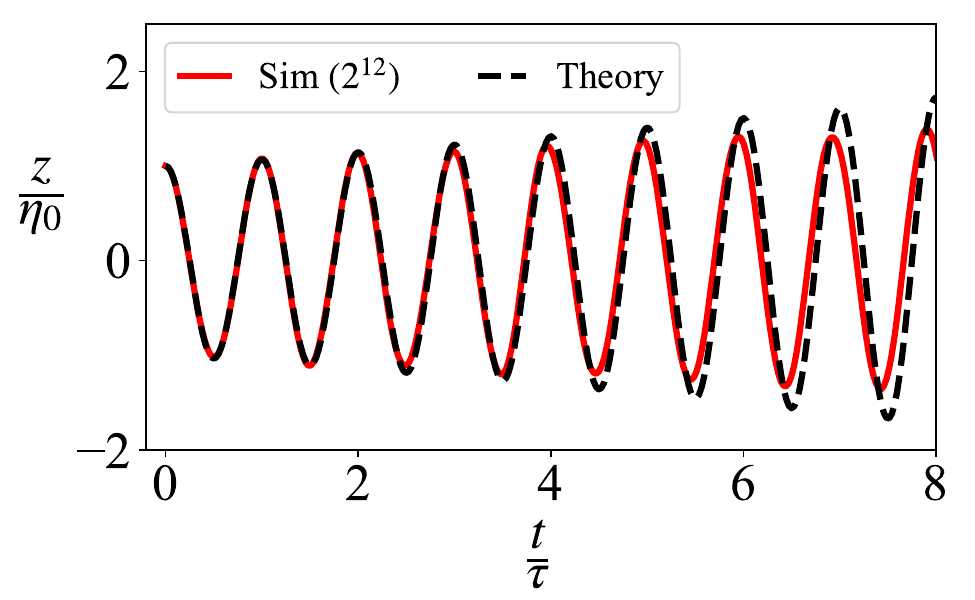}\label{fig18b}}
		\caption{$Bo=87.81,\; Fr=4,\;\delta=0.01$, \mgt{Rounded to two decimal places $\Delta/\lambda=0.12$.} Panel (a) Sub-panels $(i)-(iv)$ compare the interface shape from theory and simulations. Panel (b) The time signal obtained by tracking interface displacement at $x=0$ (horizontal mid-point of the domain). \mgt{$\tau$ is the time taken for a Fourier mode to travel one wavelength.}}
		\label{fig18}
	\end{figure}
	
	Figure \ref{fig18a} shows simulations for the case for $Fr=4, Bo=87.81$ and $\delta=0.01$. The dimensional parameters for figure \ref{fig18a} are $U_{\infty} = 199.57$ cm/s, $\Delta=2.54$ cm, $\rho_l=1$ gm/cm$^3$ and $\rho_l=0.01$ gm/cm$^3$. For these parameters, the fastest growing mode in the exponential profile has wavelength $\approx 20.9$ cm.
	We have seen earlier in figure \ref{fig15a} that the Reynolds stress vertical distribution with a sharp jump at the critical location, is nearly identical for $\delta=0.001$ (air-water) as well as $\delta=0.01$, indicative of the critical location dominating the instability. In figure \ref{fig18a}, we present the evolution of the interface from theory and simulations for this Miles regime. Note that time is measured in units of the duration ($\tau$) taken for the wave to travel its wavelength; notice the slow growth in figure~\ref{fig18b}. As the wave amplitude significantly exceeds its initial value, a mismatch with linear theory is apparent by $t/\tau=5$ (five wave periods). We note that simulations for air-water density ratio ($\delta=0.001$) were also attempted; however due to the extremely small growth rate, long simulations are necessary. Computational inaccuracies in such long runs prevented detection of any exponential growth and thus are not reported here. Figure \ref{fig18} is the smallest density ratio ($\delta=0.01$) where we were able to detect growth and compare against linear theory, while being in the critical layer dominated regime. Figure \ref{fig18b} reports the interface displacement at the center of the domain and a slow exponential growth alongwith a reasonably good match with theory is apparent. Note that simulations were carried out at three adaptive resolutions: $2^{11}$, for example, represents the upper limit of number of grid points along each direction. A reasonably good grid convergence is apparent in figure \ref{fig18a}.
	
	Figure \ref{fig19} is at the same Froude as figure \ref{fig18} but for density ratio one hundred times that of air-water. We recall that even at these elevated value of $\delta$, the Reynolds stress for the fastest growing mode in figure \ref{fig15a} retains a strong signature at the critical layer similar to the Miles instability for air-water. An interesting phenomena is apparent at $t/\tau= 5.6$ (see inset of Panel (a), sub-panel ($ v$)). A series of ripples developed on the backward face of the parent wave - some grid sensitivity is noted in their wavelengths. The inset to this sub-panel, describes the interface at a slightly later time $t/\tau=6.5$ where the ripples are more clearly seen. These are reminiscent of ripples on capillary-gravity Stokes waves, \mgt{see fig. $2$ in \cite{hung2009formation} for the vortical structure of such ripples}. For comparison, we have provided in panel (c), an irrotational, left to right propagating Stokes wave \mgt{($Bo=384.615,\; Fr=0.4389$)}, computed in the limit of $\delta\rightarrow 0$ with gravity and surface-tension. This is a non-linear, travelling wave solution to the potential flow equations and qualitatively similar structures are seen on the Stokes wave with a marked asymmetry between the forward and the backward face. Qualitative similarities may be seen between panel (c) and the inset in sub-panel $(v)$ of figure \ref{fig19}. \mgt{We have also compared numerical simulations for $\delta=0.01$ (figure \ref{fig18}) and for $\delta=0.1$ (figure \ref{fig19}), in the nonlinear regime upto the same value of $t/\tau_{_G}\approx 3.5$ (data not shown here). Here $\tau_{_G}\equiv \left(kc_i\right)^{-1}$ is the time-scale of growth of the instability (which depends on density ratio) and is distinct from $\tau = 2\pi/(kc_r)$ (as is done in figure \ref{fig18} and \ref{fig19}). We observe that the finite amplitude wave for $\delta=0.01$ has a far smaller steepness and exhibits tiny, ripple like structures on its surface, compared to that for $\delta=0.1$ where the finite amplitude wave is significantly steeper (slope) and displays larger amplitude ripples (upto grid precision).}
	
	\begin{figure}
		\centering
		\subfloat[Evolution of interface shape]{\includegraphics[scale=0.4]{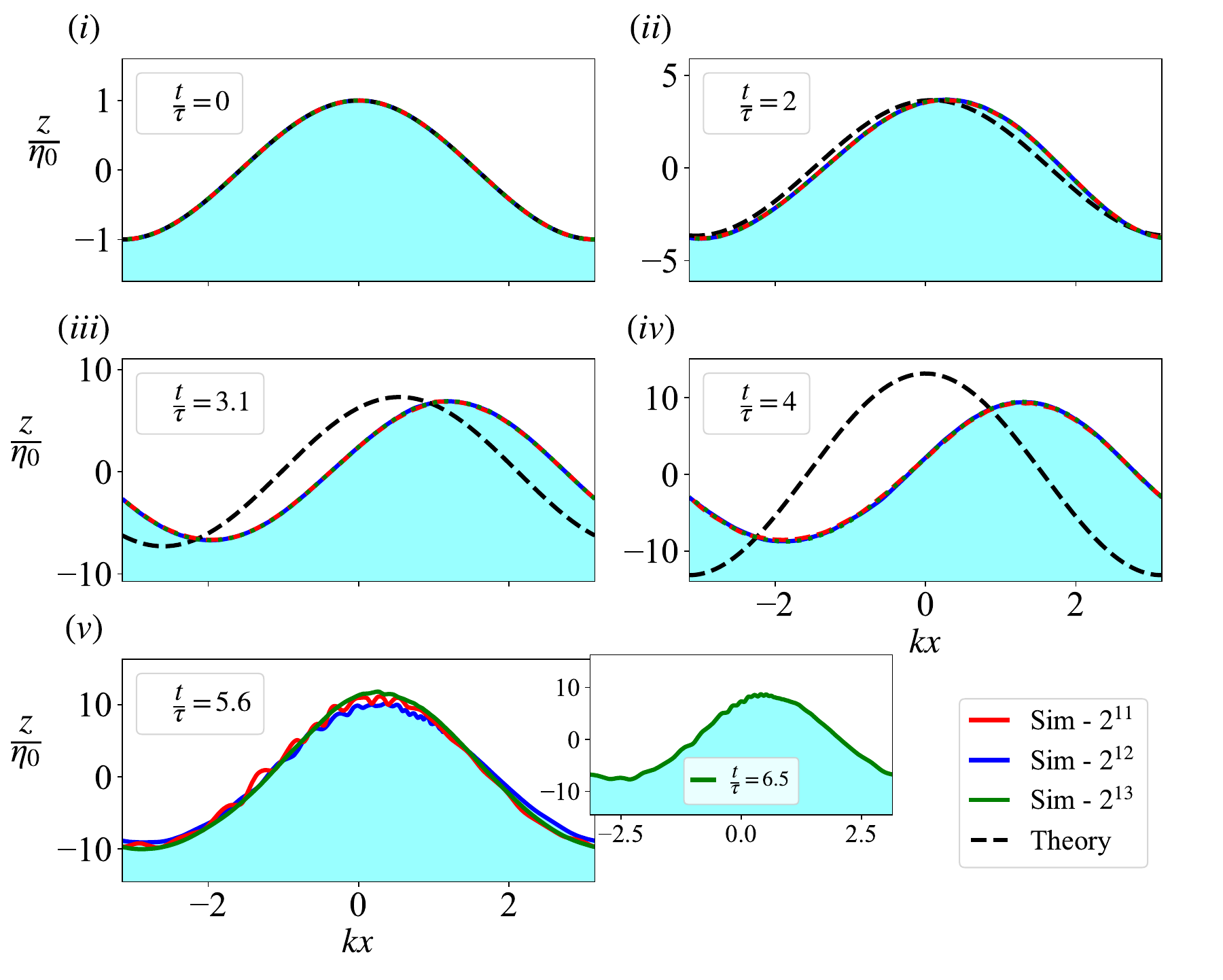}\label{fig19a}}\\
		\subfloat[Growth with time]{\includegraphics[scale=0.4]{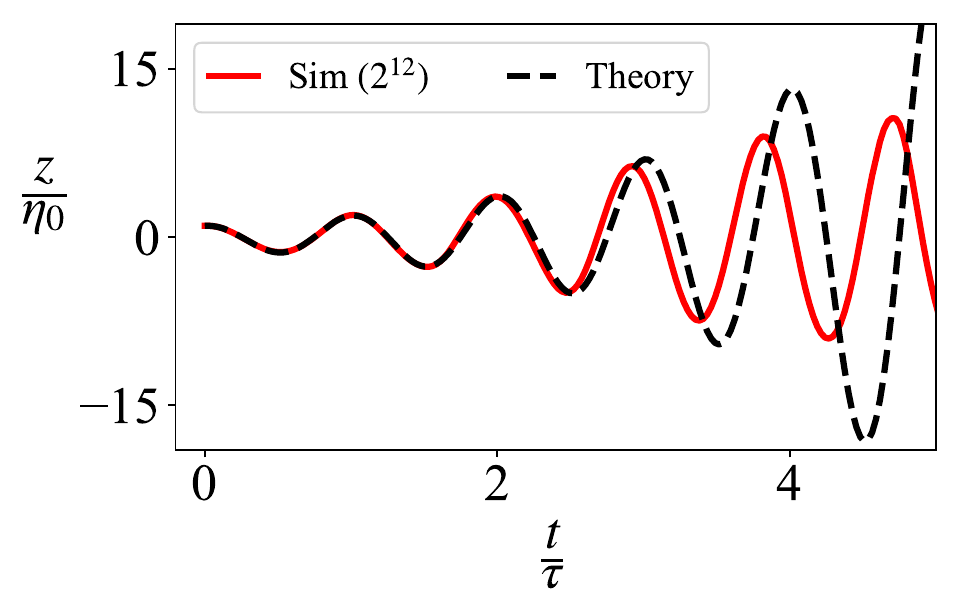}\label{fig19b}}
		\subfloat[Capillary-gravity Stokes wave]{\includegraphics[scale=0.4]{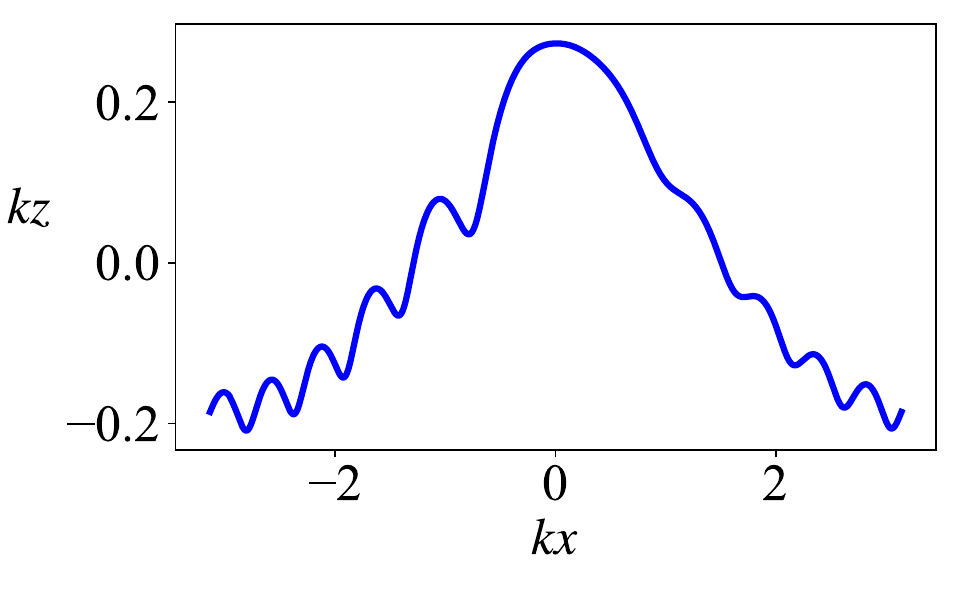}\label{fig19c}}
		\caption{$Bo=87.81,\; Fr=4,\;\delta=0.1$, \mgt{Rounded to two decimal places $\Delta/\lambda=0.12$.} Panel (a) sub-panels $(i)-(v)$ compare the interface shape from theory and simulations. Panel (b) Growth with time at $x=0$. Notice the mismatch at larger time. Panel (c) \mgt{A finite-amplitude, capillary-gravity Stokes waves, containing finer-scale ripples co-moving with the wave. The shape is computed by solving the nonlinear, potential flow equations including gravity and surface-tension following \cite{Shelton_Milewski_Trinh_2025}. For this nonlinear wave which is obtained by evolving the gravity-capillary Stokes wave in time, the initial steepness is $0.24$, Froude number $Fr \equiv \dfrac{c_p}{\sqrt{g\lambda}}=0.4389$ and $Bo \equiv \dfrac{\rho g \lambda^2}{T} = 384.615$,  with $\delta=0$ (no upper fluid). The shape of the wave as depicted here, is obtained by propagating the initial Stokes wave and taking a snapshot at $t/\tau_S=2$. Here $\lambda,c_p$ and $\tau_S$ represent the wavelength, nonlinear phase-speed and the time taken to travel one wavelength respectively, for the capillary-gravity Stokes wave.}}
		\label{fig19}
	\end{figure}
	
	\begin{figure}
		\centering
		\includegraphics[scale=0.4]{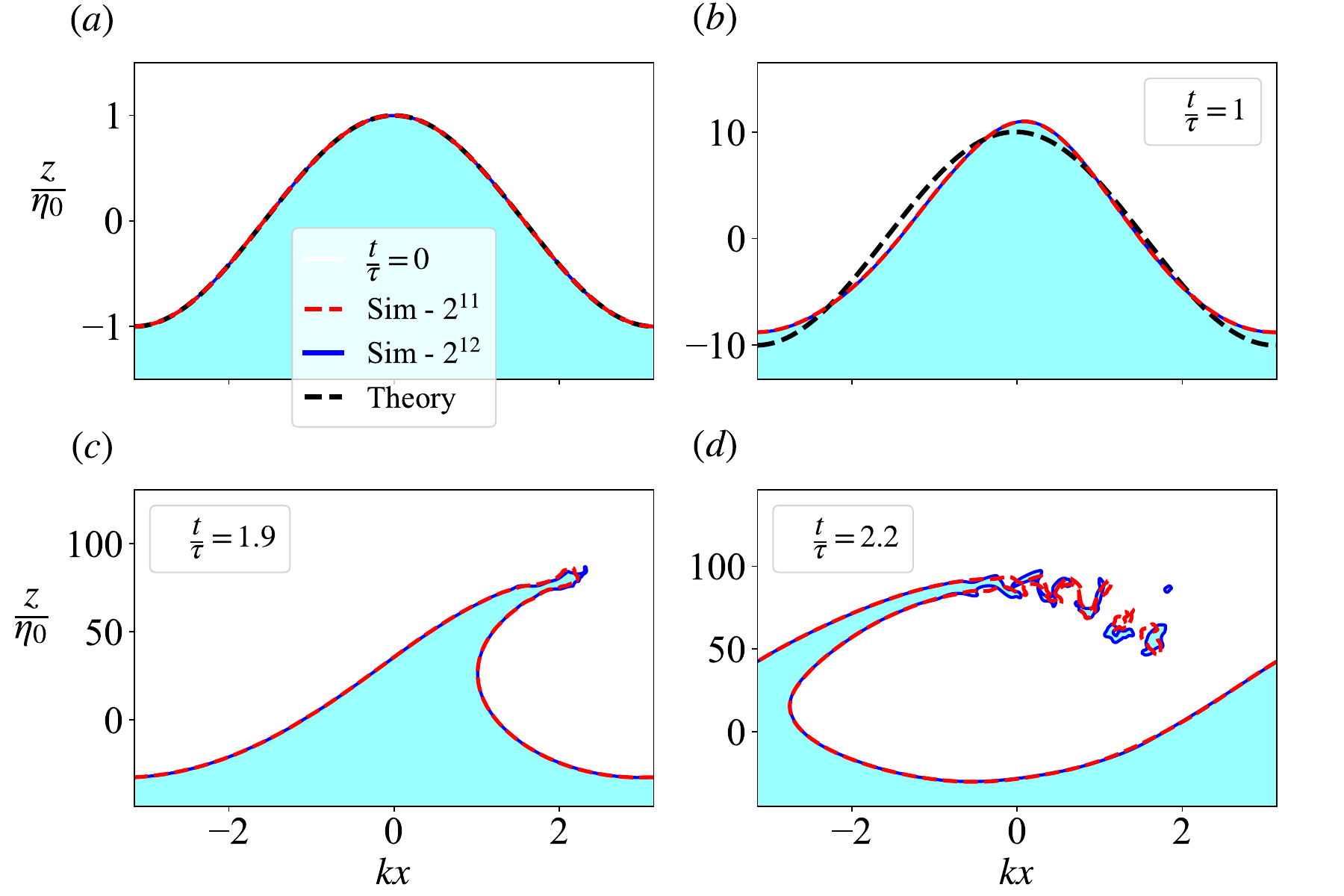}
		\caption{($Bo=87.81,\; Fr=4,\;\delta=0.5$) Panels (a)-(b) Comparison of the interface shape with linear theory. Panels (c) \& (d) - Finite amplitude regime, with Holmboe-like waves emitting spume-like droplets \citep{deike2022mass} from the tip.}
		\label{fig20}
	\end{figure}
	\begin{figure}
		\centering
		\includegraphics[scale=0.4]{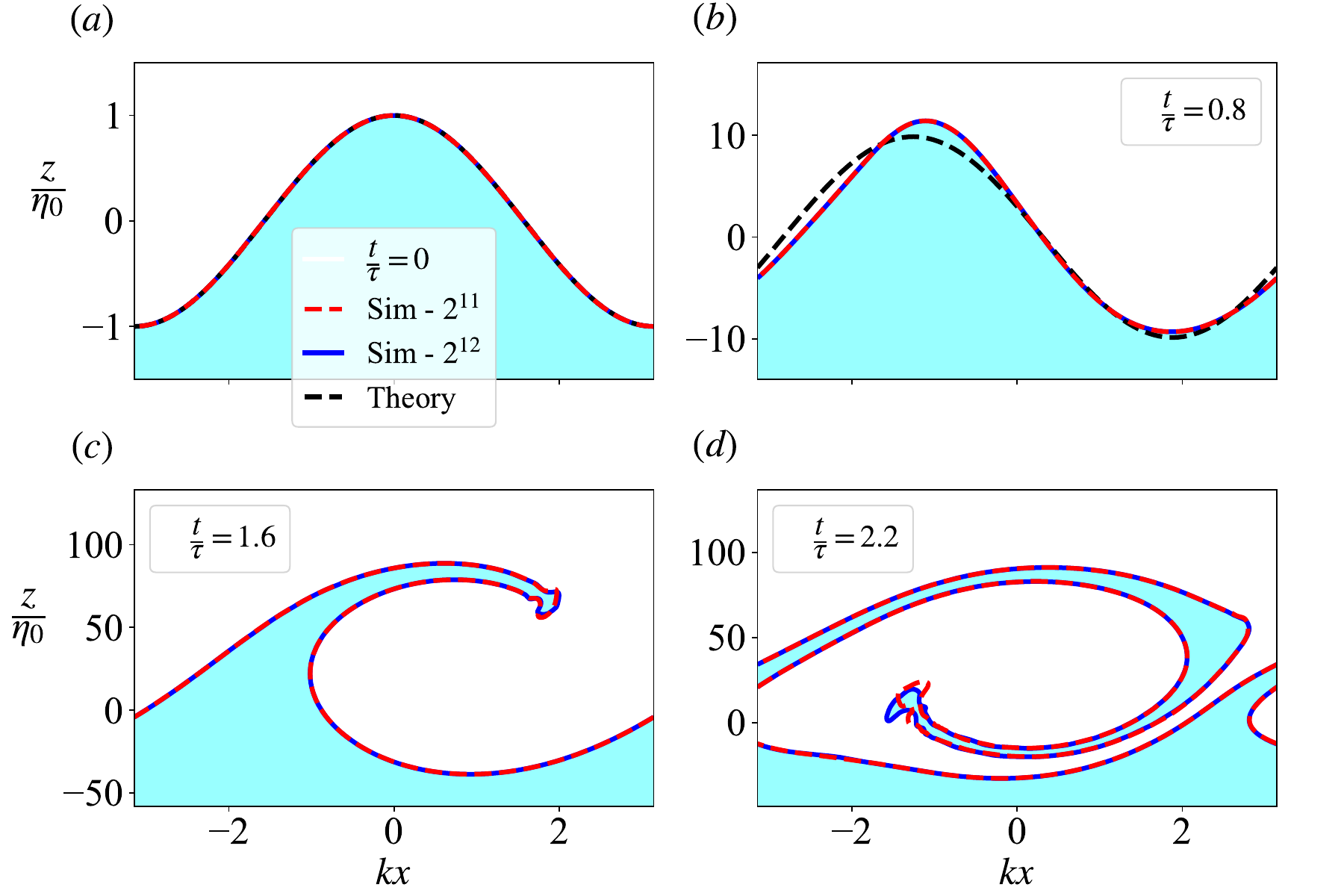}
		\caption{($Bo=87.81,\; Fr=4,\;\delta=0.9$) Panels (a)-(b) Comparison of the interface shape with linear theory. Panels (c)-(d) Finite amplitude regime, note the formation of classic KH spirals in the last two panels.}
		\label{fig21}
	\end{figure}
	
		\begin{figure}
		\centering
		\includegraphics[scale=0.4]{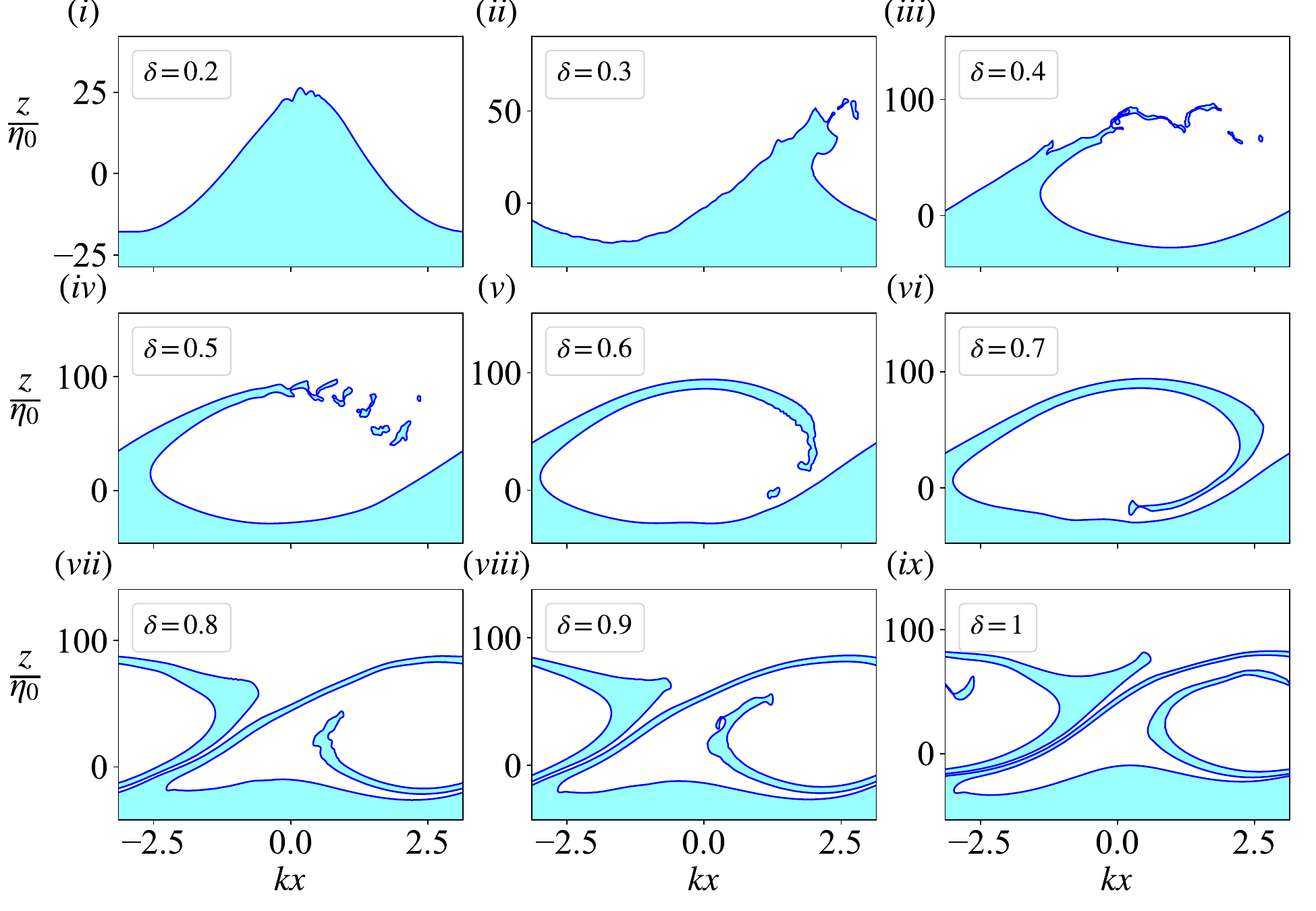}
		\caption{($Bo=87.81,\; Fr=4$) Panel (i) $t/\tau=2.863$. Rounded to two decimal places, $\Delta/\lambda=0.12$ (ii) $t/\tau=2.897$, $\Delta/\lambda=0.12$  (iii) $t/\tau=2.434$, $\Delta/\lambda=0.12$ (iv) $t/\tau=2.278$, $\Delta/\lambda=0.12$ (v) $t/\tau=2.237$, $\Delta/\lambda=0.12$ (vi) $t/\tau=2.270$, $\Delta/\lambda=0.12$ (vii) $t/\tau=2.530$, $\Delta/\lambda=0.13$ (viii) $t/\tau=2.489$, $\Delta/\lambda=0.13$ (ix) $t/\tau=2.573$, $\Delta/\lambda=0.13$. \mmgt{The final set of images at $\delta$ close to unity, bear remarkable similarity to the experimental images in fig. $14.3$ in \cite{cushmanroisin2011}. We thank an anonymous referee for pointing us to these images}.}
		\label{fig22}
	\end{figure}
\begin{figure}
		\centering
		\includegraphics[scale=0.4]{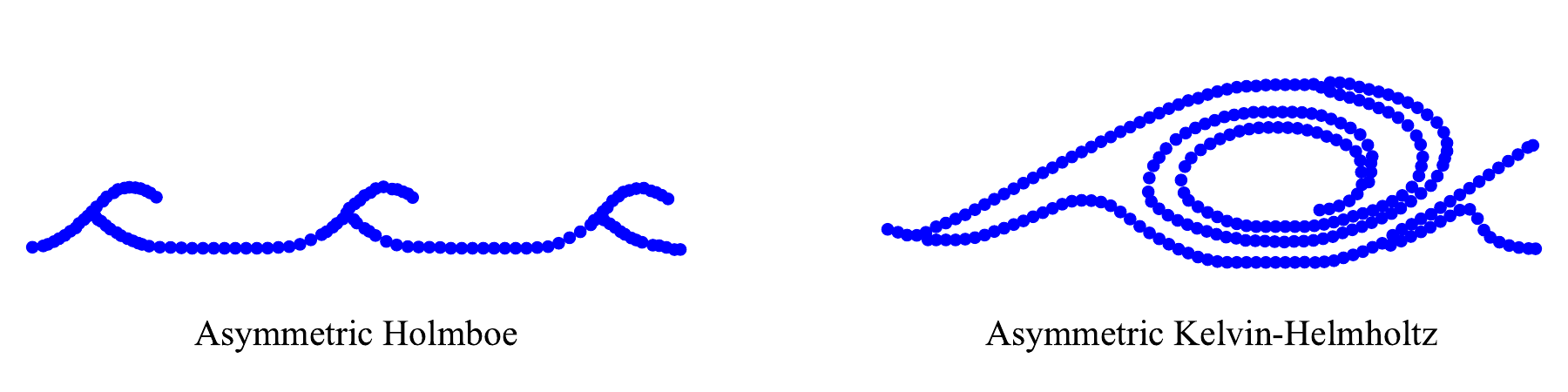}
		\caption{Data extracted from figure $9$ of \cite{lawrence1998search}. Compare the left (Holmboe) and right (KH) images of this figure with panel (d) of figures \ref{fig20} and \ref{fig21} respectively.}
		\label{fig23}
	\end{figure}

	As anticipated, a clear signature of differences in the nonlinear regime is seen between figure \ref{fig20} and \ref{fig21}, as density is increased from $\delta=0.5$ to $\delta=0.9$ respectively. \mgt{These differences are further seen readily in the collage of images provided in figure \ref{fig22}, all of them approximately at the same value of $t/\tau$. One notes transitions in the nonlinear regime around $\delta \approx 0.3$ and $\delta \approx 0.7$}. In figure \ref{fig20}, we particularly note the formation of a cusp-like structure at the crest which emits spume droplets (panel (d)) \citep{deike2022mass}; we treat this as a signature of the Holmboe instability. Compare panel (d) of figure \ref{fig20} with the shape of the interface, extracted from \cite{lawrence1998search} in figure \ref{fig23} (left figure) which is an asymmetric Holmboe instability. With further increase in density ratio $\delta=0.9$, we recover the classic KH spirals in panel (d) of figure \ref{fig21}, comparing this to the figure on the right in figure \ref{fig23}, symptomatic of the asymmetric KH instability. Figures \ref{fig20} and \ref{fig21}, \mgt{as well as the collage of images in fig. \ref{fig22},} thus show clear evidence that the nonlinear deformation regime is quite different for $\delta=0.5$ compared to $\delta=0.9$. While the former bears similarity to the asymmetric H instability of \cite{lawrence1998search} in figure \ref{fig23}, the latter is very much the asymmetric KH instability, in the same figure. We thus validate our hypothesis that the fastest growing mode in the exponential profile, indeed undergoes a transition from the Miles to the Holmboe and further to the KH instability, as the density ratio is increased.
	
\begin{figure}
	\centering
	\subfloat[$\delta=0.5$]{\includegraphics[scale=0.4]{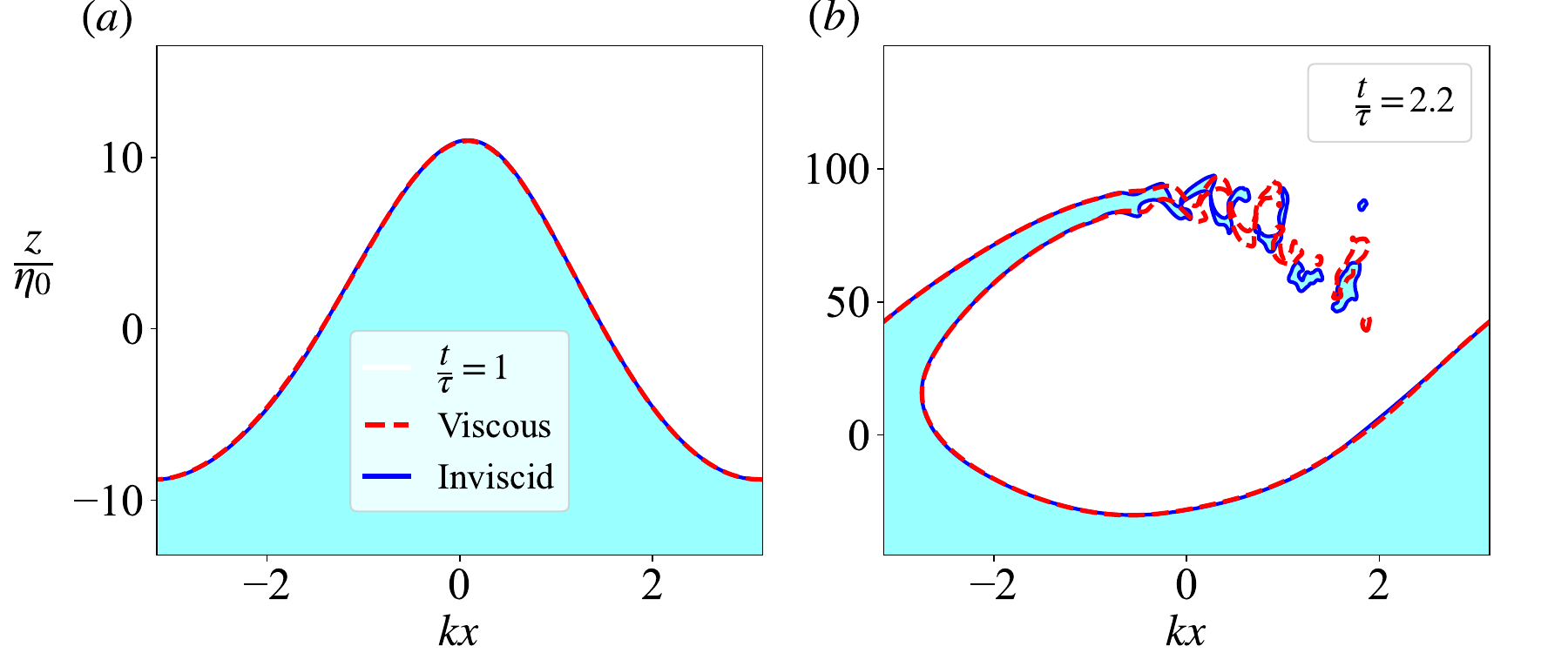}}\\
	\subfloat[$\delta=0.9$]{\includegraphics[scale=0.4]{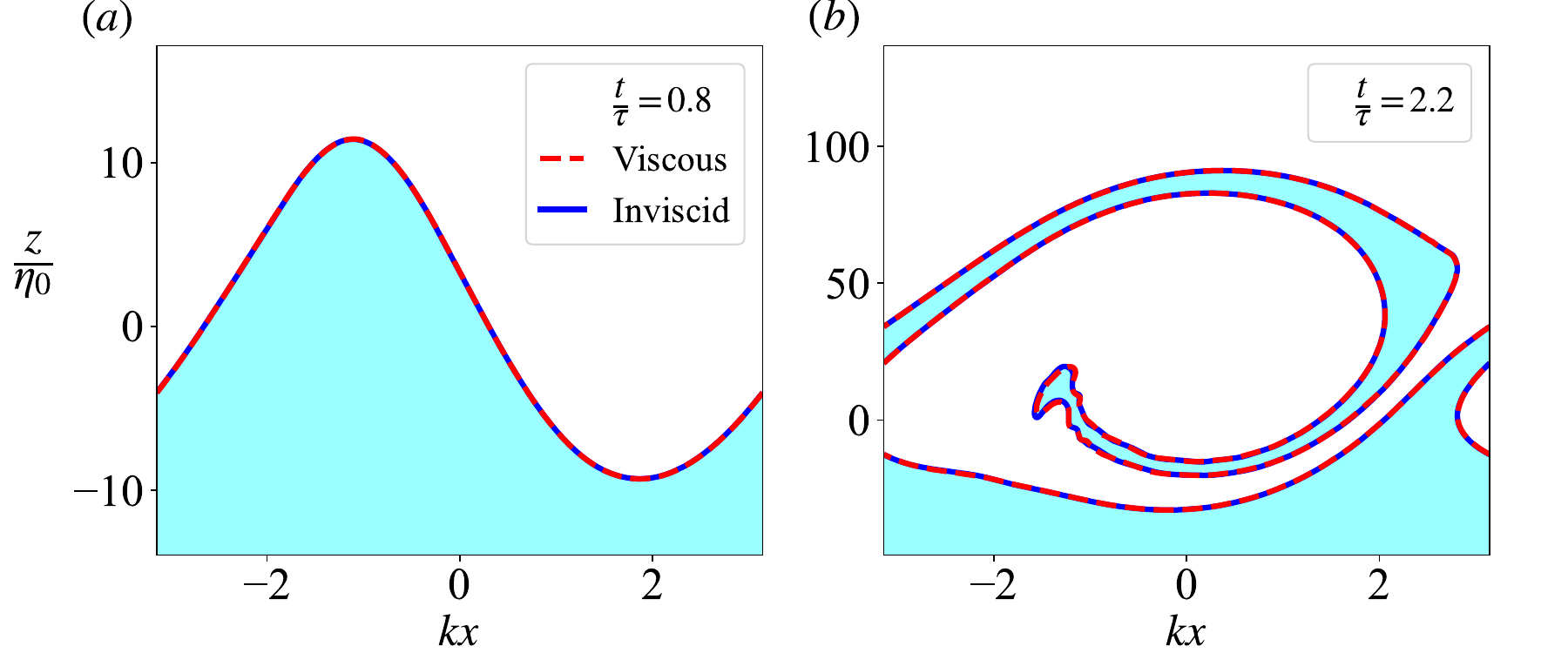}}
	\caption{\mgt{Comparison of inviscid and viscous simulations at the same $Bo$ and $Fr$. Panel (a) Inviscid simulation (as in figure~\ref{fig20}) versus a viscous case with Perfluorodecalin ($\mu=5.01$ mPa$\cdot$s, heavier phase) and water ($\mu=1$ mPa$\cdot$s, lighter phase), with density ratio $\delta = 0.5$. Panel (b) Inviscid simulation (as in figure~\ref{fig21}) versus a viscous case with brine ($\mu=1.05$ mPa$\cdot$s, heavier phase) and fresh water ($\mu=1$ mPa$\cdot$s, lighter phase), with $\delta = 0.9$. For the viscous results, we solve the incompressible Navier–Stokes equations with kinematic viscosity $\nu>0$ in both fluids. Viscosity has a relatively minor effect within the time window considered.}}
	\label{fig24}
	\end{figure}

	In order to establish the qualitative validity of these inviscid simulations, we have also compared these against viscous simulations employing realistic value of viscosity of fluids, at the density ratios of our interest. \mgt{To account for viscosity, the viscous term, $\nu\nabla^2\bm{u}$ is added to the right-hand side of eqn. (\ref{eq4.1b}). This addition converts the system described by eqns. (\ref{eq4.1}) into the two dimensional, Navier-Stokes equations, where $\nu$ is the kinematic viscosity.} Two fluid combinations (water \& Perfluorodecalin) with density ratio $= 0.5$ and fresh-water and brine with density ratio $= 0.9$ are considered. We emphasize that these viscous simulation results are preliminary and are reported for consistency - a careful comparison of the solution to the Rayleigh equation against the numerical solution to the Orr-Sommerfeld equations for the exponential profile, is necessary and is currently underway. The viscous simulations have been carried out employing free-slip boundaries thus ensuring that boundary-layers do not develop at the computational boundaries. These simulations are reported in panel (a) and (b) of figure \ref{fig24} comparing these against their inviscid counterparts; a good comparison in general is apparent.

	\section{Effect of change of $\delta$ on the background profile: the double exponential profile}\label{sec:EffectofDelta}
	\mgt{So far, we have considered inviscid linear stability of the background shear and density profiles in eqns. \ref{eq2.1}a and \ref{eq2.1}b, with shear only in the upper fluid. Recall that within the inviscid approximation, \textit{any} shear profile in either fluid (upper or lower) alongwith a flat density interface ($\eta=0$) with arbitrary $\rho_u,\rho_l > 0$ and a hydrostatic pressure field, is an exact solution to the inviscid momentum and mass equations, as well as the boundary conditions (see Appendix C). In experiments or naturally occurring situations involving \textit{viscous}, immiscible fluids, one however expects that as the density ratio $\delta \rightarrow 1^{-}$ (such as in this study), the viscous shear stress exerted by the upper fluid on the lower one, generates significant momentum transfer to the lower fluid. \footnote{we thank an anonymous referee for pointing this}. For air-water configuration, the importance of viscous shear in either fluids was studied by \cite{valenzuela1976growth}, in their solution to the Orr-Sommerfeld equation (see their fig. $3b$ for the background state). Viscosity interestingly, can also generate a viscous Holmboe instability distinct from the inviscid Holmboe instability discussed so far; this was discovered and analysed in \cite{parker2020viscous} for smooth shear and density profiles of the tan hyperbolic form. These results, in turn imply that as $\delta\rightarrow 1^{-}$, for our inviscid linear stability results to be physically relevant and realizable, we need to consider background states with shear in both fluids. This is in contrast to \ref{eq2.1} which is expected to be an accurate physical representation in the limit of $\delta\rightarrow0$ only.} 
		
\begin{figure}	    	
			\centering
			\includegraphics[scale=0.4]{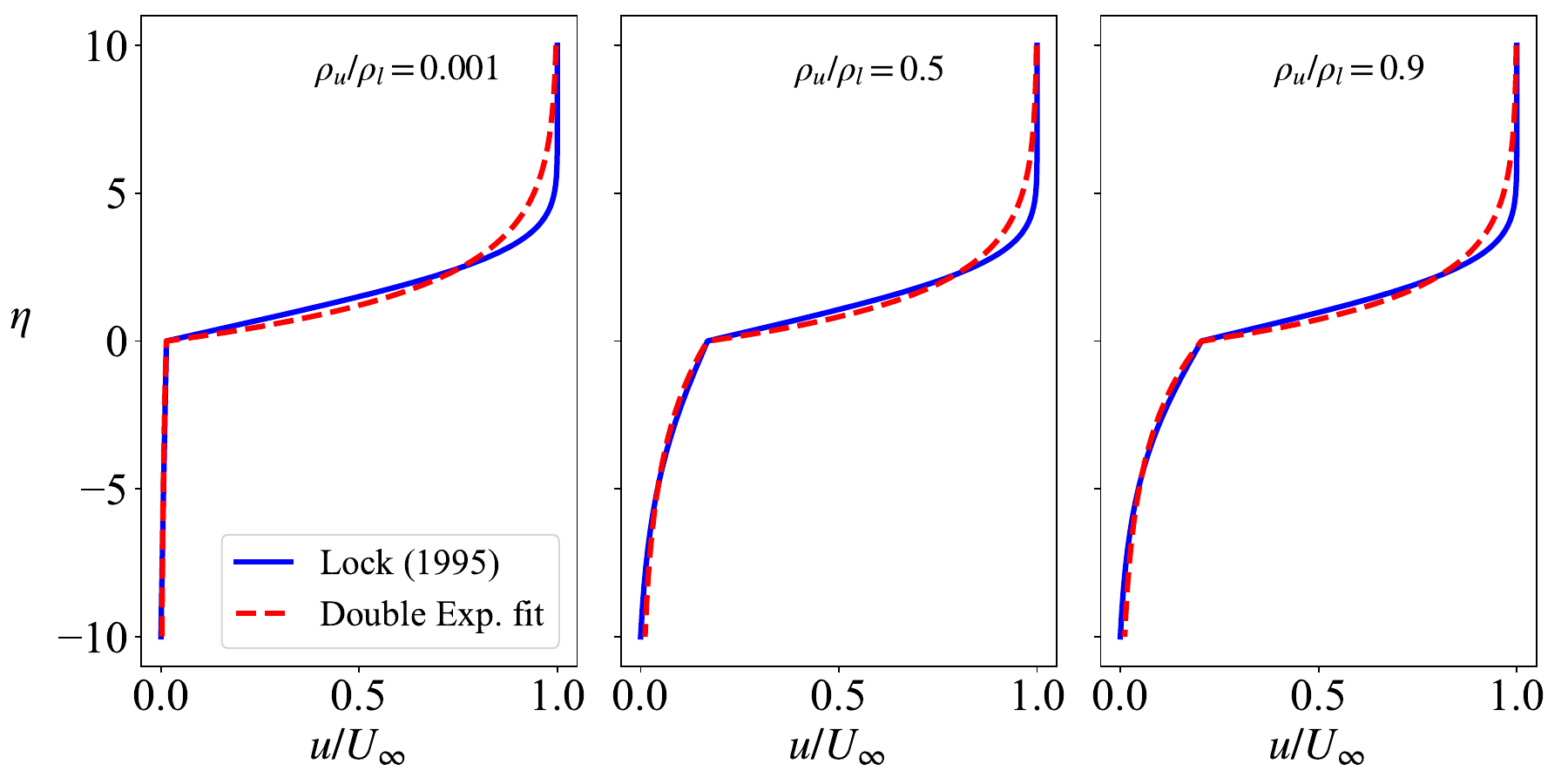}
			\caption{\mgt{Vertical variation of the horizontal velocity profile with $\delta$ i.e. $\dfrac{u_{u}}{U_{\infty}}= f_{1}(\eta_{1}),\;\dfrac{u_{l}}{U_{\infty}}= f_{2}(\eta_{2})$, for fixed viscosity ratio ($\mu^{u}/\mu^{l}=0.02$) corresponding to air-water and $Bo \rightarrow \infty$. Note that $\eta=\eta_{1}$ and $\eta_{2}$, for upper and lower fluids, respectively. The profiles (dashed) are obtained by numerically solving eqns. $9$ (upper layer) and $15$ (lower layer) in \cite{lock1951velocity} governing $f_1,f_2$, subject to boundary conditions provided therein. We choose the far field velocity in the lower fluid $U_2=0$ as $\eta_2 \rightarrow-\infty$, (\citep{lock1951velocity}) employing the shooting method. The variable $\eta_{1} \equiv y\sqrt{\frac{U_{\infty}}{\nu_u x}}$ with $\eta_2 = \sqrt{\dfrac{\nu_u}{\nu_l}}\eta_{1}$. In each case, we fit a double exponential velocity profile of the form provided in expressions \ref{eq5.1} with the (dimensional) coordinate $z$ in the upper and lower fluids being replaced by (non-dimensional) $\eta_{1}$ and $\eta_{2}$ respectively and the dimensional variables $\Delta_u$ and $\Delta_l$, replaced by (non-dimensional) $\mathcal{A}$ and $\mathcal{B}$ respectively. The fitted parameters are (a) $U_0\approx 0.01386$, $\mathcal{A}=1.1777$, $\mathcal{B}=5.766$ (b) $U_0\approx 0.1685$, $\mathcal{A}=1.607$, $\mathcal{B}=3.659$ and (c) $U_0\approx 0.2046$, $\mathcal{A}=3.408$, $\mathcal{B}=3.408$. For all three panels and the double exponential profiles in these, $Fr=4$.}}
			\label{fig25}
		\end{figure}

	\mgt{We demonstrate below that the transition in the nature of instability observed with varying $\delta$ for the background state of \ref{eq2.1}, \textit{also} occurs for profiles with background shear in both fluids. Towards this, we first compute the self-similar velocity profile at the interface of two viscous, immiscible fluids with a flat density interface; this background state seems to have been first reported by \cite{lock1951velocity}. Figure \ref{fig25} depicts the velocity profile as a function of the similarity coordinate $\eta$ (see caption to the figure for definition) obtained by solving the equations of \cite{lock1951velocity} (see figure \ref{fig25} caption for details). Expectedly, as $\delta$ is increased (for fixed viscosity ratio), the background shear in the lower fluid becomes comparable to that in the upper one. In the same figure, for comparison, the ``double exponential'' approximation (expression \ref{eq5.1a}, \citep{young2014generation}) to the viscous, self-similar velocity profile of \cite{lock1951velocity} are also provided. This double exponential profile is given by:}
		\begin{subequations}\label{eq5.1}
		\begin{align}
			& U(z) = \left\{
			\begin{array}{ll}
				U_u(z)=U_{\infty} - \left(U_{\infty} - U_{0}\right)\exp\left(-\dfrac{z}{\Delta_u}\right), & \qquad z > 0  \vspace{0.2cm} \\
				U_l(z)=U_0\exp\left(\dfrac{z}{\Delta_l}\right),                                                            & \qquad  z < 0
			\end{array}\right. \label{eq5.1a} \\
			& \rho(z) = \left\{
			\begin{array}{ll}
				\rho_u, & \qquad z > 0  \vspace{0.2cm} \\
				\rho_l, & \qquad  z < 0
			\end{array}\right. \label{eq5.1b}
		\end{align}
	\end{subequations}	
   	\mgt{where $U_0$ is the background interface velocity at $z=0$.} 
   	
   	\mgt{Expressions \ref{eq5.1} generalise the earlier background state of eqns. \ref{eq2.1}; setting $U_0 = 0$ in the former recovers the latter.
	A good qualitative match between the double exponential and the viscous profiles is apparent in figures \ref{fig25}. Note in particular, that for air-water density ratio $\delta = 0.001 << 1$, the shear in water is negligibly small ($U_0 \approx 0$, leftmost panel of figure \ref{fig25}) and the viscous as well as the double exponential profile are modelled well, by the \textit{single exponential profile} of eqns. \ref{eq2.1}. This is not so at higher $\delta\rightarrow 1^{-}$ (middle and right panels of figure \ref{fig25}) where the background shear in the lower fluid becomes substantial. Our choice of the double exponential profile as an approximation to the self-similar viscous profile \citep{lock1951velocity}, is partly motivated by the observation that the (inviscid) linear stability problem can also be analytically solved for this profile \citep{young2014generation}. This is contrast to the profile of \cite{lock1951velocity}, for which only numerical solutions to the Rayleigh equation seem feasible. An additional and important advantage is that the change in the sign of $D^2U(z)$ across $z=0$ for expression \ref{eq5.1a}, implies that the KH instability can be expected for this profile in the unstratified limit $\delta=1$, as a consequence of Rayleigh's inflection point theorem.} 
	
	\mgt{Due to the good qualitative match seen between the self-similar, viscous profiles \citep{lock1951velocity} and the double exponential profiles in figure \ref{fig25}, we proceed next with the intuitive expectation that linear, inviscid stability results, including transitions in the instability mechanism obtained for the background state \ref{eq5.1}, also apply to the viscous profiles of \cite{lock1951velocity}. We thus examine the \textit{inviscid}, linear stability of the background state \ref{eq5.1}a,b as a function of density ratio $\delta$.}
		\mgt{The stability analysis of eqns. \ref{eq5.1} is carried out in an exactly similar manner as section \ref{sec:lin_stab} earlier, the only difference being that both fluids now have shear in their background state. The Rayleigh equations in both fluids, may be solved for this background state in terms of the Gauss hypergeometric function; we omit details as these can be found in \cite{young2014generation}. Note that two additional non-dimensional numbers appear for the background state of \ref{eq5.1a}, \ref{eq5.1b} viz. $\Delta_l/\Delta_u$ and $U_s/U_{\infty}$: both these ratios were zero in our analysis so far, for the background state of \ref{eq2.1}.} 
			
\begin{figure}	    	
			\centering
			\subfloat[]{\includegraphics[scale=0.4]{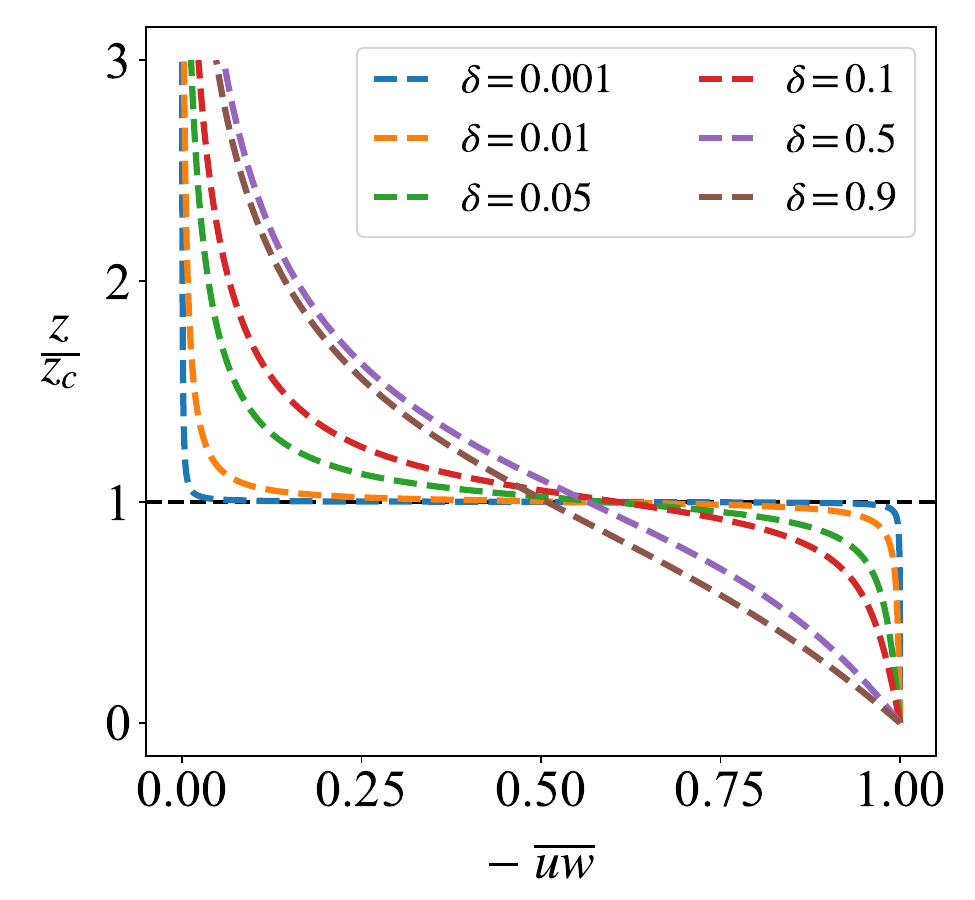}		\label{fig26a}}
			\subfloat[]{\includegraphics[scale=0.4]{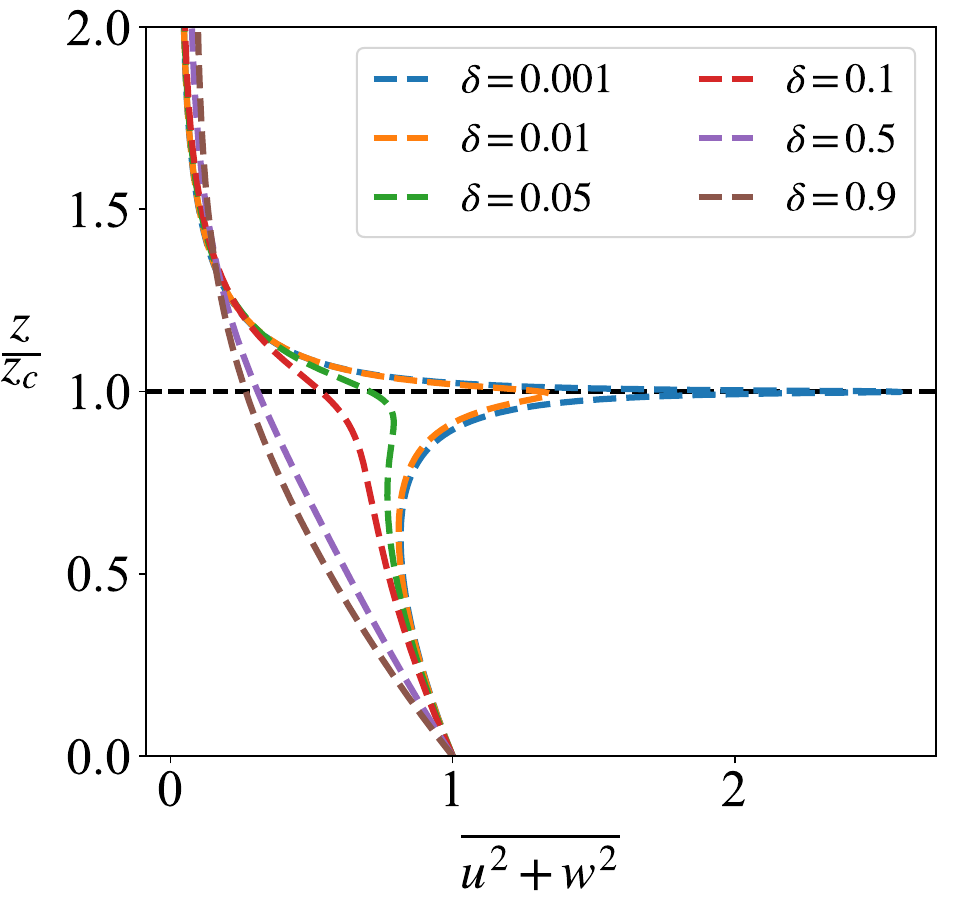}		\label{fig26b}}		
			\caption{\mgt{($Fr=4,\; Bo=87.813,\;\Delta_l/\Delta_u = 0.1,\; U_s/U_{\infty}= 0.1$) - Panel (a) Vertical variation of Reynolds stress (b) Perturbation kinetic energy for the double exponential background profile of expressions \ref{eq5.1}.}}
			\label{fig26}
	    \end{figure}

\begin{figure}	    	
	    	\centering
	    	\includegraphics[scale=0.35]{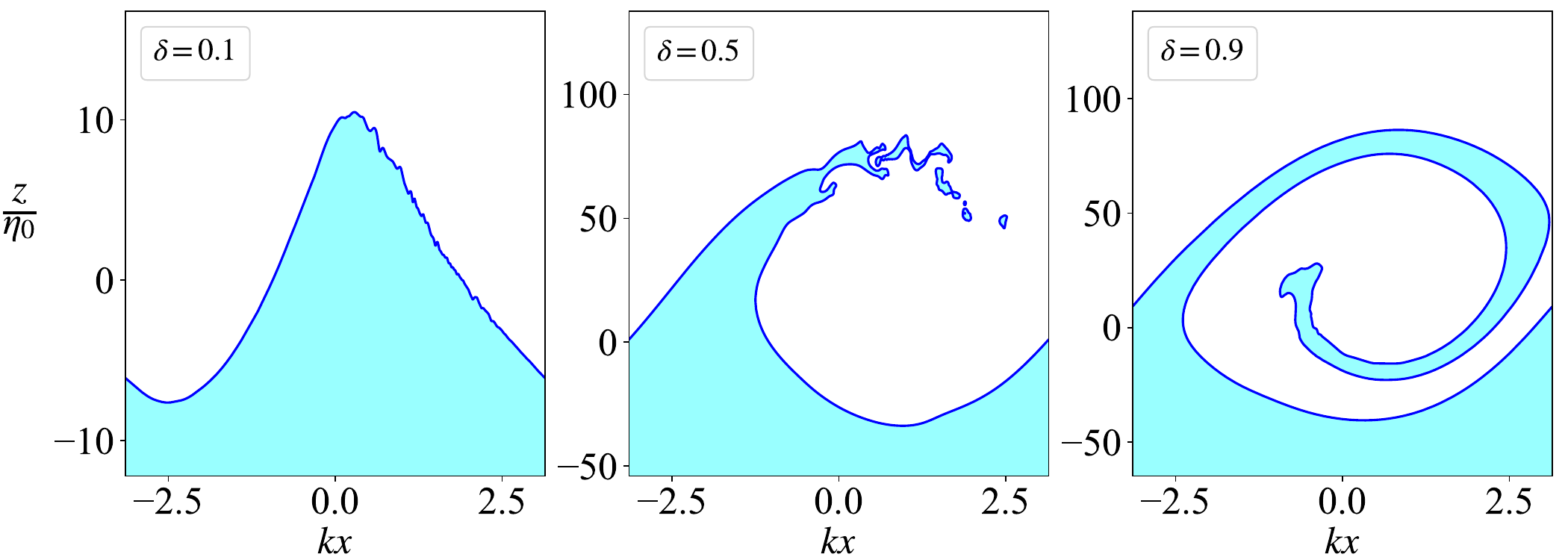}
	    	\caption{\mgt{($Fr=4,\; Bo=87.813,\;\Delta_l/\Delta_u = 0.212,\; U_s/U_{\infty}= 0.1$) Evolution of the interface initialised with eigenmodes obtained from linear stability analysis of the background state in eqn. \ref{eq5.1}. Panel (left) $t/\tau=8.317$ (middle) $t/\tau=3.0$ (right) $t/\tau=2.55$ where $\tau$ is the time taken to travel one wavelength.}}
	    	\label{fig27}
	    \end{figure}

		\mgt{The vertical variation of Reynolds stress and perturbation kinetic energy as a function of $\delta$ are depicted in figures \ref{fig26a} and \ref{fig26b} respectively. Note that the parameters of the double exponential profile were chosen such that the rippling instability \citep{young2014generation} does not manifest. In figure \ref{fig26}, one notes the same transition in the vertical variation of either metrics, as was noted in figure \ref{fig15} for the single exponential profile. Further, results from nonlinear simulations initialised with eigenfunctions obtained from linear stability analysis of the double exponential profile of eqn. \ref{eq5.1} are shown in figure \ref{fig27}. It thus becomes clear that our earlier results, notably the transition from Miles to the H and eventually to the KH instability, with increasing $\delta$ remain applicable also to the double exponential background state of \ref{eq5.1a} and \ref{eq5.1b}. The applicability of our results, spanning a range of density ratios from very low ($\delta << 1$) to close to unity ($\delta \sim O_m(1)$), is thus apparent.}
    
	\section{Conclusions}\label{sec:concl}
	In this study, we have identified a novel transition in the nature of the fastest growing mode for the exponential profile with a sharp density interface. This transition has been studied at low Froude and high Bond number, and is observed with varying density ratio. At $\delta << 1$, the fastest growing mode has features typical of the Miles instability \citep{miles1957generation} with a near-discontinuous phase change in the eigenfunction across $z_c$; the extremely low value of $c_i/c_r$ even for the fastest growing mode in this regime, makes it behave nearly like a singular neutral mode \citep{maslowe2013study}, leading to energy extraction only below the critical location and upto the density interface i.e. a subset of the shear layer. As the density ratio is increased beyond $0.1$, the value of $c_i/c_r$ increases proportionally. For a growing mode, while the semi-circle theorem always ensures that $U_u(z_c)=c_r(k)$, the sharp jump in Reynolds stress at $z_c$ gives way to a smooth variation though $z_c$, implying energy extraction throughout the shear layer. We rationalise this transition employing an analytical expression for the Reynolds stress, originally due to \cite{lin1954some}. The aforementioned smoothening is also shown to occur in the phase of the leading order term bearing the logarithmic singularity in Tollmien's inviscid, series solutions. At $\delta=0.5$, we show that the growth rate for the fastest mode, compares well against that of the PL profile. This suggests that the transition from Holmboe to the KH instability seen in the PL profile with increasing $\delta$, can also manifest for the exponential model, albeit modified by the presence of the critical location in the latter. We emphasize that for the PL profile due to lack of profile curvature (except at the corners), the unstable eigenmodes are not singular even though they have a critical layer (see similar observations for stable eigenfunctions in \cite{kadam2023wind} for the case of a water layer at $\delta=0$ with  a linear shear profile). We have further validated that for the asymmetric profile under consideration and with increasing $\delta$, the Reynolds stress and perturbation kinetic energy do not bear any signature of a potential transition from the H to the KH instability, that we anticipate. To establish this, we turn to nonlinear simulations.
	\begin{figure}
		\centering
		\includegraphics[scale=0.38]{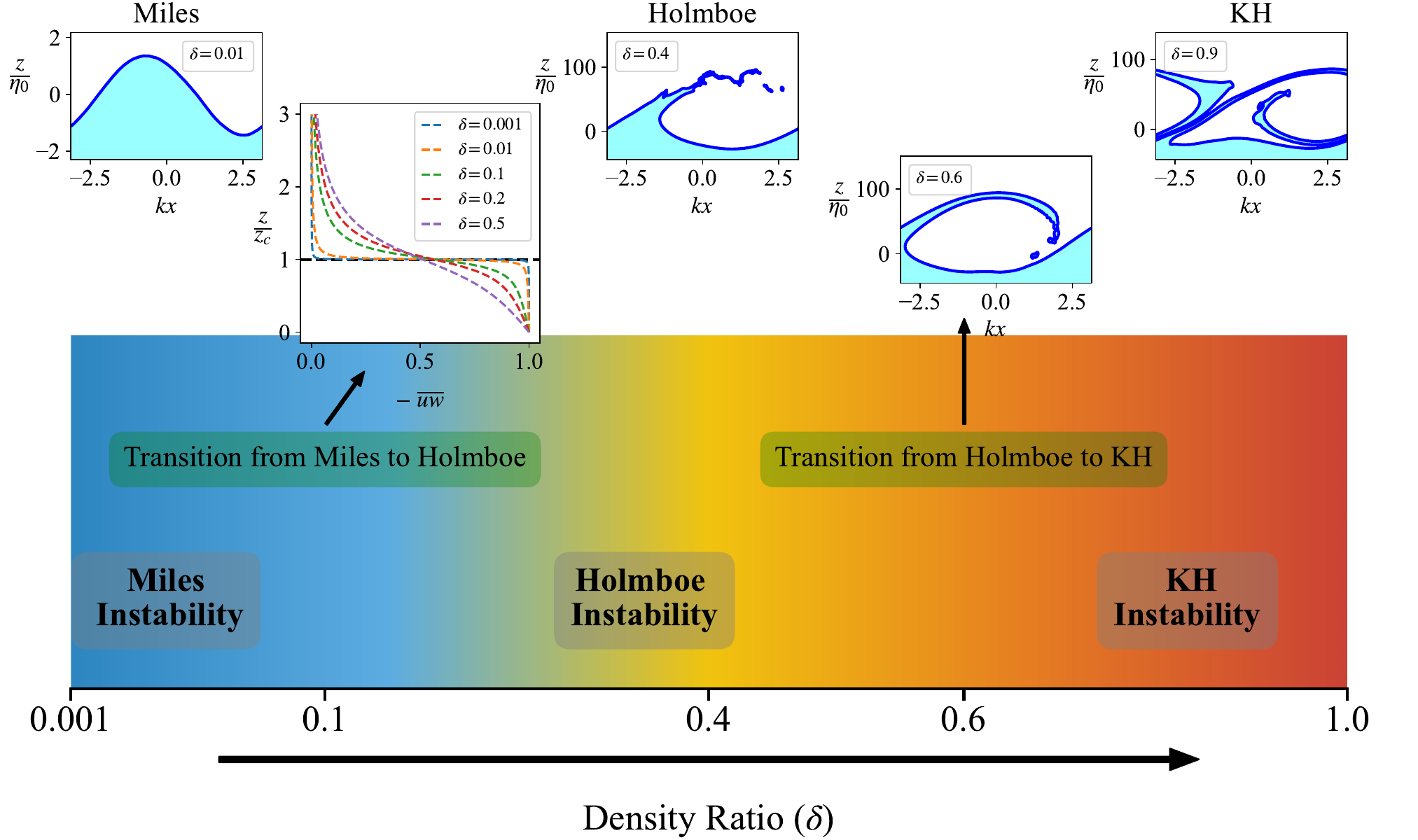}
		\caption{\mgt{A pictorial summary of key results. The color contours are informed by results obtained from theory and simulations. These have been generated synthetically, chosen so as to highlight the transition in instability mechanism, with increasing $\delta$.}}
		\label{fig28}
	\end{figure}
	In addition to the excellent validation of linear predictions at least for a few wave-periods, our simulations clearly show qualitative differences in the nonlinear regime between $\delta=0.5$ and $\delta=0.9$.The former is associated with a finite-amplitude wave with a sheared cusp emitting spume-like droplets from its tip. In marked contrast, at $\delta=0.9$ nonlinear simulations recover the classic KH spirals. We show that these differences are consistent with the classification scheme proposed in \cite{lawrence1998search}, distinguishing the Holmboe from the KH instability for asymmetric profiles.
	
	An important takeaway from our study is the observation that the Miles instability persists for $\delta=0.01$ (at low $Fr$ and high $Bo$) i.e. even at ten times the air-water ratio. Given that the growth rate of this instability scales with density ratio, this plausibly makes it easier to experimentally measure growth rates of the Miles instability, by exploring fluid combinations other than air-water. Our study also reports several comparisons against the PL profile in high and low $Fr$ regimes, establishing the qualitative role of background profile curvature. The role of viscosity requires careful analysis involving the Orr-Sommerfeld equation. This is underway and will be reported subsequently.
	\\
	
	\noindent{\bf Acknowledgements:} RD acknowledges Prof. Anubhab Roy for suggesting the Holmboe instability at the early stage of this study. Figure \ref{fig19c} was generated by Nikhil Yewale and is acknowledged. We thank the organization of the International Research
	Network (IRN) under the auspice of CNRS for the workshop held at Univ. Paris, Saclay, France in June 2025 where this
	work was presented. We thank Prof. Herbert E. Huppert for helpful comments.\\
	
	\noindent{\bf Funding\bf{:}} We gratefully acknowledge financial support from DST-SERB (Govt. of India) grants MTR/2019/001240, CRG/2020/003707, SPR/2021/000536, MoE-STARS/STARS-2/2023-0595 on various problems concerning nonlinear ocean waves, wave-breaking and spray formation. The Ph.D. tenure of AK is supported by the SERB grant (CRG/2020/003707) and is gratefully acknowledged.\\
	
	\noindent{\bf Declaration of Interests\bf{.}} The authors report no conflict of interest.
	
	\section*{Appendix A}
\begin{figure}
	\centering
	\subfloat[\cite{young2014generation}]{\includegraphics[scale=0.35]{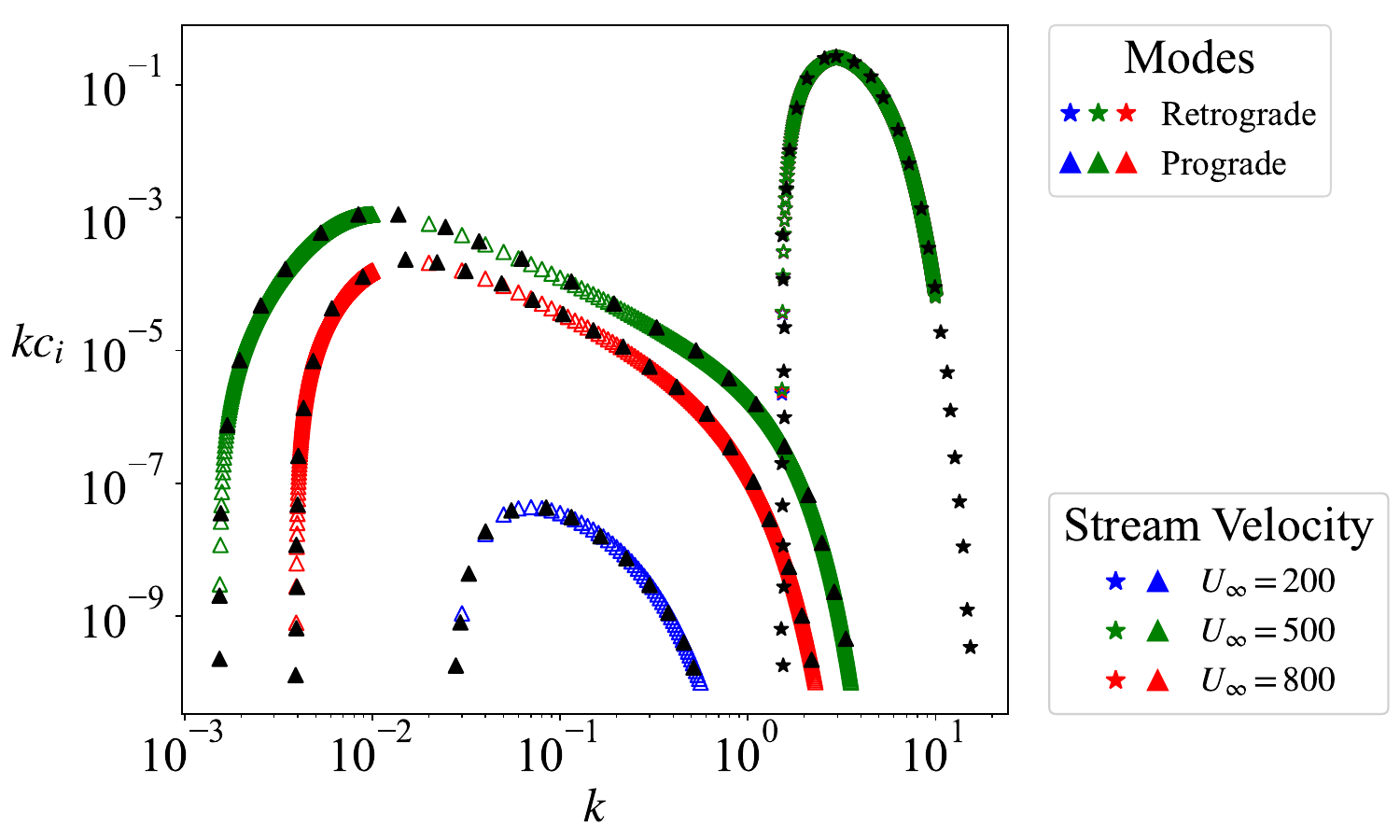}\label{figAppA-1}}\\
	\subfloat[\cite{young2014generation}]{\includegraphics[scale=0.35]{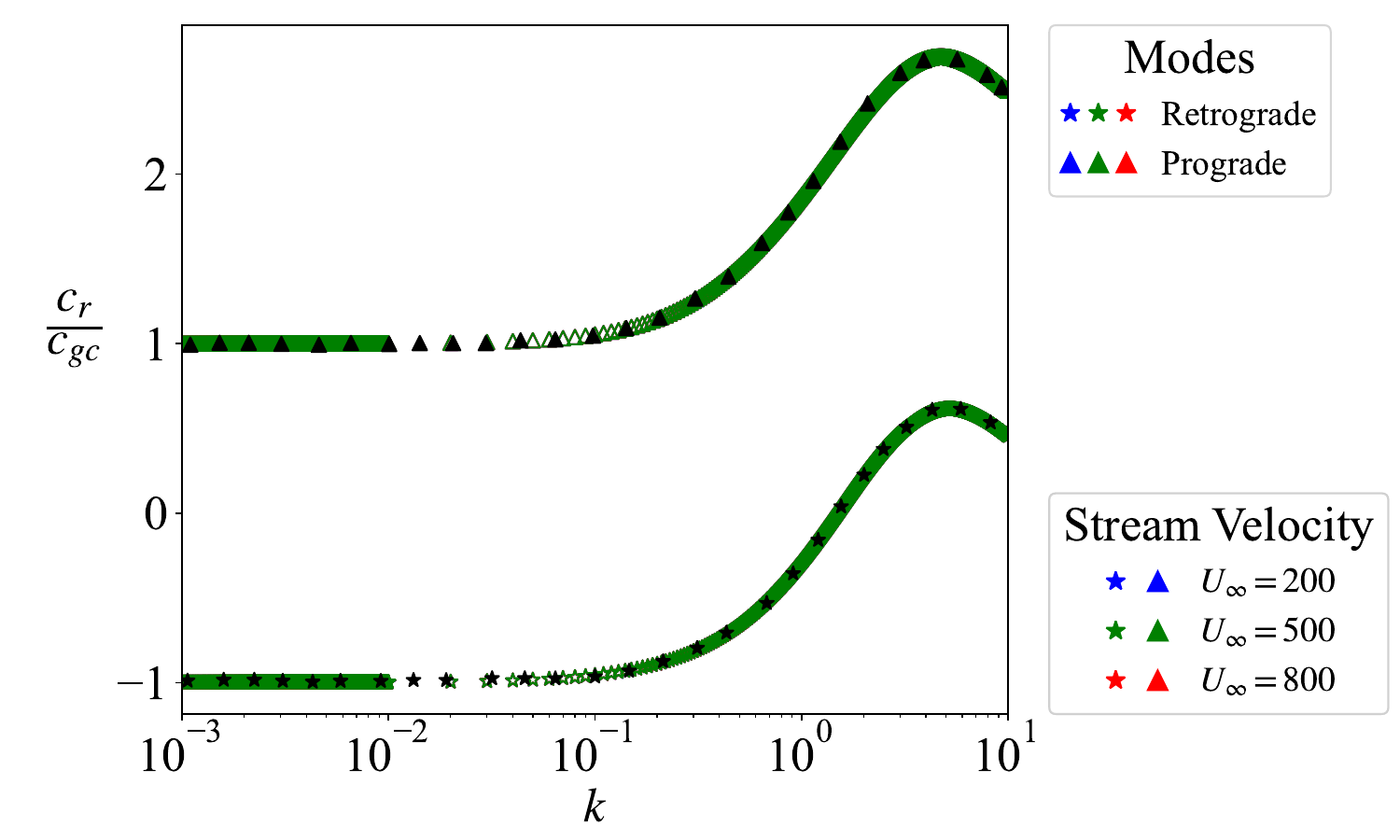}\label{figAppA-2}}\\
	\subfloat[\cite{morland1993effect}]{\includegraphics[scale=0.4]{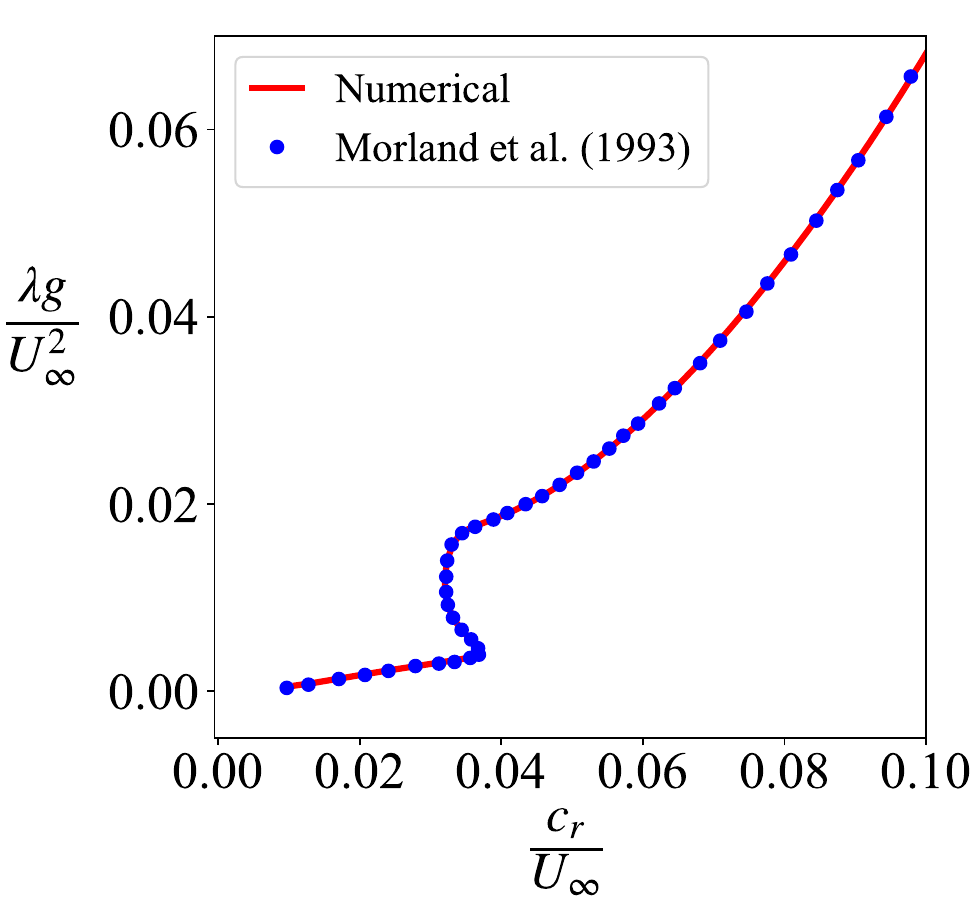}\hspace{2.8cm}\label{figAppA-3}}
	\caption{Validation of numerical solution procedure. We reproduce figure $2$ in \cite{young2014generation} and figure $3$ in \cite{morland1993effect} to validate our numerical procedure of solving the dispersion relation \ref{eq2.12}. Black symbols in panel (a) and (b) are from \cite{young2014generation}. \mgt{Coloured symbols are from our numerical solution to the dispersion relation provided in \cite{young2014generation}.}}
	\label{fig29}
	\end{figure}

	We provide here validations of our numerical solution to the dispersion relation from the work of \cite{young2014generation} and \cite{morland1993effect}. Please refer to the caption of figure \ref{fig29} for details.
	
	\section*{Appendix B}
\begin{figure}
	\centering
	\subfloat[Holmboe instability - $\delta=0.001$]{\includegraphics[scale=0.4]{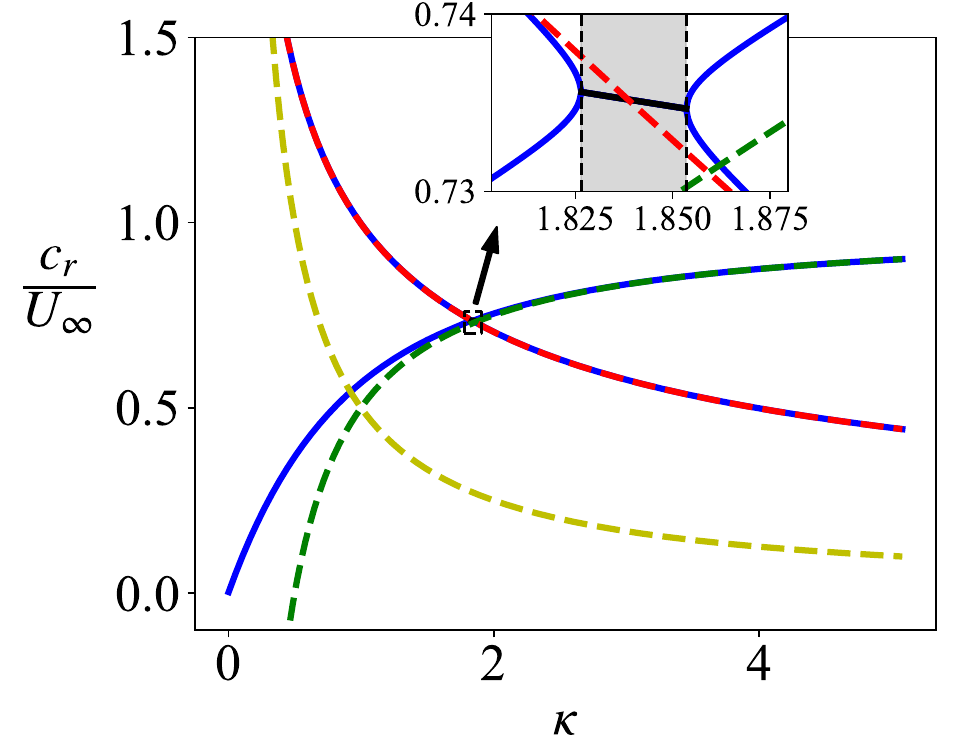}\label{figAppB-1}}
	\subfloat[Holmboe $+$ KH instability - $\delta=0.5$]{\includegraphics[scale=0.4]{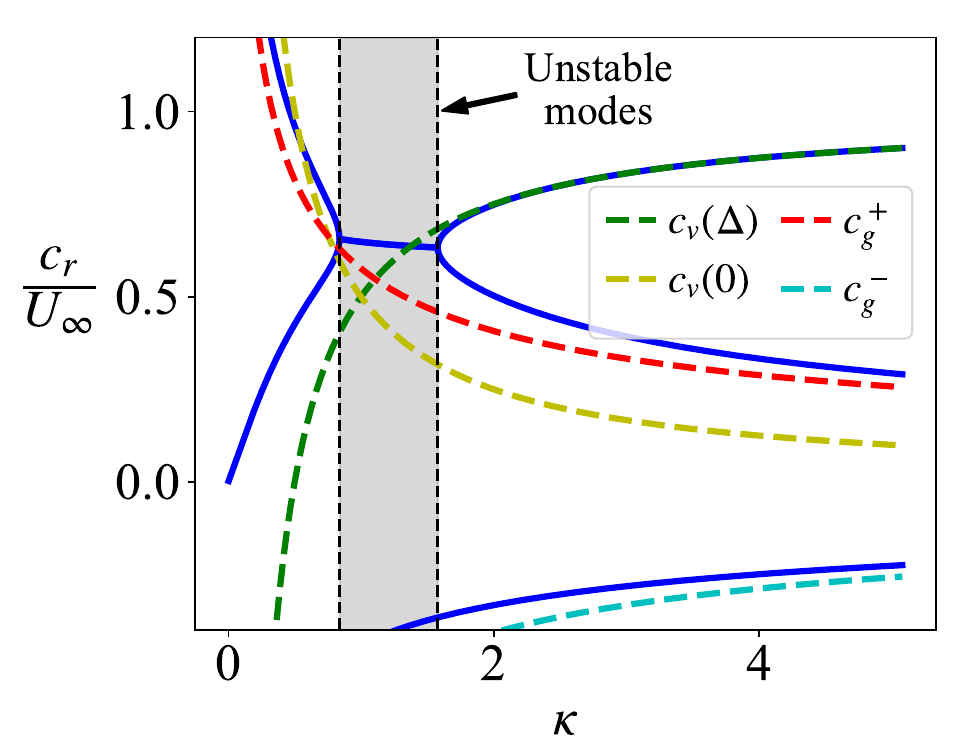}\label{figAppB-2}}\\
	\subfloat[KH instability - $\delta=0.9$]{\includegraphics[scale=0.4]{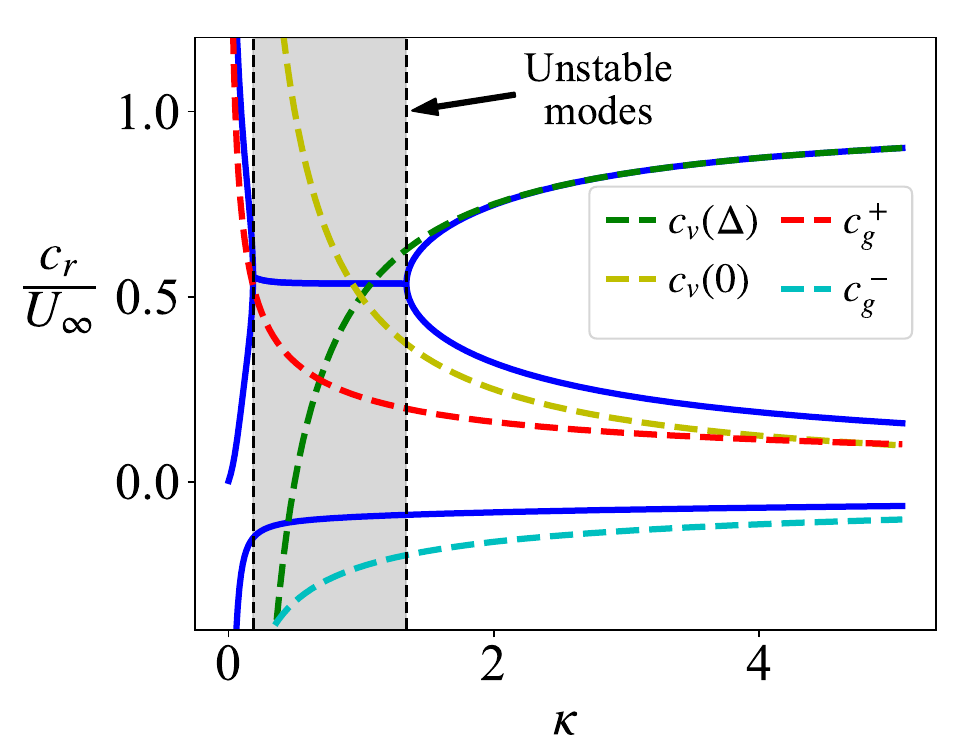}\label{figAppB-3}}
	\caption{($Fr=1,\;Bo\rightarrow\infty$) Transition from the Holmboe to the KH instability for the PL profile as the density ratio is varied. The color coding is the same as figure \ref{fig7}. At $\delta=0.5$, there is a strong overlap between the two instabilities.}
	\label{fig30}
	\end{figure}

	Figure \ref{fig30}, panels (a)-(c) demonstrate the transition in the PL profile of figure \ref{fig2a}. At low $\delta = 0.001$, the instability is of the Holmboe type whereas closer to unity ($\delta=0.9$), it is dominated by the KH instability. For $\delta=0.5$, there are contributions from both instabilities and \mgt{the transition is smooth}.
	
	\section*{Appendix C: Background shear profile in the inviscid approximation}
	 \mgt{Under the inviscid approximation, the governing equations are:     
	 \begin{subequations}\label{eq5}
		\begin{align}
			& \frac{\partial U_{i}}{\partial x} + \frac{\partial W_i}{\partial z} = 0, \tag{\theequation a,b} \\
			&\frac{\partial U_i}{\partial t} + U_i\frac{\partial U_i}{\partial x} + W_i\frac{\partial U_i}{\partial z} = -\frac{1}{\rho_i}\left(\frac{\partial P_i}{\partial x}\right), \tag{\theequation c,d} \\
			& \frac{\partial W_i}{\partial t} + U_i\frac{\partial W_i}{\partial x} + W_i\frac{\partial W_i}{\partial z} = -\frac{1}{\rho_i}\frac{\partial P_i}{\partial z} - g, \tag{\theequation e,f} \nonumber \\
			\intertext{with boundary conditions}  
			& \biggl\{\frac{\partial \eta}{\partial t} + U_i(x,z,t) \frac{\partial \eta}{\partial x} = W_i(x,z,t)\biggr\}_{z=\eta(x,t)}, \tag{\theequation g,h}\\
			\text{and}\quad &\biggl\{P_l(x,z,t) - P_u(x,z,t)\biggr\}_{z=\eta(x,t)} = -T \left[\frac{\frac{\partial^2\eta}{\partial x^2}}{\bigg\{1+ \left(\frac{\partial\eta}{\partial x}\right)^2\bigg\}^{3/2}}\right]. \tag{\theequation i,j}
		\end{align}
\end{subequations}}
	\mgt{where $i=u$ and $l$ for upper and lower fluids, respectively.} \\
	
	\mgt{It is verified that with $\eta(x,t)=0$ (flat interface) and hydrostatic pressure variation i.e. $P_u(x,z,t)=P_0 - \rho_u gz\;( z > 0)$, $P_l(x,z,t)=P_0 - \rho_u gz\;(z < 0)$ ($P_0$ is a reference pressure at the interface), any shear profile $\left[U_u(z),U_l(z),W_u=0,W_l=0,\right]$ and the discontinuous density profile of eqn. \ref{eq2.1}b with arbitrary $\rho_l,\rho_u>0$, constitute an \textit{exact solution} to eqns. \ref{eq5}(a)-(f) as well as boundary conditions \ref{eq5}(g)-(j). We validate the exact nature of the aforementioned inviscid solution, via numerical simulations. Choosing $U_u(x,z,t=0) = U_{\infty}\left[1 - \exp(-z/\Delta)\right], W_u(x,z,t=0)=0$ and $U_l(x,z,t=0)=W_l(x,z,t=0)=0$ as the initial velocity field, we initialise the simulation with a flat interface $(\eta(x,t=0))$ and boundary conditions described in figure \ref{fig17a}. Since these initial conditions, constitute exact solutions to the governing equations and boundary conditions, we expect no changes to the velocity field and the flat density interface upon time integration.}
		
\begin{figure}	    	
		\centering
		\subfloat[$\delta=0.01$]{\includegraphics[scale=0.4]{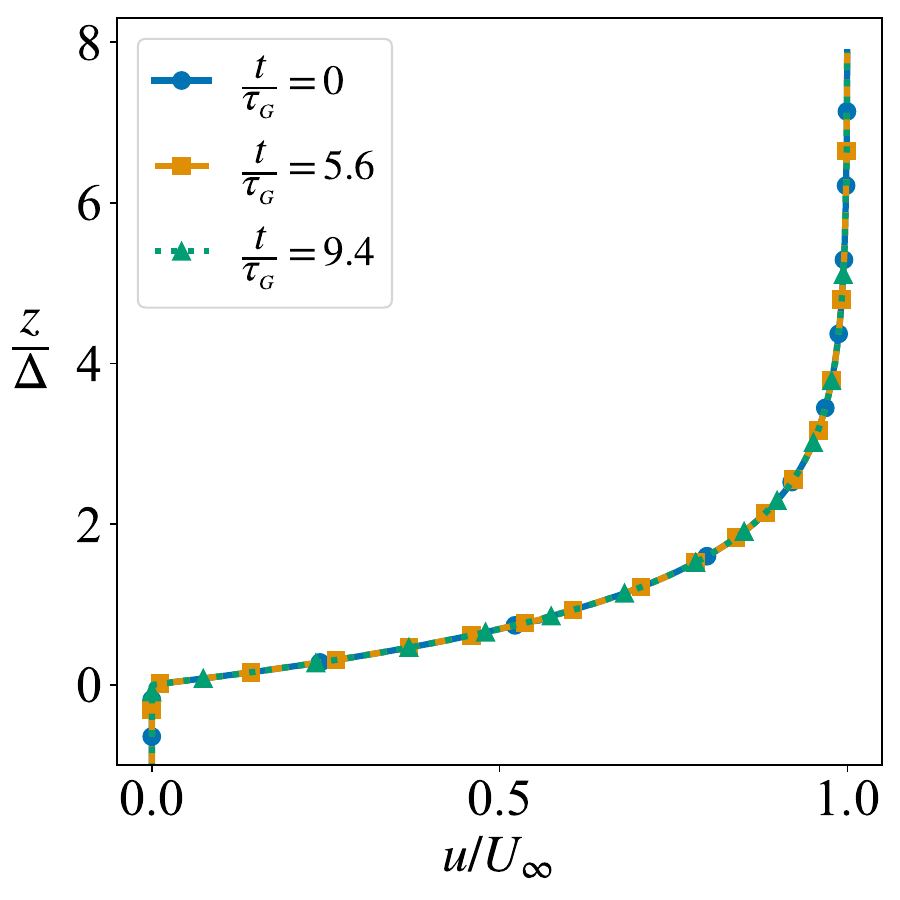}\label{AppC-4}\label{fig31a}}	\subfloat[$\delta=0.5$]{\includegraphics[scale=0.4]{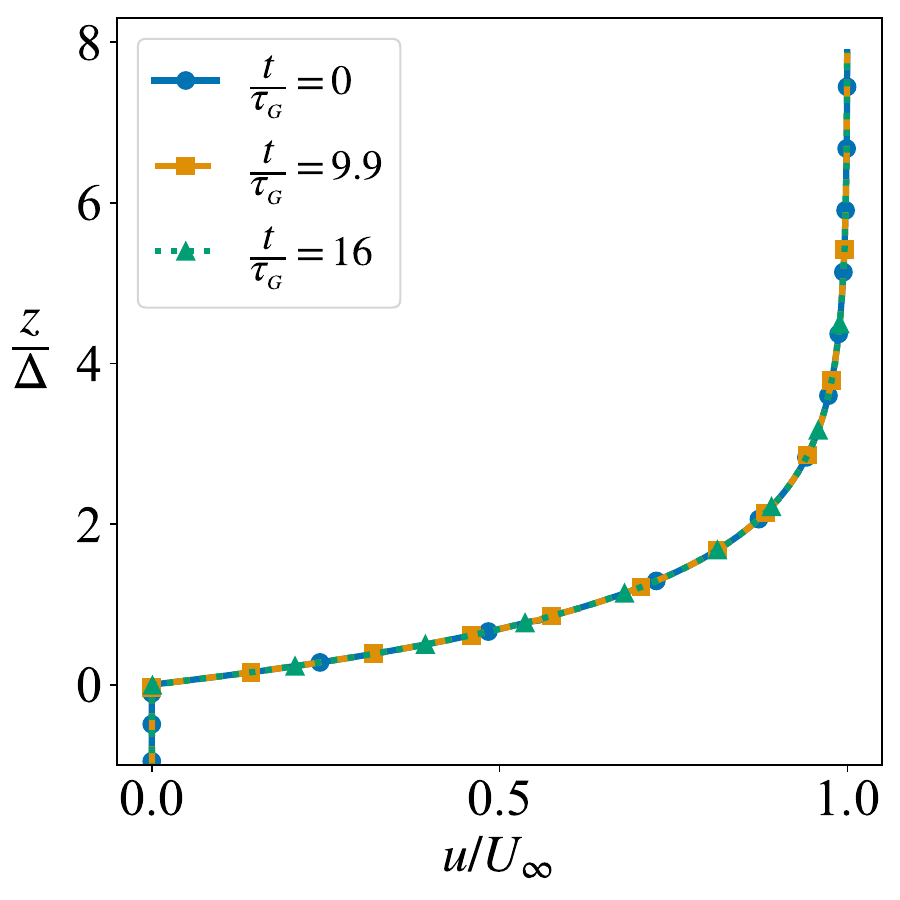}\label{AppC-5}\label{fig31b}}\\
		\subfloat[$\delta=0.9$]{\includegraphics[scale=0.4]{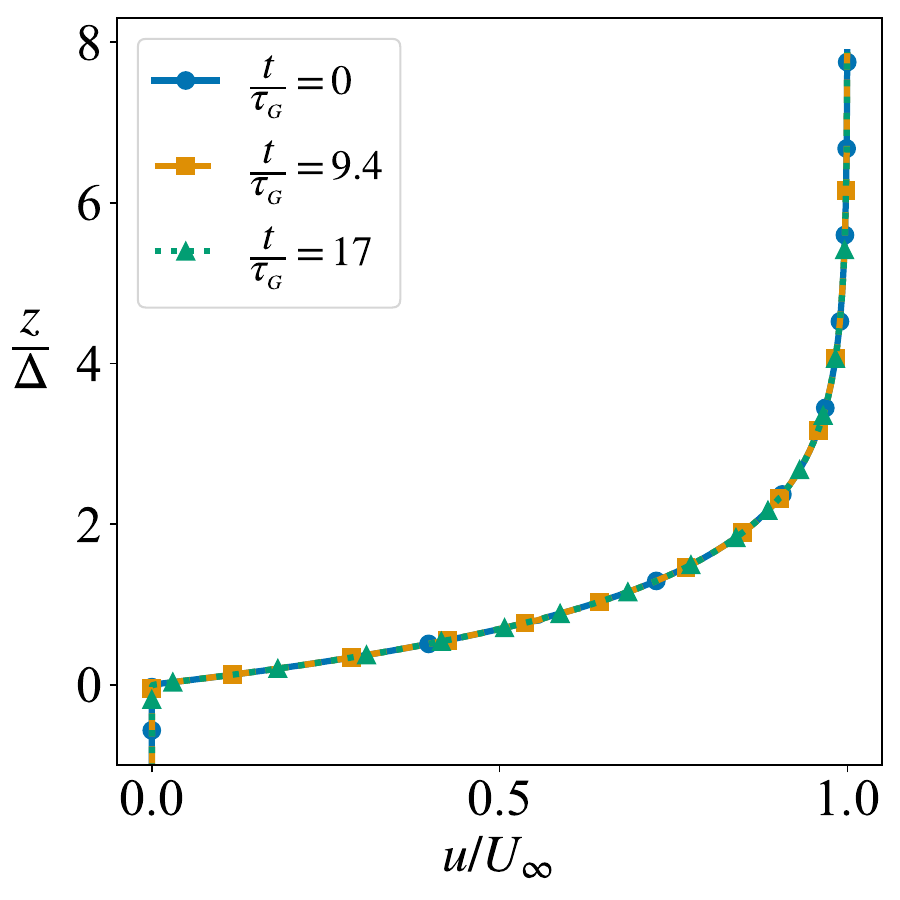}\label{AppC-6}\label{fig31c}}
		\caption{\textcolor{magenta}{Numerical validation of the background state at varying density ratios ($\delta$) and for $\text{Bo}=87.81,\text{Fr}=4$. Panel (a) $\delta=0.01$ (b) $\delta=0.5$ and (c) $\delta=0.9$. In each case, the simulation has been initialised as described in Appendix $C$, without imposing an initial perturbation i.e. only the background state is imposed initially. The horizontal velocity profile is subsequently measured at $x=0$ (horizontal mid-point of domain in figure \ref{fig17a}) at various times ($t>0$). $\tau_{_G} \equiv \dfrac{1}{kc_i}$ represents the time taken for the fastest growing Fourier mode, for each case as computed from linear theory, to grow to $\exp(1)$ times its initial value.}}
		\label{fig31}
	\end{figure}

	\mgt{Figures \ref{fig31a}-\ref{fig31c} depict the velocity profile at three distinct $\delta$, as obtained from the simulations, these having been initialised with aforementioned initial conditions. These profiles were obtained at $x=0$ at varying times (see figure captions). Similar profiles are also observed at other $x$ although this data is not shown here. It is established from figures \ref{fig31a}-\ref{fig31c} that the velocity profiles remain unchanged from their initial distribution, in all our simulations for times well beyond $\tau$. Here $\tau$ is the time taken to cover one wavelength for the fastest growing mode, obtained through linear stability analysis. Within this time window, the interface remains flat in all our simulations, thereby establishing the accuracy of the inviscid solver.}
	
	\bibliographystyle{jfm}%
	\bibliography{jfm}
	\end{document}